\documentclass[11pt]{article}
\usepackage{amsmath,amsthm,amssymb,amsfonts,fullpage,verbatim,bm,graphicx,enumerate,epstopdf,lscape,subfigure,multirow}
\usepackage{algorithm,algorithmicx}
\usepackage{algpseudocode}
\usepackage{boxedminipage,longtable}
\usepackage[dvipsnames,usenames]{color}
\usepackage[pdftex,letterpaper=true,colorlinks=true,pdfpagemode=none,urlcolor=blue,linkcolor=blue, 
   citecolor=blue,pdfstartview=FitH,plainpages=true]{hyperref}
\usepackage{tikz,verbatim,bibentry}
\usepackage{bm,mathrsfs,cite,mathabx}
\usepackage[comma,authoryear]{natbib}

\newtheorem{lemma}{Lemma}
\newtheorem{theorem}{Theorem}
\newtheorem{proposition}{Proposition}

\theoremstyle{definition}
\newtheorem{rmk}{Remark}

\newtheorem{assumption}{Assumption}

\newcommand{\utwi}[1]{\mbox{\boldmath $ #1$}}
\newcommand{\ba}{{\utwi{a}}}
\newcommand{\bb}{{\utwi{b}}}

\newcommand{\bff}{{\utwi{f}}}
\newcommand{\bg}{{\utwi{g}}}

\newcommand{\bu}{{\utwi{u}}}

\newcommand{\bz}{{\utwi{z}}}
\newcommand{\bA}{{\utwi{A}}}
\newcommand{\bB}{{\utwi{B}}}

\newcommand{\bE}{{\utwi{E}}}
\newcommand{\bF}{{\utwi{F}}}

\newcommand{\bI}{{\utwi{I}}}

\newcommand{\bQ}{{\utwi{Q}}}
\newcommand{\bR}{{\utwi{R}}}

\newcommand{\bU}{{\utwi{U}}}
\newcommand{\bV}{{\utwi{V}}}

\newcommand{\bX}{{\utwi{X}}}

\newcommand{\bLambda}{{\utwi{\mathnormal\Lambda}}}

\newcommand{\bSigma}{{\utwi{\mathnormal\Sigma}}}

\newcommand{\bPsi}{{\utwi{\mathnormal\Psi}}}

\newcommand{\cA}{{\cal A}}
\newcommand{\cB}{{\cal B}}

\newcommand{\cE}{{\cal E}}
\newcommand{\cF}{{\cal F}}

\newcommand{\cR}{{\cal R}}

\newcommand{\cX}{{\cal X}}
\newcommand{\cY}{{\cal Y}}
\newcommand{\cZ}{{\cal Z}}

\newcommand{\bel}{\begin{eqnarray}\label}
\newcommand{\eel}{\end{eqnarray}}
\newcommand{\bes}{\begin{eqnarray*}}
\newcommand{\ees}{\end{eqnarray*}}

\def\mat{\hbox{\rm mat}}
\def\vec{\hbox{\rm vec}}

\def\I{\mathrm{I}}
\def\II{\mathrm{II}}

\def\E{\mathbb{E}}
\def\P{\mathbb{P}}
\def\RR{\mathbb{R}}
\def\R{\mathbb{R}}

\DeclareMathOperator*{\argmin}{\arg\min}

\def\ideal{(\text{\footnotesize ideal})}
\def\vec{\hbox{\rm vec}}
\def\cpca{{\rm\tiny cpca}}
\def\iso{{\rm\tiny iso}}

\begin{document}

\title{\textbf{CP Factor Model for Dynamic Tensors}\thanks{\footnotesize{ Yuefeng Han is Assistant Professor, Department of Applied and Computational Mathematics and Statistics, University of Notre Dame, Notre Dame, IN 46556. Email: yuefeng.han@nd.edu.
    Dan Yang is Associate Professor, Business School, The University of Hong Kong, Hong Kong. Email: dyanghku@hku.hk.
    Cun-Hui Zhang is Professor, Department of Statistics, Rutgers University, Piscataway, NJ 08854. E-mail:
    czhang@stat.rutgers.edu.  Rong Chen is Professor, Department of
    Statistics, Rutgers University, Piscataway, NJ 08854. E-mail:
    rongchen@stat.rutgers.edu.
    Rong Chen is the corresponding author.
    Han's research was supported in part by National Science Foundation grant IIS-1741390. Yang's research was supported in part by NSF grant IIS-1741390, Hong Kong grant GRF 17301620, Hong Kong grant CRF C7162-20GF and Shenzhen grant SZRI2023-TBRF-03. Zhang's research was supported in part by NSF grants DMS-1721495, IIS-1741390, CCF-1934924, DMS-2052949 and DMS-2210850. Chen's research was supported in part by National Science Foundation grants DMS-1737857, IIS-1741390, CCF-1934924, DMS-2027855 and DMS-2319260. }} }

\author{Yuefeng Han$^1$, Dan Yang$^2$, Cun-Hui Zhang$^3$ and Rong Chen$^3$}

\date{$^1$Department of Applied and Computational Mathematics and Statistics, University of Notre Dame, Notre
Dame, IN 46556, USA \\
$^2$Business School, The University of Hong Kong, Hong Kong\\
$^3$Department of Statistics, Rutgers University, Piscataway, NJ 08854, USA }

\maketitle

\begin{abstract}
Observations in various applications are frequently represented as a time series of multidimensional arrays, called tensor time series, preserving the inherent multidimensional structure. In this paper, we present a factor model approach, in a form similar to tensor CP decomposition, to the analysis of high-dimensional dynamic tensor time series. As the loading vectors are uniquely defined but not necessarily orthogonal, it is significantly different from the existing tensor factor models based on Tucker-type tensor decomposition. The model structure allows for a set of uncorrelated one-dimensional latent dynamic factor processes, making it much more convenient to study the underlying dynamics of the time series. A new high order projection estimator is proposed for such a factor model, utilizing the special structure and the idea of
the higher order orthogonal iteration procedures commonly used in Tucker-type tensor factor model and general tensor CP decomposition procedures. Theoretical investigation provides statistical error bounds for the proposed methods, which shows the significant advantage of utilizing the special model structure. Simulation study is conducted to further demonstrate the finite sample properties of the estimators. Real data application is used to illustrate the model and its interpretations.
\end{abstract}


\section{Introduction} \label{section:introduction}

In recent years, information technology has made tensors or high-order arrays observations routinely available in applications. For example, 
such data arises naturally from genomics \citep{alter2005, omberg2007}, neuroimaging analysis \citep{zhou2013tensor, sun2017store}, recommender systems \citep{bi2018}, computer vision \citep{liu2012}, community detection
\citep{anandkumar2014community}, longitudinal data analysis \citep{hoff2015multilinear}, among others. Most of the developed tensor-based methods were designed for independent and identically distributed (i.i.d.) tensor data or tensor data with i.i.d. noise.

On the other hand, in many applications, the tensors are observed over time, and hence form a tensor-valued time series. For example, the monthly import export volumes of multi-categories of products (e.g. Chemical, Food, Machinery and Electronic, and Footwear and Headwear) among countries naturally form a dynamic sequence of 3-way tensor-variates, each of which representing a weighted directional transportation network. Another example is functional MRI, which typically consists hundreds of thousands of voxels observed over time. 
A sequence of 2-D or 3-D images can also be modeled as matrix or tensor time series to preserve temporal structure.
Development of statistical methods for analyzing such large scale tensor-valued time series is still in its infancy.

In many 
settings, although
the observed tensors are of high order and high dimension, there is often hidden low-rank
structures in the tensors that can be exploited to facilitate the data analysis. Such a low-rank condition provides convenient decomposable structures and has been widely used in tensor data analysis. Two common choices of low-rank tensor structures are CANDECOMP/PARAFAC (CP) structure and multilinear/Tucker structure, and each of them has their respective benefits; see the survey in \cite{kolda2009tensor}.

In dynamic data, the low-rank structures are often realized through factor models, one of the most effective and popular dimension reduction tools. Over the past few decades, there has been a large body of literature in the statistics and econometrics communities on factor models for vector time series. An incomplete list of the publications includes
\cite{chamberlain1983, bai2002, stock2002, bai2003, fan2011, fan2013, forni2000, forni2004generalized, forni2005generalized, fan2016, pena1987identifying, pan2008, lam2011, lam2012}. Recently, the factor model approach has been developed for analyzing high dimensional dynamic tensor time series \citep{wang2019, chen2019constrained, chen2023statistical, chen2024semi, chen2022factor, han2020iterative, han2022rank, chang2023modelling}. These existing works utilize the Tucker low-rank structure in formulating the factor models.
Such Tucker type tensor factor model is also closely related to separable factor analysis in \cite{fosdick2014separable} under the array Normal distribution of \cite{hoff2011separable}.

In this paper, we investigate a tensor factor model with a CP type low-rank structure, called TFM-cp. Specifically, let $\cX_t$ be an order $K$
tensor of dimensions $d_1\times d_2\times\ldots\times d_K$. We assume
\begin{equation}\label{model}
\cX_t=\sum_{i=1}^r w_if_{it}\ba_{i1}\otimes\ba_{i2}\otimes\cdots\otimes\ba_{iK}+\cE_t, \:\:\: t=1,\ldots, T,
\end{equation}
where $\otimes$ denotes tensor product, $w_i>0$ represents the signal strength, $\ba_{ik}$, $i=1,\ldots, r$, are unit vectors of dimension ${d_k}$, with $\|\ba_{ik}\|_2=1$,
$\cE_t$ is a noise tensor of the same dimension as $\cX_t$, and \{$f_{it}$, $i=1,\ldots,r$\} is a set of uncorrelated univariate latent factor processes. That is, the signal part of the observed tensor at time $t$ is a linear combination of $r$ rank-one tensors, $w_i\ba_{i1}\otimes\ba_{i2}\otimes\cdots\otimes\ba_{iK}$.
These rank-one tensors are fixed and do not change over time. Here, $\{\ba_{ik}, 1\le i\le r, 1\le k \le K\}$ are called loading vectors and the loading vectors for each mode, $\{\ba_{ik}, 1\le i\le r\}$, are not necessarily orthogonal. The dynamics of the tensor time series are driven by the $r$ univariate latent processes $f_{it}$. By stacking the fibers of the tensor $\cX_t$ into a vector, the TFM-cp can be written as a vector factor model, with $r$ factors and a $d\times r$ (where $d=d_1\ldots d_K$) loading matrix of a special structure induced by the TFM-cp. More detailed discussion of the model is given in Section \ref{section:model}.


A standard approach for dynamic factor model estimation is through the analysis of the covariance or autocovariance of the observed process. The autocovariance of a TFM-cp process in \eqref{model} is also a tensor with a low-rank CP structure. Hence, potentially the estimation of \eqref{model} can be done with a tensor CP decomposition procedure. However, tensor CP decomposition is well known to be a notoriously challenging problem as it is in general NP hard to compute and the CP rank is not lower semi-continuous \citep{haastad1990tensor, kolda2009tensor, hillar2013most}. There are a number of works on tensor CP decomposition, which is often called tensor principal component analysis (PCA) in the literature, including alternating least squares \citep{comon2009tensor}, robust tensor power methods with orthogonal components \citep{anandkumar2014tensor}, tensor unfolding approaches \citep{montanari2014statistical, wang2017tensor}, rank-one alternating least squares \citep{anandkumar2014guaranteed, sun2017provable}, and simultaneous matrix diagonalization \citep{kuleshov2015tensor}.
See also \cite{zhou2013tensor,wangmy2017tensor,hao2020sparse, wang2020learning, auddy2023perturbation, han2023guaranteed}, among others.
Although these methods can be used directly to obtain the low-rank CP components of the autocovariance tensors, they have been designed for general tensors and do not utilize the special structure embedded in the TFM-cp.

In this paper, we develop a new estimation procedure, named as {\bf H}igh-{\bf O}rder {\bf P}rojection {\bf E}stimators (HOPE), for TFM-cp in \eqref{model}.
The procedure includes a warm-start initialization using a newly developed composite principal component analysis (cPCA), and an iterative simultaneous orthogonalization scheme to refine the estimator. The procedure is designed to take the advantage of the special structure of TFM-cp whose autocovariance tensor has a specific CP structure with components close to being orthogonal
and of a high-order coherence in a multiplicative form.
The proposed cPCA takes advantage of this feature
so the initialization is better than using random projection initialization often used in generic CP decomposition algorithms.
The refinement step makes use of the multiplicative coherence again and is better than the alternating least squares, the iterative projection algorithm \citep{han2020iterative}, and other forms of the high order orthogonal iteration (HOOI) \citep{de2000, liu2014, zhang2018tensor}.
Our theoretical analysis provides details of these improvements.

In the theoretical analysis, we establish statistical upper bounds on the estimation errors of the factor loading vectors for the proposed algorithms. 
The cPCA yields useful and good initial estimators with less restrictive conditions,
and the iterative algorithm provides faster statistical error rates under weaker conditions than the generic CP decomposition algorithms.
For cPCA, the number of factors $r$ can increase with the dimensions of the tensor time series and is allowed to be larger than $\max_k d_k$. 
We also derive the statistical guarantees of the iterative algorithm under the settings where the tensor is (sufficiently) undercomplete ($r \ll \min_k d_k$). 
It is worth noting that the iterative refinement algorithm has much sharper upper bounds for the statistical error than the cPCA initial estimators.

The TFM-cp in \eqref{model} can also be written as a tensor factor model with a Tucker form (TFM-tucker) of a special structure. 
See \eqref{eq:tucker-model} for the definition of TFM-tucker and Remark \ref{rmk:TFMcompare} below for the comparison between TFM-cp and TFM-tucker from the perspectives of modeling assumptions and interpretations. Potentially, the iterative estimation procedures designed for TFM-tucker can also be used here \citep{han2020iterative}, ignoring the special TFM-cp structure. However, HOPE {has lower computational complexity per iteration}, requires less restrictive conditions and exhibits faster convergence rate, by fully utilizing the structure of TFM-cp. See Remark \ref{rmk:TFMcompare2} for further discussion. They also share the nice properties that the increase in either the dimensions $d_1,\ldots, d_k$, or the sample size can improve the estimation of the factor loading vectors or spaces.

The rest of the paper is organized as follows. After a brief introduction of the basic notations and preliminaries of tensor analysis in Section \ref{section:notation}, we introduce a tensor factor model with CP low-rank structure in Section \ref{section:model}. The estimation procedures of the factors and the loading vectors are presented in Section \ref{section:estimation}. Section \ref{section:theories} investigates the theoretical properties of the proposed methods. Section \ref{section:simulation} develops some alternative algorithms to tensor factor models, which extend existing popular CP methods to the auto-covariance tensors with cPCA as initialization, and provides some simulation studies to demonstrate the numerical performance of all the estimation procedures. Section \ref{section:application} illustrates the model and its interpretations in real data applications. Section \ref{section:discussion} provides a short concluding remark. 
All technical details and more simulation results are relegated to the supplementary materials.

\subsection{Notations and preliminaries} \label{section:notation}

The following basic notations and preliminaries will be used throughout the paper.
Define $\|x\|_q = (x_1^q+...+x_p^q)^{1/q}$, $q\ge 1$, for any vector $x=(x_1,...,x_p)^\top$.
The matrix 
spectral norm is denoted as
$$\|A\|_{\rm S}=  \max_{\|x\|_2=1,\|y\|_2= 1} \|x^\top A y\|_2.$$
For two sequences of real numbers $\{a_n\}$ and $\{b_n\}$, write $a_n=O(b_n)$ (resp. $a_n\asymp b_n$) if there exists a constant $C$ such that $|a_n|\leq C |b_n|$ (resp. $1/C \leq a_n/b_n\leq C$) holds for all sufficiently large $n$, and write $a_n=o(b_n)$ if $\lim_{n\to\infty} a_n/b_n =0$. Write $a_n\lesssim b_n$ (resp. $a_n\gtrsim b_n$) if there exists a constant $C$ such that $a_n\le Cb_n$ (resp. $a_n\ge Cb_n$).


For any two tensors $\cA\in\RR^{m_1\times m_2\times \cdots \times m_K}, \cB\in \RR^{r_1\times r_2\times \cdots \times r_N}$, denote the tensor product $\otimes$ as $\cA\otimes \cB\in \RR^{m_1\times \cdots \times m_K \times r_1\times \cdots \times r_N}$, such that
$$(\cA\otimes\cB)_{i_1,...,i_K,j_1,...,j_N}=(\cA)_{i_1,...,i_K}(\cB)_{j_1,...,j_N} .$$
The $k$-mode product of $\cA\in\RR^{r_1\times r_2\times \cdots \times r_K}$ with a matrix $U\in\RR^{m_k\times r_k}$ is an order $K$-tensor of size $r_1\times \cdots \times r_{k-1} \times m_k\times r_{k+1} \times \cdots \times r_K$ and will be denoted as $\cA\times_k U$, such that
$$ (\cA\times_k U)_{i_1,...,i_{k-1},j,i_{k+1},...,i_K}=\sum_{i_k=1}^{r_k} \cA_{i_1,i_2,...,i_K} U_{j,i_k}.  $$

Given $\cA\in \R^{m_1\times\cdots\times m_K}$ and $m=\prod_{j=1}^K m_j$,
let ${\rm{vec}}(\cA)\in \R^m$ be vectorization of the matrix/tensor $\cA$,
$\mat_k(\cA)\in \R^{m_k\times(m/m_k)}$ the mode-$k$ matrix unfolding of $\cA$, and 
$\mat_k(\vec(\cA))=\mat_k(\cA)$.



\section{A tensor factor model with a CP low rank structure} \label{section:model}
Again, we specifically consider the following tensor factor model with CP low-rank structure (TFM-cp) for observations $\cX_t\in\R^{d_1\times \cdots \times d_K}$, $1\le t\le T$,
\begin{equation*} 
\cX_t=\sum_{i=1}^r w_if_{it}\ba_{i1}\otimes\ba_{i2}\otimes\cdots\otimes\ba_{iK}+\cE_t,
\end{equation*}
where $f_{it}$ is the unobserved latent factor process and $\ba_{ik}$ are the fixed unknown factor loading vectors. We assume without loss of generality, $\E f_{it}^2=1$, $\|\ba_{ik}\|_2=1$, for all $1\le i\le r$ and $1\le k\le K$. Then, all the signal strengths are contained in $w_i$.
A key assumption of TFM-cp is that the factor process $f_{it}$ is assumed to be 
uncorrelated across different factor processes, e.g., $\E f_{it-h}f_{jt}=0$ for $i\neq j$ and $h\ge 1$.
In addition, we assume that the noise tensor $\cE_t$ are uncorrelated (white) across time, but with an arbitrary contemporary covariance structure, following \cite{lam2012}, \cite{chen2022factor}. In this paper, we consider the case that the order of the tensor $K$ is fixed but the dimensions $d_1,...,d_K\to\infty$ and rank $r$ can be fixed or diverging.


\begin{rmk}
By incorporating time, we may stack $\cX_t$ into an order-$(K + 1)$ tensor $\cY\in\R^{d_1\times\cdots\times d_K\times T}$, with time $t$ as the $(K + 1)$-th mode, referred to as the time-mode. Subsequently, model \eqref{model} can be reformulated as
\begin{equation} \label{eq:cp-model}
\cY=\sum_{i=1}^r w_i\ba_{i1}\otimes\ba_{i2}\otimes\cdots\otimes\ba_{iK}\otimes \bff_i+\cE,
\end{equation}
where $\bff_i=(f_{i1},...,f_{iT})^\top$. While it is enticing to directly estimate the signal part in \eqref{eq:cp-model} with standard tensor CP decomposition approaches based on the assumed CP structure,
the dynamics and dependencies in the time direction (auto-dependency) are pivotal and warrant a distinct treatment. In our model, the component in the time direction is deemed latent and random. Consequently, it is crucial to examine the unique role of the time-mode and the (auto)-covariance structure in the time direction. The assumptions and interpretations inherent in our model, along with the corresponding estimation procedures and theoretical properties, markedly diverge from those of using the standard CP decompositions.

\end{rmk}

\begin{rmk}
Ignoring the random noise $\cE$, the CP decomposition in \eqref{eq:cp-model} is unique up to scaling and permutation indeterminacy if $\sum_{k=1}^K\cR(\bA_k)+\cR(\bF)\ge 2r+K$, where $\bA_k=(\ba_{1k},...,\ba_{rk}), \bF=(\bff_1,...,\bff_r)$ and {$\cR(\bA) = \max\{s: \text{any $s$ columns of the matrix $\bA$ are linearly independent}\}$.} 
Such a requirement provides a sufficient condition for uniqueness as per \cite{kolda2009tensor}. In the subsequent estimation procedure, we delve into the estimation of the auto-covariance tensor $\bSigma_h$ in \eqref{eq:sigmah} below. The sufficient identifiability condition for the CP decomposition of the auto-covariance tensor becomes $2\sum_{k=1}^K\cR(\bA_k)\ge 2r+2K-1$. This condition is significantly milder compared to the condition necessary to ensure statistical convergence.
\end{rmk}

\begin{rmk}[{\bf Comparison of TFM-cp with a Tucker low-rank structure}] \label{rmk:TFMcompare}
\cite{chen2022factor, chen2023statistical, han2020iterative, han2022rank}
studied the following tensor factor models with a Tucker low-rank structure (TFM-tucker):
\begin{equation} \label{eq:tucker-model}
\cX_t= \cF_{t}\times_1 \bA_{1}\times\cdots\times\bA_{K}+\cE_t,
\end{equation}
where the core tensor $\cF_t\in \R^{r_1\times\cdots \times r_K}$ is the latent factor process in a tensor form, and $\bA_i$'s are $d_i\times r_i$ loading matrices. For example, when $K=2$ (matrix time series), the TFM-cp can be rewritten as a TFM-tucker,
\begin{equation}
\bX_t=\bA_1\bF_t\bA_2^\top+\bE_t, \label{eq:tucker2}
\end{equation}
where $\bF_t={\rm diag}(f_{1t},\ldots,f_{rt})$, and
$\bA_1=(\ba_{11},\ldots, \ba_{r1})$ and $\bA_2=(\ba_{12},\ldots, \ba_{r2})$
are matrices 
with the column vectors being $\ba_{ik}$'s. There are four major differences between TFM-tucker and TFM-cp.
First, TFM-tucker suffers from a severe identification problem, as the model remains equivalent if $\cF_t$ is replaced by $\cF_t\times_k \bR$ and $\bA_k$ replaced by $\bA_k\bR^{-1}$ for any invertible $r_k\times r_k$ matrix $\bR$. For the $K=2$ case, $\bX_t=(\bA_1\bR_1^{-1})(\bR_1\bF_t\bR_2^\top)(\bA_2\bR_2^{-1})^\top+\bE_t$ are all equivalent under TFM-tucker.
Such ambiguity makes it difficult to find an `optimal' representation of the model, which often leads to {\it ad hoc} and convenient representations that are difficult to interpret \citep{Bekker1986, Neudecker1990, bai2014identification, bai2015identification}. On the other hand, TFM-cp is uniquely defined up to sign changes, under an ordering of the signal strengths $w_1\ge w_2\ge \ldots\ge w_r$. As a result, the interpretation of the model becomes much easier. Second, although TFM-cp can be rewritten in the form
of \eqref{eq:tucker-model} with a {\it diagonal} core latent tensor consisting of the individual $f_{it}$'s, it is not under a typical Tucker form since TFM-tucker typically adopts the representation that the loading matrices $\bA_k$'s are orthonormal,
due to its identification problem. In TFM-cp, the loading vectors $\{\ba_{ik}, 1\le i\le r\}$ are not necessarily orthogonal vectors. 
In the $K=2$ example in \eqref{eq:tucker2}, if we find rotation matrices $\bR_1$ and $\bR_2$ so that $\bA_1\bR_1^{-1}$ and $\bA_2\bR_2^{-1}$ are orthonormal, then the corresponding core factor process in \eqref{eq:tucker2} becomes 
$\bR_1\bF_t\bR_2^\top$, no longer diagonal and with $r^2$ heavily correlated components, rather than $r$ uncorrelated components. Third, TFM-cp separates the factor processes into a set of univariate time series, which enjoys great advantages over the tensor-valued factor processes
in TFM-tucker. Modelling univeriate time series are much easier and more flexible due to the 
vast repository of linear and nonlinear options. 
Lastly, TFM-cp is often much more parsimonious due to its restrictions, while enjoying great flexibility. Note that TFM-tucker is also a special case of TFM-cp, as it can be written as a sum of $r=r_1\ldots r_K$ rank-one tensors, albeit with many repeated loading vectors. With its condensed formulation, in practical applications, the number of factors $r$ needed under TFM-cp is typically much smaller than the total number $r_1\ldots r_K$ of factors needed in TFM-tucker.
\end{rmk}

\begin{rmk}
There are two different types of factor model assumptions in the literature. One type of factor models assumes that the common factors must have impact on `most' (defined asymptotically) of the time series, but allows the idiosyncratic noise ($\cE_t$) to have weak cross-correlations and weak auto-correlations; see, e.g., \cite{forni2000, bai2002,stock2002, fan2011, fan2013, chen2023statistical}. 
PCA of the sample covariance matrix is typically used to estimate the factor loading space, with various extensions. The other type of factor models assumes that the factors accommodate all dynamics, making the idiosyncratic noise `white' with no auto-correlation, but allows substantial contemporary cross-correlation among the error process; see, e.g., \cite{pena1987identifying,pan2008,lam2011,lam2012,wang2019}. Under such assumptions, PCA is applied to the non-zero lagged autocovariance matrices. In this paper, we adopt the 
latter type of assumptions in our model development.
\end{rmk}

\section{Estimation procedures} \label{section:estimation}
In this section, we focus on the estimation of the factors and loading vectors of model \eqref{model}. The proposed procedure includes two steps: an initialization step using a new composite PCA (cPCA) procedure, presented in Algorithm \ref{algorithm:initial}, and an iterative refinement step using a new iterative simultaneous orthogonalization (ISO) procedure, presented in Algorithm \ref{algorithm:projection}. We call this two-step procedure HOPE ({\bf H}igh-{\bf O}rder {\bf P}rojection {\bf E}stimators) as it repeatedly 
perform high order projections on high order moments of the tensor observations. It utilizes the special structure of the model and leads to higher statistical and computational efficiency, which will be demonstrated later.


For $\cX_t$ following \eqref{model}, the lagged cross-product operator, denoted by $\bSigma_h$, is the $(2K)$-tensor satisfying
\begin{eqnarray}  \label{eq:sigmah}
\bSigma_h & = & \E \left[\sum_{t=h+1}^T \frac{\cX_{t-h}\otimes \cX_t}{T-h} \right] \nonumber \\
& = &  \sum_{i=1}^r \lambda_{i,h} (\ba_{i1}\otimes\ba_{i2}\otimes\cdots\otimes \ba_{iK})^{\otimes2} \in\R^{d_1\times \cdots \times d_K \times d_1 \times\cdots \times d_K},
\label{eq:bsigma}
\end{eqnarray}
for a given $h\ge 1$, where $\lambda_{i,h}=w_i^2\E f_{i,t-h} f_{i,t}$. Note that the tensor $\bSigma_h$ is expressed in a CP-decomposition form with each $\ba_{ik}$ used twice.
Let $\widehat\bSigma_h$ be the sample version of $\bSigma_h$,
\begin{align} \label{eq:hatsigmah}
\widehat\bSigma_h=\sum_{t=h+1}^T \frac{\cX_{t-h}\otimes \cX_t}{T-h} .
\end{align}
When $\cX_t$ is weakly stationary and $\cE_t$ is white noise, a natural approach to estimating the loading vectors is via minimizing the empirical squared loss
\begin{align} \label{problem:ls}
(\ba_{i1},\ba_{i2},...,\ba_{iK}, 1\le i\le r) = \argmin_{\substack{\ba_{i1},\ba_{i2},...,\ba_{iK},1\le i\le r, \\ \|\ba_{i1}\|_2=\ldots=\|\ba_{iK}\|_2 =1 } } \left\| \widehat\bSigma_h -\sum_{i=1}^r \lambda_{i,h} (\ba_{i1}\otimes\ba_{i2}\otimes\cdots\otimes \ba_{iK})^{\otimes2} \right\|_{\rm HS}^2,
\end{align}
where the Hilbert Schmidt norm for a tensor $\cA$ is defined as $ \|\cA\|_{{\rm HS}}=\| \vec(\cA)\|_2.$
In other words, $\ba_{i1}\otimes\ba_{i2}\otimes\cdots\otimes \ba_{iK}$ can be 
estimated by the leading {\it principal component} of the sample auto-covariance tensor $\widehat\bSigma_h$.
However, due to the non-convexity of \eqref{problem:ls} or its variants, a straightforward implementation of many local search algorithms, such as gradient descent and alternating minimization, may easily get trapped into local optima and result in sub-optimal statistical performance.
As shown by \cite{auffinger2013random}, there could be an exponential number of local optima and the great majority of these local optima are far from the best low rank approximation. However, if we start from an appropriate initialization not too far from the global optimum, then a local optimum reached may be as good an estimator as the global optimum. A critical task in estimating the factor loading vectors is thus to obtain good initialization.

We develop a warm initialization procedure, the composite PCA (cPCA) procedure. Note that, if we unfold $\bSigma_h$ into a $d\times d$ matrix $\bSigma_h^*$, where $d=d_1\ldots d_K$, then \eqref{eq:bsigma} implies that
\begin{equation}\label{eq:bsigmah*}
\bSigma_h^*=\sum_{i=1}^r \lambda_{i,h}\ba_i\ba_i^\top,
\end{equation}
a sum of $r$ rank-one matrices, each of the form $\ba_i\ba_i^\top$, where $\ba_i = \vec(\otimes_{k=1}^K \ba_{ik})$. This is very close to the principal component decomposition of $\bSigma_h^*$, except that $\ba_i$'s are not necessarily orthogonal in this case. However, the following intuition provides a solid justification of using PCA to obtain an estimate of $\ba_i$. We call this estimator the cPCA estimator.

The accuracy of using the principal components of $\bSigma_h^*$ as the estimate of $\ba_i$ heavily depends on the coherence of the components, defined as $\vartheta = \max_{1\le i < j\le r}| \ba_{i}^\top \ba_{j}|$, the maximum pairwise correlation among the $\ba_i$'s. When the components are orthogonal ($\vartheta=0$), there is no error in using PCA. The main idea of cPCA is to take advantage of the special structure of
TFM-cp, which leads to a multiplicative high-order coherence of the CP components.
In the following, we provide an analysis of $\vartheta$ under TFM-cp.

Let $\bA_k = (\ba_{1k},\ldots,\ba_{rk})\in \R^{d_k\times r}$ be the matrix with $\ba_{ik}$ as its columns, and $\bA_k^\top \bA_k = (\sigma_{ij,k})_{r\times r}$. As $\sigma_{ii,k}=\|\ba_{ik}\|_2^2=1$, the correlation among columns of $\bA_k$ can be measured by
\begin{align}\label{corr-k}
\vartheta_k = \max_{1\le i < j\le r}|\sigma_{ij,k}|,\ \
\delta_k = \|\bA_k^\top \bA_k - I_{r}\|_{\rm S},\ \
\eta_{jk} = \Big( \sum_{i\in [r]\setminus \{j\}}\sigma_{ij,k}^2 \Big)^{1/2}.
\end{align}
Similarly we use
\begin{align}\label{corr-all}
\vartheta = \max_{1\le i < j\le r}| \ba_{i}^\top \ba_{j}|,\quad
\delta = \| \bA^\top \bA - I_{r}\|_{\rm S},
\end{align}
to measure the correlation of the matrix $\bA = (\ba_1,\ldots,\ba_r)\in \R^{d\times r}$ with $\ba_i = \vec(\otimes_{k=1}^K \ba_{ik})$ and
$d=\prod_{k=1}^K d_k$. It can be seen that the coherence $\vartheta$ has the bound $\vartheta\le \prod_{k=1}^K\vartheta_k\le \vartheta_{\max}^K$, due to $\ba_{i}^\top \ba_{j} = \prod_{k=1}^K \ba_{ik}^\top \ba_{jk} =\prod_{k=1}^K \sigma_{ij,k}$.
The spectrum norm $\delta$ is also bounded by the multiplicative of correlation measures in \eqref{corr-k}. More specifically, we have the following
proposition.

\begin{proposition}\label{prop:delta}
Define $\mu_* = \max_j \min_{k_1,k_2}
\max_{i\neq j} \prod_{k\neq k_1,k\neq k_2,k\in [K]}
\sqrt{r}|\sigma_{ij,k}|/\eta_{jk}\in~[1, r^{K/2-1}]$
as the (leave-two-out) mutual coherence of $\bA_1,\ldots, \bA_K$.
Then, $\delta\le\min_{1\le k\le K}\delta_k$  and
\begin{align}\label{eq:prop}
\delta &\le (r-1)\vartheta, \mbox{\ \ and \ \ } \vartheta \le \prod_{k=1}^K \vartheta_k \le \vartheta_{\max}^K,  \\
\delta &\le \mu_* r^{1-K/2}\max_{j\le r}\prod_{k=1}^K\eta_{jk}
\le \mu_* r^{1-K/2}\prod_{k=1}^K\delta_k.
\end{align}
\end{proposition}
When (most of) the quantities in \eqref{corr-k} are small, the products in \eqref{eq:prop} would be very small so that the $\ba_i$'s are nearly orthogonal. For example, if $(\ba_{11},\ba_{21})$ and $(\ba_{12},\ba_{22})$ both have i.i.d. bi-variate random rows with correlation coefficients $\rho_1$ and $\rho_2$, and independent, then the population correlation coefficient of $\vec(\ba_{11}\otimes\ba_{12})$ and $\vec(\ba_{21}\otimes\ba_{22})$ is $\rho_1\rho_2$, though the variation of the sample correlation coefficient depends on the length of the $\ba_{ik}$'s.

\begin{rmk}
Let polylog denote the polynomial of the logarithm. The incoherence condition such as $\vartheta_{\max} \lesssim \text{ploylog} (d_{\min})/\sqrt{d_{\min}}$ is commonly imposed in the literature for generic CP decomposition; see e.g. \cite{anandkumar2014guaranteed, anandkumar2014tensor, sun2017provable, hao2020sparse}. Proposition \ref{prop:delta} establishes a connection between $\delta$, $\delta_k$ and
the $\vartheta_k$ in the same framework of incoherence considerations.
The parameters $\delta_k$ and $\delta$ quantify the non-orthogonality of the factor loading vectors, and play a key role in our theoretical analysis, as the performance bound of cPCA estimators involves $\delta$. Differently from the existing literature depending on $\vartheta_{\max}$, the cPCA exploits $\delta$ or the much smaller $\vartheta$ (comparing to $\vartheta_{\max}$), thus has better properties when $K\ge 2$.
Note that the idea of using tensor unfolding to enhance incoherence can be traced back to \cite{huang2015provable, jain2014learning, allman2009identifiability}, though their incoherence measure is slightly different from ours. The most notable advances from these studies is that we establish a non-asymptotic bound for the estimated loading vectors in the presence of noise (c.f. Theorem \ref{thm:initial}).
\end{rmk}

The pseudo-code of cPCA is provided in Algorithm \ref{algorithm:initial}. Though $\bSigma_{h}^*$ is symmetric, its sample version $\widehat\bSigma_{h}^*$ in general is not.
We use $(\widehat\bSigma_{h}^*+\widehat\bSigma_{h}^{*\top})/2$
to ensure symmetry and reduce the noise.
The cPCA produces definitive initialization vectors up to the sign change.

\begin{algorithm}[htbp]
\caption{Initialization based on composite PCA (cPCA)}\label{algorithm:initial}
\begin{algorithmic}[1]
\Require The observations $\cX_t\in\RR^{d_1\times\cdots\times d_K}$, $t=1,...,T$, the number of factors $r$, and the time lag $h$.

\State Evaluate $\widehat\bSigma_h$ in \eqref{eq:hatsigmah}, and unfold it to
$d\times d$ matrix $\widehat\bSigma_{h}^*$.

\State Obtain $\widehat\bu_i, 1\le i\le r$, 
the top $r$ eigenvectors of $(\widehat\bSigma_{h}^*+\widehat\bSigma_{h}^{*\top})/2$.


\State Compute $\widehat\ba_{ik}^{\cpca}$ as the top left singular vector of $\mat_k(\widehat \bu_i)\in\R^{d_k\times (d/d_k)}$, for all $1\le k\le K$.

\Ensure
$\:\:\: \widehat\ba_{ik}^{\cpca}, \quad i=1,...,r, \ \ k=1,...,K.$
\end{algorithmic}
\end{algorithm}



After obtaining a warm start via cPCA (Algorithm \ref{algorithm:initial}), we 
engage an iterative simultaneous orthogonalization (ISO) algorithm (Algorithm \ref{algorithm:projection}) to refine the solution of $\ba_{ik}$ and obtain estimations of the factor process $f_{it}$ and the signal strength $w_i$. Algorithm \ref{algorithm:projection} can be viewed as an extension of HOOI \citep{de2000,zhang2018tensor} and the iterative projection algorithm in \cite{han2020iterative} to undercomplete ($r < d_{\min}$) and non-orthogonal CP decompositions. It is motivated by the following observation. Define $\bA_k=(\ba_{1k},\ldots,\ba_{rk})$ and $\bB_k = \bA_k(\bA_k^{\top} \bA_k)^{-1} = (\bb_{1k},...,\bb_{rk}) \in\R^{d_k\times r}$. Let
\begin{align}
\cZ_{t,ik}=&\; \cX_t \times_1 \bb_{i1}^\top \times_2 \cdots \times_{k-1} \bb_{i,k-1}^\top \times_{k+1} \bb_{i,k+1}^\top \times_{k+2}\cdots\times_K \bb_{iK}^\top , \label{eq:z}\\
\cE_{t,ik}^*=&\; \cE_t \times_1 \bb_{i1}^\top \times_2 \cdots \times_{k-1} \bb_{i,k-1}^\top \times_{k+1} \bb_{i,k+1}^\top \times_{k+2}\cdots\times_K \bb_{iK}^\top . \label{eq:e*}
\end{align}
Since $\ba_{jk}^\top\bb_{ik}=I_{\{i=j\}}$, model \eqref{model} implies that
\begin{align} \label{eq:cp-ideal}
\cZ_{t,ik}=w_if_{it} \ba_{ik} + \cE_{t,ik}^*.
\end{align}
Here $\cZ_{t,ik}$ is a vector, and \eqref{eq:cp-ideal} is in a factor model form with a univariate factor. The estimation of $\ba_{ik}$ can be done easily and much more accurately than dealing with the much larger $\cX_t$.
The operation in \eqref{eq:z} achieves two objectives. First, by multiplying a vector on every mode except the $k$-th mode to $\cX_t$, it reduces the tensor to a vector. It also serves as an averaging operation to reduce the noise variation. Second, as $\bb_{ik}$ is orthogonal to all
$\ba_{jk}$ except $\ba_{ik}$, it is an orthogonal projection operation that
eliminates all $\otimes_{k=1}^K\ba_{jk}$ terms in \eqref{model} except the $i$-th term, resulting in \eqref{eq:cp-ideal}.
If the matrix $\bA_k^{\top} \bA_k$ is not ill-conditioned, \emph{i.e.} $\{\ba_{ik}, 1\le i\le r\}$ are not highly correlated, then $\bB_k$ and all individual $\bb_{jk}$ are well defined and this procedure shall work well.
Under proper conditions on the combined noise tensor $\cE_{t,ik}^*$, estimation of
the loading vectors $\ba_{ik}$ based on $\cZ_{t,ik}$ can be made significantly
more accurate, as the statistical error rate now depends on
$d_k$ rather than $d_1d_2\ldots d_k$. Intuitively, $\bb_{ik}$ can also be viewed as 
a form of (normalized) residuals of $\ba_{ik}$ projected onto the space spanned by $\{\ba_{jk}, j\neq i, 1\le j\le r \}$.

In practice we do not know $\bb_{il}$, for $1\le i\le r$, $1\le l\le K$ and $l\neq k$. Similar to
back-fitting algorithms, we iteratively estimate the loading vector $\ba_{ik}$ at iteration number $m$ based on
\begin{align*}
&\cZ_{t,ik}^{(m)}=\cX_t \times_1 \widehat \bb_{i1}^{(m)\top} \times_2 \cdots \times_{k-1} \widehat \bb_{i,k-1}^{(m)\top} \times_{k+1} \widehat \bb_{i,k+1}^{(m-1)\top} \times_{k+2}\cdots\times_K \widehat \bb_{iK}^{(m-1)\top} ,
\end{align*}
using the estimate $\widehat \bb_{il}^{(m-1)},~ k<l\le K$, obtained in the previous iteration and the estimate $\widehat \bb_{il}^{(m)},~ 1\le l< k$, obtained in the current iteration. As we shall show in the next section, such an iterative procedure leads to a much improved statistical rate in the high dimensional tensor factor model scenarios, as if all $\bb_{il}$, $1\le i\le r$, $1\le l\le K$, $l\neq k$, are known and we indeed observe $\cZ_{t,ik}$ that follows model \eqref{eq:cp-ideal}.
Note that the projection error is
\[
\cZ_{t,ik}^{(m)}-\cZ_{t,ik}
=\sum_{j=1}^r w_{j}f_{j,t}\xi_{ij}^{(m)}\ba_{jk} +\cE_{t,ik}^{*(m)} -
\cE_{t,ik}^*
\]
where
\begin{equation}
\xi_{ij}^{(m)}=
\prod_{\ell =1}^{k-1} [\ba_{j\ell}^\top \widehat \bb_{i\ell}^{(m)}]~\prod_{\ell =k+1}^K [\ba_{j\ell}^\top \widehat \bb_{i\ell}^{(m-1)}] -I_{\{i=j\}},  \label{eq:xi}
\end{equation}
and
$\cE_{t,ik}^{*(m)}$ is that in \eqref{eq:e*} with $\bb_{ik}$
replaced with $\widehat\bb_{ik}^{(m-1)}$ or $\widehat\bb_{ik}^{(m)}$. The multiplicative measure of projection error $|\xi_{ij}^{(m)}|$
decays rapidly since, for $j\neq i$, $\ba_{j\ell}^\top \widehat\bb_{i\ell}^{(m)}$ goes to zero quickly as the iteration $m$ increases, and $\xi_{ij}^{(m)}$ is a product of $K-1$ such terms. In fact, the higher the tensor order $K$ is, the faster the error goes to zero.

\begin{algorithm}[htbp]
\caption{Iterative Simultaneous Orthogonalization (ISO)} \label{algorithm:projection}
\begin{algorithmic}[1]
\Require The observations $\cX_t\in\RR^{d_1\times\cdots\times d_K}$, $t=1,...,T$, the number of factors $r$, the warm-start initial estimates $\widehat\ba_{ik}^{(0)}$, $1\le i\le r$ and $1\le k\le K$, the time lag $h$, the tolerance parameter $\epsilon>0$, and the maximum number of iterations $M$.

\State Compute $\widehat\bB_k^{(0)} = \widehat\bA_k^{(0)}(\widehat\bA_k^{(0)\top}\widehat\bA_k^{(0)})^{-1} = (\widehat\bb_{1k}^{(0)},...,\widehat\bb_{rk}^{(0)})$ with $\widehat\bA_k^{(0)}=(\widehat\ba_{1k}^{(0)},\ldots,\widehat\ba_{rk}^{(0)})\in\R^{d_k\times r}$ for $k=1,\ldots, K$. 
Set $m=0$.
\Repeat 

\State Let $m=m+1$.
\For $k=1$ to $K$.

\For $i=1$ to $r$.


\State Given previous estimates $\widehat\ba_{ik}^{(m-1)}$, calculate
\begin{align*}
&\cZ_{t,ik}^{(m)}=\cX_t \times_1 \widehat \bb_{i1}^{(m)\top} \times_2 \cdots \times_{k-1} \widehat \bb_{i,k-1}^{(m)\top} \times_{k+1} \widehat \bb_{i,k+1}^{(m-1)\top} \times_{k+2}\cdots\times_K \widehat \bb_{iK}^{(m-1)\top} ,
\end{align*}
for 
$t=1,...,T$. Let
\begin{align*}
&\widehat\bSigma_h \left( \cZ_{1:T,ik}^{(m)} \right)=\frac{1}{T-h}\sum_{t=h+1}^T \cZ_{t-h,ik}^{(m)} \otimes\cZ_{t,ik}^{(m)}.
\end{align*}
Compute $\widehat\ba_{ik}^{(m)}$ as the top eigenvector of $\widehat\bSigma_h ( \cZ_{1:T,ik}^{(m)} )/2+\widehat\bSigma_h ( \cZ_{1:T,ik}^{(m)} )^\top/2$. 

\EndFor
\State Compute $\widehat\bB_k^{(m)} = \widehat\bA_k^{(m)}(\widehat\bA_k^{(m)\top}\widehat\bA_k^{(m)})^{-1} = (\widehat\bb_{1k}^{(m)},...,\widehat\bb_{rk}^{(m)})$ with $\widehat\bA_k^{(m)}=(\widehat\ba_{1k}^{(m)},\ldots,\widehat\ba_{rk}^{(m)})$.

\EndFor
\Until $m=M$ or
\begin{equation*}
\max_{1\le i\le r}\max_{1\le k\le K}\| \widehat\ba_{ik}^{(m)} \widehat\ba_{ik}^{(m)\top} - \widehat\ba_{ik}^{(m-1)} \widehat\ba_{ik}^{(m-1)\top} \|_{\rm S}\le \epsilon,
\end{equation*}

\Ensure Estimates
\begin{align*}
\widehat \ba_{ik}^{\iso}&=\widehat \ba_{ik}^{(m)},\quad i=1,...,r,\ \ k=1,...,K, \\
\widehat w_{i}^{\iso} &= \left( T^{-1}\sum_{t=1}^T \left(\cX_t\times_{k=1}^K \widehat \bb_{ik}^{(m)\top}\right)^2 \right)^{1/2}, \quad \quad i=1,...,r, \\
\widehat f_{it}^{\iso}&= \left( \widehat w_{i}^{\iso}  \right)^{-1} \cdot \cX_t\times_{k=1}^K \widehat \bb_{ik}^{(m)\top}, \quad \quad i=1,...,r, \ \ t=1,...,T, \\
\widehat\cX_t^{\iso} &= \sum_{i=1}^r \cX_t\times_{k=1}^K \widehat \bb_{ik}^{(m)\top} \times_{k=1}^K \widehat \ba_{ik}^{(m)} , \quad \quad  t=1,...,T.
\end{align*}

\end{algorithmic}
\end{algorithm}

\begin{rmk}[{\bf Comparison with alternating least square}]
The updates in Algorithm \ref{algorithm:projection} can be viewed as
a variant of the standard alternating least squares procedure. For example, suppose that $\widehat\bb_{ik}^{(m-1)}, 1\le i\le r, 2\le k\le K,$ are fixed. Then the optimization problem to update $\ba_{i1}^{(m)}$ for each $1\le i\le r$ can be rewritten as
\begin{align*}
\arg\min_{\ba_{i1}\in\R^{d_1}} \left\|\widehat\bSigma_h\times_{k=2}^K\widehat\bb_{ik}^{(m-1)} \times_{k=K+2}^{2K}\widehat\bb_{i,k-K}^{(m-1)} -w_i \ba_{i1}\ba_{i1}^\top  \right\|_{\rm F}    .
\end{align*}
This is a least-squares problem. However, the algorithm cannot be viewed as an alternating least square procedure since we do not have an over-arching (least square) objective function such that every iteration is done to minimize the objective function given other components. This is due to the construction and involvement of $\bb_{ik}$ in the algorithm. As a matter of fact, if one uses standard ALS to minimize the objective function in \eqref{problem:ls}, to update mode $k$, it would involve the inverse of the Hadamard product of $\bA_{k'}^\top\bA_{k'}$ for $k'\ne k$. In contrast, due to the nice property of $\cZ_{t,ik}$ (defined based on $\bB_{k'}$ for $k'\ne k$), we only need to compute the inverse of $\bA_{k'}^\top\bA_{k'}$ for each $k'\ne k$, not their Hadamard product.
\end{rmk}

\begin{rmk}[{\bf The role of $h$}] \ \
In Algorithm \ref{algorithm:initial}, we use a fixed $h\ge 1$. Let $\widehat\lambda_{1,h}\ge \widehat\lambda_{2,h}\ge ... \ge \widehat \lambda_{d,h}$ be the eigenvalues of $\widetilde \bSigma_h^*:= (\widehat\bSigma_{h}^*+\widehat\bSigma_{h}^{*\top})/2$. In practice, we may select $h$ to maximize the fraction of the explained variance $\sum_{i=1}^r \widehat\lambda_{i,h}^2/\sum_{i=1}^d \widehat\lambda_{i,h}^2$ under different lag values $1\le h\le h_0$, given some pre-specified maximum allowed lag $h_0$. 
Step 2 in Algorithm \ref{algorithm:initial} can be improved by accumulating information from different time lags. For example, let $\widehat \bU^{(0)}\in R^{d\times r}$ be a matrix with its columns $\widehat \bu_i$'s being the top $r$ eigenvectors of $\widetilde \bSigma_h^*$. 
With such $\widehat \bU^{(0)}$ as the initialization, we may iteratively refine $\widehat \bU^{(m)}$ to be the top $r$ eigenvectors of
$\sum_{h=1}^{h_0}  \widetilde\bSigma_h^* \widehat \bU^{(m-1)} \widehat \bU^{(m-1)\top} \widetilde\bSigma_h^* . $
\end{rmk}

\begin{rmk}[{\bf Condition number of $\widehat\bA_k^{(m)\top}\widehat\bA_k^{(m)}$}] \label{rmk:cond_num}
Our theoretical analysis assumes that the condition number of the matrix $\bA_k^{\top} \bA_k$ is bounded. However, in practice, the condition number of $\widehat\bA_k^{(m)\top} \widehat\bA_k^{(m)}$ in Algorithm \ref{algorithm:projection} may be very large, especially when $m=0$. We suggest a simple regularized strategy. Define the eigen decomposition $\widehat\bA_k^{(m)\top} \widehat\bA_k^{(m)}=\bV_k^{(m)} \bLambda_k^{(m)} \bV_k^{(m)\top}$.
For all eigenvalues in $\bLambda_k^{(m)}$ that are smaller than a numeric constant $c$ (e.g., $c=0.1$), we set them to $c$. Denote the resulting matrix as $\widetilde\bLambda_k^{(m)}$ and get
the corresponding $\widehat\bB_{k}^{(m)}$ 
by $\widehat\bA_k^{(m)\top} (\bV_k^{(m)} \widetilde \bLambda_k^{(m)} \bV_k^{(m)\top})^{-1}$. Many alternative empirical methods can also be applied to bound the condition number.
\end{rmk}

\begin{rmk}
Algorithm \ref{algorithm:initial} requires that $\delta<1$ in order to obtain reasonable estimates. And it can accommodate the case that
$r\ge d_{\max}$.
In contrast, Algorithm \ref{algorithm:projection} needs 
stronger conditions that $\delta_k<1$ and $r \le d_{\min}$ to rule out the possibility of co-linearity, as $\bA_k^\top \bA_k$ needs to be invertible. It may not hold under certain situations. For example, $\ba_{1k}=\ba_{2k}$ would lead
to an ill-conditioned $\bA_k^{\top} \bA_k$. In such cases, the incoherence condition commonly required in the literature, e.g., $\vartheta_{\max}\ll 1$, is also violated.
It is possible to extend our approach to a more sophisticated projection scheme so the conditions can be weakened. As it requires more sophisticated analysis both on the methodology and on the theory, we do not purse this direction in this paper.
\end{rmk}


\begin{rmk}
As mentioned before, $\ba_{i1}\otimes\ba_{i2}\otimes\cdots\otimes \ba_{iK}$ can be regarded as the {\it principal component} of the auto-covariance tensor $\bSigma_h$. Hence, our HOPE estimators (Algorithms \ref{algorithm:initial} and \ref{algorithm:projection} together) can also be characterized as a procedure of {\it principal component analysis} for order $2K$ auto-covariance tensor, albeit with a special structure in \eqref{eq:bsigma}.
\end{rmk}

\begin{rmk}[\bf{The number of factors}] \label{rmk:rank}
Here the estimators are constructed with given rank $r$, though in the theoretical analysis it is allowed to diverge. Determining the number of factors in a data-driven way has been an important research topic in the factor model literature. \cite{bai2002,bai2007,hallin2007} proposed consistent estimators in the vector factor models based on the information criteria approach. \cite{lam2012,ahn2013} developed an alternative approach to study the ratio of each pair of adjacent eigenvalues. Recently, \cite{han2022rank} established a class of rank determination approaches for the factor models with Tucker low-rank structure, based on both the information criterion and the eigen-ratio criterion. Those procedures can be extended to TFM-cp.
\end{rmk}

\section{Theoretical Properties} \label{section:theories}
In this section, we shall investigate the statistical properties of the proposed algorithms described in the last section. Our theories provide theoretical guarantees for consistency and present statistical error rates in the estimation of the factor loading vectors $\ba_{ik}$, $1\le i\le r, 1\le k\le K$, under proper regularity conditions. As the loading vector $\ba_{ik}$ is identifiable only up to the sign change, we use
\begin{align*}
\|\widehat\ba_{ik}\widehat\ba_{ik}^\top  -\ba_{ik}\ba_{ik}^\top \|_{\rm S}=\sqrt{1-(\widehat\ba_{ik}^\top \ba_{ik})^2 } = \sup_{\bz\perp \ba_{ik}}|\bz^\top \widehat \ba_{ik}|
\end{align*}
to measure the distance between $\widehat\ba_{ik}$ and $\ba_{ik}$. 

Recall
$\bSigma_{h} = \E\widehat\bSigma_{h} =\sum_{i=1}^r \lambda_{i,h} (\ba_{i1}\otimes\ba_{i2}\otimes\cdots\otimes \ba_{iK})^{\otimes2},
$
as in \eqref{eq:bsigma} and $\lambda_{i,h}= w_i^2\E f_{i,t-h}f_{i,t}$.
We will also continue to use the notations 
$\bA_k=(\ba_{1k},...,\ba_{rk})\in\R^{d_k\times r}$, and $\bB_k = \bA_k(\bA_k^{\top} \bA_k)^{-1} = (\bb_{1k},...,\bb_{rk}) \in\R^{d_k\times r}$. Let $d=\prod_{k=1}^K d_k$, $d_{\min}=\min\{d_1,...,d_K\}$, $d_{\max}=\max\{d_1,...,d_K\}$ and $d_{-k}=\prod_{j\ne k}d_j$.

To present theoretical properties of the proposed procedures, we impose the following assumptions.

\begin{assumption}\label{asmp:error}
The error process $\cE_t$ are independent Gaussian tensors, conditioning on the factor process $\{f_{it}, 1\le i\le r,t\in\mathbb Z\}$. In addition, there exists some constant $\sigma>0$, such that
\begin{equation*}
\E (u^\top \text{vec}(\cE_t))^2\le \sigma^2 \|u\|_2^2, \quad u\in\RR^d.
\end{equation*}
\end{assumption}

\begin{assumption}\label{asmp:mixing}
Assume the factor process $f_{it}, 1\le i\le r$
is stationary and strong $\alpha$-mixing in $t$, with $\E f_{it}^2=1$, $\E f_{it-h}f_{it}\neq 0$, $\E f_{it-h} f_{jt}=0$ for all $i\neq j$ and
$h \geq 1$. Let $F_t=(f_{1t},...,f_{rt})^\top$. For any $v\in\R^{r}$ with $\|v\|_2=1$,
\begin{align}\label{cond2}
\max_t\P\left( \left| v^\top F_{t} \right| \ge x \right) \le c_1 \exp\left( -c_2x^{\gamma_2} \right),
\end{align}
where $c_1,c_2$ are some positive constants and $0<\gamma_2\le 2$. In addition, 
the mixing coefficient satisfies
\begin{align}\label{cond1}
\alpha(m) \le \exp\left( - c_0 m^{\gamma_1} \right)
\end{align}
for some constant $c_0>0$ and $0<\gamma_1\le 1$, where
\begin{align*}
\alpha(m) = \sup_t\Big\{\Big|\P(A\cap B) - \P(A)\P(B)\Big|:
A\in \sigma(f_{is}, 1\le i\le r, s\le t), B\in \sigma(f_{is}, 1\le i\le r, s\ge t+m)\Big\}.
\end{align*}
\end{assumption}

\begin{assumption}\label{asmp:eigenvalue}
Assume $h\le T/4$ is fixed, and $\lambda_{1,h},...,\lambda_{r,h}$ are all distinct. Without loss of generality, let $\lambda_{1,h}>\lambda_{2,h}>\cdots>\lambda_{r,h}>0$.
Here, we emphasize that $\lambda_{i,h}$ depends on $h$, though
in other places when $h$ is fixed we will omit $h$ in the notation.
\end{assumption}

Assumption \ref{asmp:error} is similar to those on the noise imposed in \cite{lam2011}, \cite{lam2012}, \cite{han2020iterative}. It accommodates general patterns of dependence among individual time series fibers, but also allows a presentation of the main results with manageable analytical complexity. The normality assumption, which ensures fast statistical error rates in our analysis, is imposed for technical convenience.
In theory, it can be supplanted by a sub-Gaussian condition, or replaced with more general distributions with heavier tails. However, adopting such weaker conditions would significantly complicate the formulae, statistical outcomes, and requisite conditions within our time series framework. This added complexity would likely detract from the paper's readability, without providing additional statistical insights. Our primary goal is to maintain the readers' focus on the core content of the paper, and thus, we have chosen to assume additive Gaussian errors. This choice simplifies the exposition without compromising the fundamental tenets and the findings of our study.

Assumption \ref{asmp:mixing} is standard. It allows a very general class of time series models, including causal ARMA processes with continuously distributed innovations; see also \cite{tong1990non, bradley2005, tsay2005analysis, fan2008nonlinear, rosenblatt2012markov, tsay2018nonlinear}, among others. The restriction $\gamma_1\le 1$ is introduced only for presentation convenience. Assumption \ref{asmp:mixing} requires that the tail probability of $f_{it}$ decays exponentially fast. In particular, when $\gamma_2=2$, $f_{it}$ is sub-Gaussian.

Assumption \ref{asmp:eigenvalue} is sufficient to guarantee that all the factor loading vectors $\ba_{ik}$ can be uniquely identified up to the sign change. The parameters $\lambda_i$ can be viewed as an analogue of eigenvalues in the order-$2K$ tensor $\bSigma_{h}$. Similar to the eigen decomposition of a matrix, if some $\lambda_i$ are equal, estimation of the loading vectors $\ba_{ik}$ may suffer from label shift across $i$.
When $h$ is fixed and $\E f_{i,t} f_{i,t-h}\asymp 1$, $\lambda_i\asymp w_i^2$.
The signal strength of each factor is measured by $\lambda_i$.

Let us first study the behavior of the cPCA estimators in Algorithm \ref{algorithm:initial}. Theorem \ref{thm:initial} presents the performance bounds, which depends on the coherence (the degree of non-orthogonality) of the factor loading vectors.

Let
\begin{equation}\label{gap}
\lambda_*=\min_{1\le i\le r+1} \{\lambda_{i-1} -\lambda_{i}\}
\end{equation}
with $\lambda_0=\infty$, $\lambda_{r+1}=0$, be the minimum gap between the signal strengths of the factors.

\begin{theorem}\label{thm:initial}
Suppose Assumptions \ref{asmp:error}, \ref{asmp:mixing}, \ref{asmp:eigenvalue} hold. Let $1/\gamma=1/\gamma_1+2/\gamma_2$, $h\le T/4$ and $\delta<1$ with $\delta$ defined in \eqref{corr-all}. In an event with probability at least $1-(Tr)^{-C_1}-e^{-d}$,
the following error bound holds for the estimation of the loading vectors $\ba_{ik}$ using Algorithm \ref{algorithm:initial} (cPCA).
\begin{align}\label{thm:initial:eq1}
\|\widehat\ba_{ik}^{\cpca}\widehat\ba_{ik}^{\cpca\top}  -\ba_{ik}\ba_{ik}^\top \|_{\rm S} &\le \left(1+\frac{2\lambda_1}{\lambda_*}\right)\delta+
\frac{C_2 R^{(0)} }{\lambda_*},
\end{align}
for all $1\le i\le r$, $1\le k\le K$, where $C_1,C_2$ are some positive constants, and
\begin{align}\label{thm:initial:eq2}
R^{(0)} &= \max_{1\le i \le r} w_i^2 \left(\sqrt{\frac{r+ \log T}{T}} + \frac{(r+ \log T)^{1/\gamma}}{T}  \right)+ \sigma^2\sqrt{\frac{d}{T}} + \sigma \max_{1\le i\le r} w_i \sqrt{\frac{d}{T}}.
\end{align}
\end{theorem}

\begin{rmk}\label{rmk:tied_eigen}
We note that the eigengap $\lambda_*$ in Algorithm \ref{algorithm:initial} (cPCA) is not a requisite for the iterative Algorithm \ref{algorithm:projection} (ISO). Algorithm \ref{algorithm:initial} (cPCA) necessitates a significant separation among the different singular values $\lambda_i$'s. This condition can be relaxed through the use of random slicing \citep{anandkumar2014guaranteed,auddy2023large}, a widely recognized method for initialization in tensor CP decomposition. 
In our framework, the core step of random slicing is the construction of a projected auto-covariance tensor $\bSigma_h\times_k g_k \times _{K+k} \tilde g_k$, where $g_k$ and $\tilde g_k$ are independently generated Gaussian random vectors. Due to the randomness of $g_k$ and  $\tilde g_k$, even if all $\lambda_i$'s are equal, the singular values of $\bSigma_h\times_k g_k \times _{K+k} \tilde g_k$ will inherently differ. Furthermore, we can generate a sufficiently large eigengap between the top two singular values of the projected auto-covariance tensor through multiple rounds of random slicing, so that the leading component of the projected auto-covariance tensor is identifiable.
Since $\bSigma_h\times_k g_k \times _{K+k} \tilde g_k$ is an order $2K-2$ tensor, we can still employ cPCA and utilize the benefits of higher-order coherence in Proposition \ref{prop:delta} when $K>2$. The existing results in Theorem \ref{thm:initial} can be extended to such settings, although it would require more sophisticated theoretical analysis.
\end{rmk}

The first term in the upper bound \eqref{thm:initial:eq1} is induced by the non-orthogonality of the loading vectors $\ba_{ik}$, which can be viewed as bias. The second term in \eqref{thm:initial:eq1} comes from a concentration bound for the random noise, and thus can be interpreted as stochastic error.
By Proposition \ref{prop:delta}, it implies that a larger $K$ (e.g. higher order tensors)
leads to smaller bias and higher statistical accuracy of cPCA.
If $\delta\gtrsim R^{(0)}/\lambda_1$, then the error bound \eqref{thm:initial:eq1} is dominated by the bias related to $\delta$, otherwise it is dominated by the stochastic error.
Equation \eqref{thm:initial:eq2} shows that $R^{(0)}$ in the stochastic error comes from
the fluctuation of the factor process $f_{it}$ (the first two terms) and the noise $\cE_t$ in \eqref{model} (the other two terms).
When $\sum_{i=1}^r\lambda_i\asymp r\lambda_1 \asymp r w_1^2$ and $R^{(0)}/\lambda_1 +T/d \lesssim  1$, the terms related to the noise becomes
$\sqrt{r/T}/\sqrt{{\rm SNR}}$, where the signal-to-noise ratio (SNR) is
\begin{align}\label{def:SNR}
\hbox{\rm SNR}
:= \E\big\|\sum_{i=1}^r w_i f_{it}\otimes_{k=1}^K \ba_{ik}\big\|_{\rm HS}^2/\E\|\cE_t\|_{\rm HS}^2 = \sum_{i=1}^r w_i^2 /(\sigma^2 d)\asymp r\lambda_1/(\sigma^2d).
\end{align}

Roughly speaking though not completely correct, the term $\lambda_i-\lambda_{i+1}$
can be viewed as the gap of $i$-th and $(i+1)$-th largest eigenvalues of $\bSigma_{h}^*$ with $\bSigma_{h}^*$ given in \eqref{eq:bsigmah*}.
In particular, if $\lambda_1\asymp ...\asymp \lambda_r \asymp w_1^2$, then $\lambda_*\asymp w_1^2/r$. In this case, the
bound \eqref{thm:initial:eq1} can be simplified to
\begin{eqnarray}
\lefteqn{\|\widehat\ba_{ik}^{\cpca}\widehat\ba_{ik}^{\cpca\top}  -\ba_{ik}\ba_{ik}^\top \|_{\rm S}} \nonumber \\
&\le & C_3r \delta + C_4 r\left(\sqrt{\frac{r+\log T}{T}} + \frac{(r+\log T)^{1/\gamma}}{T}  \right) + \frac{C_4\sigma^2r\sqrt{d}}{w_1^2\sqrt{T}} + \frac{C_4\sigma r\sqrt{d}}{w_1\sqrt{T}}. \label{thm:initial:eq3}
\end{eqnarray}
Then, by \eqref{thm:initial:eq3} and Proposition \ref{prop:delta}, the consistency of the cPCA estimators only requires the
incoherence parameter to be at most $\vartheta_{\max}\lesssim r^{-2/K}$.

Next, let us consider the statistical performance of the iterative algorithm (Algorithm \ref{algorithm:projection}) after cPCA initialization, \emph{i.e.} HOPE estimators. As discussed earlier, the operation in \eqref{eq:z} achieves
dimension reduction by projecting $\cX_t$ into a vector
and retains only one of the $r$ factor terms, hence eliminates the interaction effects between different factors. As we update the estimation of each individual loading vector $\ba_{ik}$ separately in the algorithm, ideally this would remove the bias part in \eqref{thm:initial:eq1}
which is due to the non-orthogonality of the loading vectors, and replace the eigengap $\lambda_*$
in \eqref{thm:initial:eq1} by $\lambda_i$, as \eqref{eq:cp-ideal}
only involves one eigenvector. It also leads to the elimination of the first two terms of $R^{(0)}$.
As mentioned in Section \ref{section:estimation}, when updating $\widehat \ba_{ik}^{(m)}$, we take advantages of the multiplicative nature of the project error $\xi_{ij}^{(m)}$ in \eqref{eq:xi},
and the rapid growth of such benefits as the iteration number $m$ grows.
Thus we expect that the rate of HOPE estimators would become
\begin{align}\label{thm-projection:eq}
\max_{1\le i\le r} \max_{1\le k\le K} \|\widehat\ba_{ik}\widehat\ba_{ik}^\top  -\ba_{ik}\ba_{ik}^\top \|_{\rm S} &\le C_{0,K} R^{\ideal},
\end{align}
where 
\begin{align}
R_{k,i}^{\ideal}=  \frac{\sigma^2}{\lambda_i} \sqrt{\frac{d_k}{T}} + \sqrt{\frac{\sigma^2d_k}{\lambda_i T}}, \quad \text{and} \quad R^{\ideal} =\max_{1\le i\le r}\max_{1\le k\le K} R_{k,i}^{\ideal}.
\end{align}
Note that $R_{k,i}^{\ideal}$ replaces all $d=d_1\ldots d_K$ in the noise component of the stochastic error in \eqref{thm:initial:eq1} by $d_k$ due to dimension reduction. 
The following theorem provides conditions under which this ideal rate is indeed achieved.

Let the statistical error bound of the initialization used in Algorithm \ref{algorithm:projection} be $\psi_0$. For cPCA,
\begin{equation}
    \psi_0=\frac{\lambda_1\delta+R^{(0)}}{\lambda_*},  \label{eq:phi0}
\end{equation}
where $\lambda_*$ is the eigengap defined in \eqref{gap}
and $R^{(0)}$ is defined in \eqref{thm:initial:eq2}.

\begin{theorem}\label{thm:projection}
Suppose Assumptions \ref{asmp:error}, \ref{asmp:mixing}, 
\ref{asmp:eigenvalue} hold. Assume that $\delta_{\max}=\max_{k\le K}\delta_k<1$ with $\delta_k$ defined in \eqref{corr-k}, and $r=O(T)$. Let $1/\gamma=1/\gamma_1+2/\gamma_2$, $h \le T/4$, and $d=d_1\cdots d_K$. Suppose that for a proper numeric constant $C_{1,K}$ depending on $K$ only, we have
\begin{align}
&\sqrt{1-\delta_{\max}}-(r^{1/2}+1)\psi_0/\sqrt{1-1/(4r)}>0, \label{thm-projection:eq1a} \\
&C_{1,K}\left(\frac{\lambda_1}{\lambda_r} \right) \psi_0^{2K-3} \label{thm-projection:eq1b}
+ C_{1,K}\sqrt{\frac{\lambda_1}{\lambda_r} }\left(\sqrt{\frac{r+\log T }{T}} + \frac{(r+\log T)^{1/\gamma}}{T}  \right) \psi_0^{K-2} \le \rho <1
\end{align}
Then, after at most $M=O(\log \log (\psi_0/R^{\ideal}))$ 
iterations of Algorithm \ref{algorithm:projection}, in an event with probability at least $1-(Tr)^{-C}-\sum_{k} e^{-d_k}$, the HOPE estimator satisfies
\begin{align}\label{thm-projection:eq2}
\|\widehat\ba_{ik}^{\iso}\widehat\ba_{ik}^{\iso\top}  -\ba_{ik}\ba_{ik}^\top \|_{\rm S} &\le C_{0,K} R^{\ideal},   
\end{align}
for all $1\le i\le r$, $1\le k\le K$, where $C_{0,K}$ is a constant depending on $K$ only and $C$ is a positive numeric constant.
\end{theorem}

The detailed proof of the theorem is in Appendix \ref{section:proof}. The key idea of the analysis of HOPE is to show that the iterative estimator has an error contraction effect in each iteration.
Theorem \ref{thm:projection} implies that HOPE will achieve a faster statistical error rate than the typical $O_{\P}(T^{-1/2})$ whenever $\lambda_r\gg \sigma^2 \max_k d_k$. As $\vec(\cX_t)$ has $d$ elements, the strong factors setting in the literature \citep{lam2011,chen2022factor,han2020iterative}
typically assumes ${\rm SNR}\asymp 1$. In our case it is similar to assuming the signal strength $\E\big\|\sum_{i=1}^r w_i f_{it}\otimes_{k=1}^K \ba_{ik}\big\|_{\rm HS}^2 \asymp \sigma^2 d$.
When $r$ is fixed and $\lambda_1 \asymp \cdots \asymp \lambda_r$, the statistical error rate will be reduced to $O_{\P}(T^{-1/2}d_{-k}^{-1/2})$, where $d_{-k}=\prod_{j\ne k}d_j$.

\begin{rmk}[{\bf Iteration complexity}] \label{rmk:iteration}
Theorem \ref{thm:projection} implies that Algorithm \ref{algorithm:projection} achieves the desired estimation error $R^{\ideal}$ after at most $M=O(\log \log (\psi_0/R^{\ideal}))$ number of iterations. In this sense, after at most double-logarithmic number of iterations, the iterative estimator in Algorithm \ref{algorithm:projection} converges to a neighborhood of the true parameter $\ba_{ik}$, up to a statistical error with a rate $O(R^{\ideal})$. We observe that Algorithm \ref{algorithm:projection} typically converges within very few steps in 
practical implementations.
\end{rmk}

\begin{rmk}\label{cond:initial}
Condition \eqref{thm-projection:eq1a} requires $r^{1/2}\psi_0$ to be small. It is a relatively strong condition due to the extra multiplier $r^{1/2}$ on the error of the initial estimators. This is a technical issue due to the need to invert the estimated $\bA_k^\top \bA_k$ in our analysis to construct the mode-$k$ projection in Algorithm \ref{algorithm:projection}.
In fact the $r^{1/2}$ term may be eliminated by applying a shrinkage procedure on the singular values of $\widehat\bA_k$ after obtaining the updates of $\widehat\ba_{ik}$, $1\le i\le r$, similar to the procedure proposed by \cite{anandkumar2014guaranteed}.

Furthermore, the condition given by \eqref{thm-projection:eq1b} originates from the multiplicative nature 
of the projection error $\xi_{ij}^{(m)}$, as seen in \eqref{eq:xi} for $i\neq j$. If $\psi_0$ signifies the error bound for cPCA estimators, then condition \eqref{thm-projection:eq1b} is satisfied when $(\lambda_1/\lambda_r)\psi_0^{2K-3}\lesssim 1$. In comparison, the iterative algorithm of \cite{anandkumar2014guaranteed} requires that the initialization fulfills $\psi_0\lesssim \lambda_r/\lambda_1 + 1/\sqrt{d_{\min}}$, a condition
that is more stringent than \eqref{thm-projection:eq1b}.
The ratio $\lambda_1/\lambda_r$ in \eqref{thm-projection:eq1b} is unavoidable. When updating the estimates of $\ba_{ik}$ in Algorithm \ref{algorithm:projection}, we need to remove the effect of other factors ($j\neq i$) on the $i$-th factor, which introduces the ratio of factor strengths $\lambda_1/\lambda_r$ in the analysis.

In particular, if $\lambda_1\asymp \cdots\asymp \lambda_r$, 
the shrinkage procedure 
can reduce conditions \eqref{thm-projection:eq1a} and \eqref{thm-projection:eq1b} to
\begin{align}\label{thm-projection:eq4}
&C_{1,K}\psi_0   <1  ,
\end{align}
where $\psi_0$ is the cPCA error bound in \eqref{eq:phi0}. It ensures that, with high probability, $\|\widehat\ba_{ik}^{(0)} \widehat\ba_{ik}^{(0)\top} -\ba_{ik} \ba_{ik}^\top \|_{\rm S}$ are sufficiently small, so that the cPCA initialization is sufficiently close to the ground truth as in \eqref{thm-projection:eq4}.
\end{rmk}

\begin{rmk}[{\bf Comparison with general tensor CP-decomposition methods}]\label{rmk:initial0}
To estimate $\ba_{ik}$ in \eqref{eq:sigmah}, one can use the standard tensor
CP-decomposition algorithms, such as those in \cite{anandkumar2014guaranteed, hao2020sparse, sun2017provable}, without utilizing the special features of TFM-cp. The randomized initialization estimators in these algorithms
typically require the incoherence condition $\vartheta_{\max} \lesssim {\rm poly}\log(d_{\min})/\sqrt{d_{\min}}$. In contrast, the condition for ISO needs $\vartheta_{\max} \lesssim r^{-5/(2K)}$, which is weaker when $r=o(d_{\min}^{K/5})$. Similarly, we prove that the cPCA yields useful estimates when $r^{2}\vartheta_{\max}^K$ is small, or $\vartheta_{\max} \lesssim r^{-2/K}$. In other words, as long as $r$ is not exceedingly large (e.g. $r=o(d_{\min}^{K/5})$), both cPCA and ISO permit a more lenient incoherence condition among the CP basis.
Furthermore, the high-order coherence in TFM-cp leads to an impressive computational super-linear convergence rate of Algorithm \ref{algorithm:projection}, which is faster than the computational linear convergence rate of the iterative projection algorithm in \cite{han2020iterative} or other variants of alternating least squares approaches in the literature,
that are at most linear with the required number of iterations $M=O(\log (\psi_0/R^{\ideal}))$.
\end{rmk}


\begin{rmk}[{\bf Comparison between TFM-cp and TFM-tucker Models}]\label{rmk:TFMcompare2}
As discussed in Remark \ref{rmk:TFMcompare}, TFM-cp can be written as a TFM-tucker with a special structure. One can ignore the special structure and treat it a generic TFM-tucker in \eqref{eq:tucker-model} and estimate the loading spaces spanned by $\{\ba_{ik},1\le i\le r\}$ using the iterative estimation algorithm in \cite{han2020iterative}.
In fact, ISO (as detailed in Algorithm \ref{algorithm:projection}) can be viewed as an enhancement of the iterative algorithm presented in \cite{han2020iterative} to utilize the special structure of TFM-cp. This is achieved by permitting non-orthogonality in $\bA_k$ and estimating each $\ba_{ik}, 1\le i\le r,$ individually.

Here we provide a brief comparison in the estimation accuracy between the estimators under these two settings to show the impact of the additional structure in TFM-cp. Note that for TFM-tucker, only the linear space spanned by $\bA_k$ can be estimated hence the estimation accuracy is based on a specific space representation, different from that for the TFM-cp.
For simplicity, we consider the case $\lambda_1\asymp \cdots\asymp \lambda_r$.

(i) The iterative refinement algorithm (Algorithm \ref{algorithm:projection}) for TFM-cp requires similar conditions on the initial estimators as the iterative projection algorithms for TFM-tucker.
Under many situations, both methods only require the initialization to retain a large portion of the signal, but not the consistency. 

(ii) The statistical error rate of HOPE in \eqref{thm-projection:eq2} is the same as the upper bound of the iterative projection algorithms for estimation of the fixed rank TFM-tucker, c.f. Corollary 3.1 and 3.2 in \cite{han2020iterative}, which is shown to have the minimax optimality.
It follows that HOPE also achieves the minimax rate-optimal estimation error under fixed $r$.

(iii) When the rank $r$ diverges and SNR $\asymp 1$ where SNR is defined in \eqref{def:SNR},
the estimation error of the loading spaces by the iterative estimation procedures, iTOPUP and TIPUP-iTOPUP procedures in \cite{han2020iterative} applied to the specific TFM-tucker model implied by the TFM-cp model, is of the order
$O_{\P}(\max_k r^{3K/2-1}T^{-1/2}d_{-k}^{-1/2})$, a rate that is always larger than $O_{\P}(\max_k r^{1/2}T^{-1/2}d_{-k}^{-1/2})$, the error rate of HOPE for TFM-cp model. The iTIPUP procedure for TFM-tucker model is $O_{\P}(\max_k r^{1/2+(K-1)\zeta}T^{-1/2}d_{-k}^{-1/2})$ where $\zeta$
controls the level of signal cancellation (see \cite{han2020iterative} for details). When there is no signal cancellation, $\zeta=0$, the rate of the two procedures are the same. Note that iTIPUP only estimates the loading space, while HOPE provides estimates of the unique loading vectors. The error rate of HOPE is better when $\zeta>0$. This demonstrates that HOPE is able to utilize the specific structure in TFM-cp to achieve more accurate estimation than simply applying the estimation procedures designed for general TFM-tucker.


(iv) It can be {seen} that, computationally, {the complexities for the initialization of both TFM-cp and TFM-tucker are the same, yet,}
the {per iteration} complexity of TFM-cp is lower than that of TFM-tucker by a factor of $r^{2K-2}$ where $r_1=\ldots=r_K=r$
in TFM-tucker model.
\end{rmk}

\begin{theorem}\label{thm:factors}
Suppose Assumptions \ref{asmp:error}, \ref{asmp:mixing}, 
\ref{asmp:eigenvalue} hold. Assume that $\delta_k<1$ with $\delta_k$ defined in \eqref{corr-k}, $\sigma^2\lesssim \lambda_r$ and condition \eqref{thm-projection:eq1a} holds. Let $d_{\max}=\max_k d_k$. Then the HOPE estimator in Algorithm \ref{algorithm:projection} using a specific $h$ satisfies:

\begin{align}
w_i^{-1} \left| \widehat w_i^{\iso} \widehat f_{it}^{\iso} - w_i f_{it} \right| =O_{\P} \left( \sqrt{\frac{\sigma^2}{\lambda_r}} + \sqrt{\frac{\sigma^2d_{\max}}{\lambda_r T}} \right)  \label{thm-factors:eq1}
\end{align}
and
\begin{align}
w_i^{-1} w_j^{-1}\left|\frac{1}{T-h_*} \sum_{t=h_*+1}^T \widehat w_i^{\iso} \widehat w_j^{\iso} \widehat f_{it-h_*}^{\iso} \widehat f_{jt}^{\iso} - \frac{1}{T-h_*} \sum_{t=h_*+1}^T w_i w_j f_{it-h_*} f_{jt}  \right|    =  O_{\P} \left(  \sqrt{\frac{\sigma^2d_{\max}}{\lambda_r T}}  \right)     \label{thm-factors:eq2}
\end{align}
for $1\le i,j\le r, 1\le t\le T$ and all $1< h_*\le T/4$.
\end{theorem}

Theorem \ref{thm:factors} specifies the convergence rate for the estimated factors $f_{it}$. When $\lambda_r \gg \sigma^2 d_{\max} +T$, $w_i^{-1} | \widehat w_i^{\iso} \widehat f_{it}^{\iso} - w_i f_{it}|$ is much smaller than the parametric rate $T^{-1/2}$. If all the factors are strong \citep{lam2011} such that $\lambda_1\asymp\lambda_r \asymp \sigma^2 d$, \eqref{thm-factors:eq1} implies that $w_i^{-1} | \widehat w_i^{\iso} \widehat f_{it}^{\iso} - w_i f_{it} |=O_{\P}(d^{-1/2}+d_{\max}^{1/2}d^{-1/2}T^{-1/2})$. Then, as long as $d_k\to \infty$ and $K\ge 2$, the estimated factors are consistent, even under a fixed $T$. In comparison, the convergence rate of the estimated factors in Theorem 1 of \cite{bai2003} for vector factor models is $O_{\P}(d^{-1/2}+T^{-1})$.
Moreover, \eqref{thm-factors:eq2} shows that the error rates for the sample auto-cross-moment of the estimated factors to the true sample auto-cross-moment is also $o_{\P}(T^{-1/2})$ when $\lambda_r \gg \sigma^2 d_{\max}$. This implies that it is a valid option to use the estimated factor processes as the true factor processes to model the dynamics of the factors. When the estimation of these time series models only 
{required} auto-correlation and partial auto-correction functions, the results are expected to be the same as using the true factor process, without loss of efficiency. The statistical rates in Theorem \ref{thm:factors} lay a foundation for further modeling of the estimated factor processes with vast repository of linear and nonlinear options.

\section{Simulation Studies} \label{section:simulation}

\subsection{Alternative algorithms for estimation of TFM-cp}

Here we present two alternative estimation algorithms for TFM-cp, by
extending the
popular rank one alternating least square (ALS) algorithm of \cite{anandkumar2014guaranteed} and orthogonalized alternating least square (OALS) of \cite{sharan2017orthogonalized} designed for CP decomposition of
noisy tensors, because $\widehat \bSigma_h$ in \eqref{eq:bsigma} is indeed
in a CP form, but with repeated components.
In addition, we use cPCA estimates for initialization, instead of randomized initialization used for general CP decomposition. We
will denote the algorithms as cALS (Algorithm \ref{algorithm:als})
and cOALS (Algorithm \ref{algorithm:oals}), respectively.
The simulation study below shows that, although cALS and cOALS perform better than the straightforward implementation of ALS and OALS with randomized
initialization, they do not perform as well as the proposed HOPE algorithm.
Hence we do not investigate their theoretical properties in this paper.



\begin{algorithm}[htbp]
\caption{cPCA-initialized Rank One Alternating Least Square (cALS)} \label{algorithm:als}


\begin{algorithmic}[1]
\Require  Observations $\cX_t\in\RR^{d_1\times\cdots\times d_K}$ for $t=1,...,T$, the number of factors $r$,
the time lag $h$,
the cPCA initial estimate $(\widehat \ba_{i1}^{\cpca}, ..., \widehat \ba_{iK}^{\cpca})$, $1\le i\le r$, the tolerance parameter $\epsilon>0$, and the maximum number of iterations $M$.


\State Compute $\widehat \bSigma_{h}$ 
defined in \eqref{eq:hatsigmah}. 

\State
Initialize unit vectors $\widehat \ba_{i k}^{(0)}=\widehat \ba_{ik}^{\cpca}$ for $1\le k\le K$, $1\le i \le r$. Set $m=0$.

\For $i=1$ to $r$.
\Repeat
\State Set $m=m+1$.
\For $k=1$ to $K$.

\State Compute
$\widetilde \ba_{i k}^{(m)} = \widehat \bSigma_{h} \times_{\ell=1}^{k-1}  \widehat  \ba_{i\ell}^{(m)\top} \times_{\ell=k}^{K}  \widehat  \ba_{i\ell}^{(m-1)\top} \times_{\ell=K+1}^{K+k-1}  \widehat  \ba_{i\ell}^{(m)\top} \times_{\ell=K+k+1}^{2K}  \widehat  \ba_{i\ell}^{(m-1)\top}, $
where $\widehat  \ba_{i\ell}^{(m-1)}= \widehat  \ba_{i\ell-K}^{(m-1)}$ for $\ell>K$.
\State Compute $\widehat \ba_{i k}^{(m)} = \widetilde\ba_{i k}^{(m)}/\| \widetilde\ba_{i k}^{(m)} \|_2$.
\EndFor

\Until $m=M$ or
$\max_k\| \widehat\ba_{ik}^{(m)} \widehat\ba_{ik}^{(m)\top}  - \widehat\ba_{ik}^{(m-1)} \widehat\ba_{ik}^{(m-1)\top} \|_{\rm S}\le \epsilon .$
\State 
Let $\widehat\ba_{ik}^{{\rm \tiny cALS}}= \widehat\ba_{ik}^{(m)}$, $1\le k\le K$.
\EndFor


\Ensure
$\widehat\ba_{ik}^{{\rm \tiny cALS}},
\quad i=1,...,r, \ \ k=1,...,K.$
\end{algorithmic}
\end{algorithm}

\begin{algorithm}[htbp]
\caption{cPCA-initialized Orthogonalized Alternating Least Square (cOALS)}  \label{algorithm:oals}
\begin{algorithmic}[1]
\Require Observations $\cX_t\in\RR^{d_1\times\cdots\times d_K}$ for $t=1,...,T$, the number of factors $r$, the time lag $h$, the cPCA initial estimate $(\widehat \ba_{i1}^{\cpca}, ..., \widehat \ba_{iK}^{\cpca})$, $1\le i\le r$, the tolerance parameter $\epsilon>0$, the maximum number of iterations $M$.

\State Compute $\widehat \bSigma_{h}$ defined in \eqref{eq:hatsigmah}.

\State
Initialize unit vectors $\widehat \ba_{i k}^{(0)}=\widehat \ba_{ik}^{\cpca}$ for $1\le k\le K$, $1\le i \le r$. Set $\widehat \bA_{k}^{(0)}=(\widehat \ba_{1k}^{(0)}, ..., \widehat \ba_{rk}^{(0)})$ and $m=0$.

\Repeat
\State Set $m=m+1$.
\State Find QR decomposition of $\widehat \bA_{k}^{(m-1)}$, set $\widehat \bA_{k}^{(m-1)}= \bQ_{k}^{(m-1)} \bR_k^{(m-1)}$ for $1\le k\le K$.
\For $k=1$ to $K$.

\State Compute $\widehat \bA_k^{(m)}=\mat_k(\widehat \bSigma_{h}) (\ast_{\ell\neq k}^{2K} \bQ_{\ell}^{(m-1)}) $, where $\bQ_{\ell}^{(m-1)}=\bQ_{\ell-K}^{(m-1)}$ for $\ell>K$ and $\ast$ is the Khatri–Rao product.
\EndFor

\Until $m=M$ or
$\max_i\max_k\| \widehat\ba_{ik}^{(m)} \widehat\ba_{ik}^{(m)\top}  - \widehat\ba_{ik}^{(m-1)} \widehat\ba_{ik}^{(m-1)\top} \|_{\rm S}\le \epsilon .$

\Ensure
$\widehat\ba_{ik}^{{\rm \tiny cOALS}}=\widehat\ba_{ik}^{(m)}, \quad i=1,...,r, \ \ k=1,...,K.$

\end{algorithmic}
\end{algorithm}

\subsection{Simulation}

In this section, we compare the empirical performance of different procedures of estimating the loading vectors of TFM-cp, under various simulation setups. We consider the cPCA initialization (Algorithm \ref{algorithm:initial}) alone, the iterative procedure HOPE, and the intermediate output from the iterative procedure when the number of iteration is 1 after initialization. The one step procedure will be denoted as 1HOPE. We also check the performance of the alternative algorithms ALS, OALS, cALS, and cOALS as described above.
The estimation error shown is given by $\max_{i,k}\|\widehat \ba_{ik}\widehat \ba_{ik}^\top  - \ba_{ik} \ba_{ik}^\top \|_{\rm S}$.

We demonstrate the performance of all procedures under
TFM-cp with $K=2$ (matrix time series) with
\begin{equation}
\cX_t=\sum_{i=1}^r w f_{it} \ba_{i1} \otimes \ba_{i2}+ \cE_t. \label{eq:simu}
\end{equation}

\noindent
For $K=2$ with model \eqref{eq:simu},
we consider the following three experimental configurations:
\begin{enumerate}
\item[I.] Set $r=2$, $d_1=d_2=40$, $T=400$, $w=6$ and vary $\delta$ in the set $[0,0.5]$. The purpose of this setting is to verify the theoretical bounds of cPCA and HOPE in terms of the coherence parameter $\delta$.

\item[II.] Set $r=2$, $d_1=d_2=40$, 
$\delta=0.2$. We vary the sample size $T$ and the signal strength $w$
to investigate the impact of $\delta$ against signal strength and sample size.

\item[III.] Set $r=3$, $d_1=d_2=40$, $T=400$, $w=8$ and vary $\delta$ to check the sensitivities of $\delta$ for all the proposed algorithms and compare with randomized initialization.

\end{enumerate}

Results from an additional simulation settings under $K=2$ and $K=3$ cases are given in Appendix \ref{section:additional_simulation}.
We repeat all the experiments 100 times. For simplicity, we set $h=1$.

The loading vectors are generated as follows. First, the elements of matrices $\widetilde \bA_{k}=(\widetilde \ba_{1k},..., \widetilde \ba_{rk})\in \R^{d_k\times r}$, $1\le k\le K$, are generated from i.i.d. $N(0,1)$ and then
orthonormalized through QR decomposition. Then if $\delta=0$, set $\bA_{k}=\widetilde \bA_{k}$, otherwise, set $\ba_{1k}=\widetilde \ba_{1k}$ and $\ba_{ik}= (\widetilde \ba_{1k}+\theta \widetilde \ba_{ik})/\|\widetilde \ba_{1k}+\theta \widetilde \ba_{ik}\|_2$ for all $i\ge 2$ and $1\le k\le K$, with $\vartheta=\delta/(r-1)$ and $\theta=(\vartheta^{-2/K}-1)^{1/2}$.
The commonly used incoherence measure \citep{anandkumar2014guaranteed,hao2020sparse} under this construction is
$\vartheta_{\max}=(1+\theta^2)^{-1/2}=\vartheta^{1/K}$.

The noise $\cE_t$ in 
the model is white 
$\cE_t\perp \cE_{t+h},h>0$, and generated according to $\cE_t=\Psi_1^{1/2} Z_t\Psi_2^{1/2}$
where all of the elements in the $d_1\times d_2$ matrix $Z_t$ 
are i.i.d. $N(0,1)$. Furthermore, $\Psi_1, ~\Psi_2$
are the covariance matrices along each mode with
the diagonal elements being $1$ and all the off-diagonal elements being $\psi_1,~\psi_2$. 
Throughout this section, we set the off-diagonal entries of the covariance matrices of the noise as $\psi_1=\psi_2$.

Under Configurations I and II with $r=2$, the factor processes $f_{1t}$ and $f_{2t}$ are generated as two independent AR(1) processes, following $f_{1t}=0.8f_{1t-1}+e_{1t}$, $f_{2t}=0.6f_{2t-1}+e_{2t}$. Under Configuration III and Configurations IV and V in Appendix \ref{section:additional_simulation}, with $r=3$, $f_{1t},f_{2t},f_{3t}$ are generated as independent AR(1) processes, with $f_{1t}=0.8f_{1t-1}+e_{1t}, f_{2t}=0.7f_{2t-1}+e_{2t},  f_{3t}=0.6f_{3t-1}+e_{3t}$. Here, all of the innovations follow i.i.d. $N(0,1)$.
The factors are not normalized.

\begin{figure}[htbp]
\centering
\includegraphics[width=5.5in]{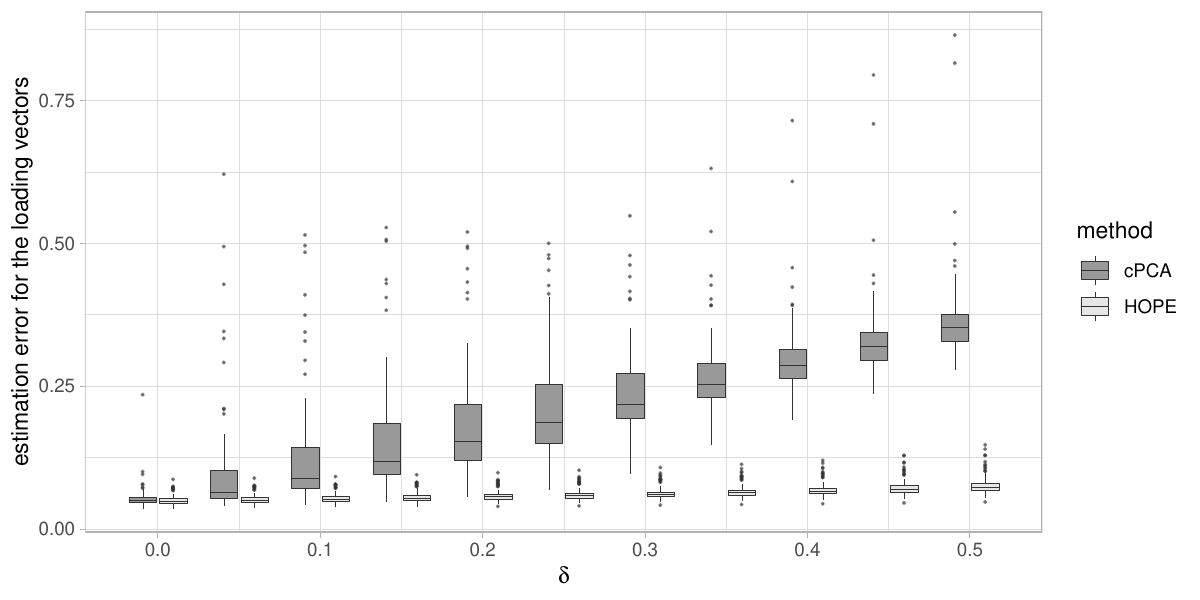}
\caption{Boxplots of the estimation error over 100 replications under experiment configuration I with different $\delta$.} \label{fig:delta}
\end{figure}

Figure \ref{fig:delta} shows the boxplots of the estimation errors for cPCA and HOPE under configuration I, for different $\delta$. It can be seen that the performance of cPCA deteriorates as $\delta$ increases, while that of HOPE remains almost unchanged. The median of the cPCA estimation errors increases almost linearly with $\delta$, with a $R^2$ of $0.977$.
This 
linear effect of $\delta$ on the performance bounds of cPCA
is confirmed by
the theoretical results in \eqref{thm:initial:eq1}.

\begin{figure}[htbp]
\centering
\includegraphics[width=5.5in]{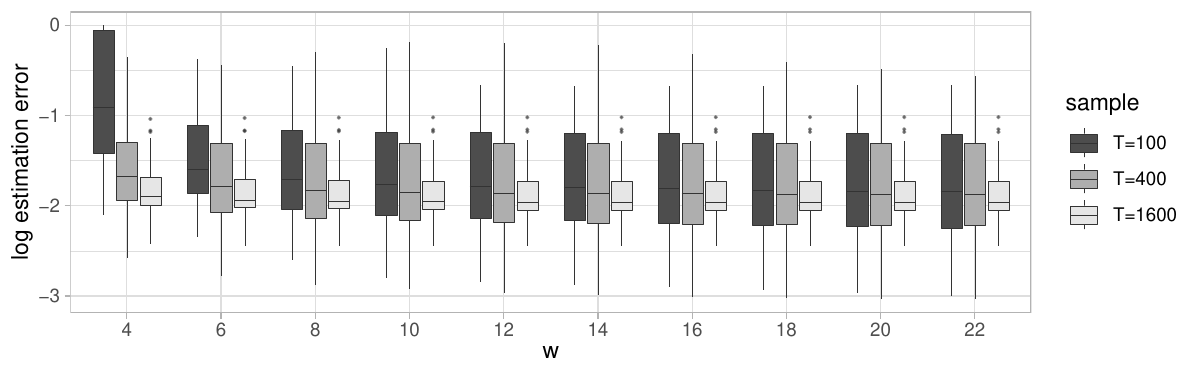}
\caption{Boxplots of the logarithm of the estimation error for cPCA under experiment configuration II.} \label{fig:rate_cpca}
\end{figure}

\begin{figure}[htbp]
\centering
\includegraphics[width=5.5in]{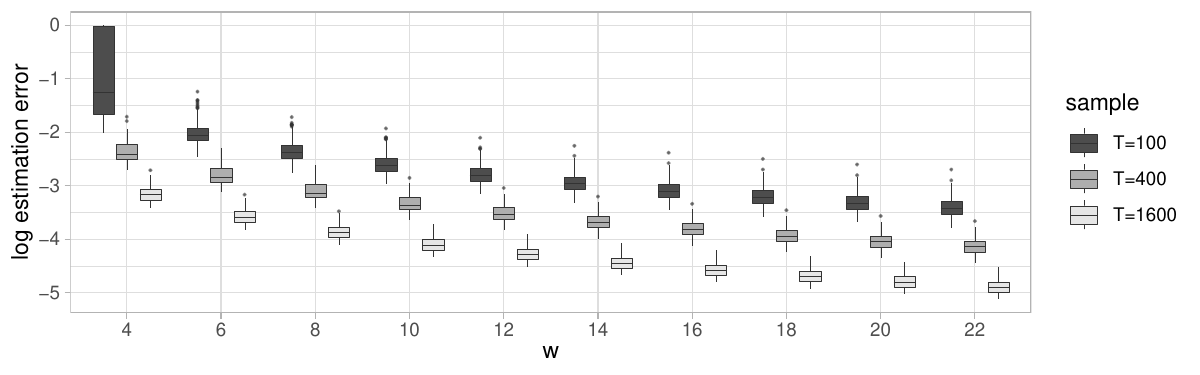}
\caption{Boxplots of the logarithm of the estimation error for HOPE under experiment configuration II.} \label{fig:rate_hope}
\end{figure}

\begin{figure}[htbp]
\centering
\includegraphics[width=\textwidth]{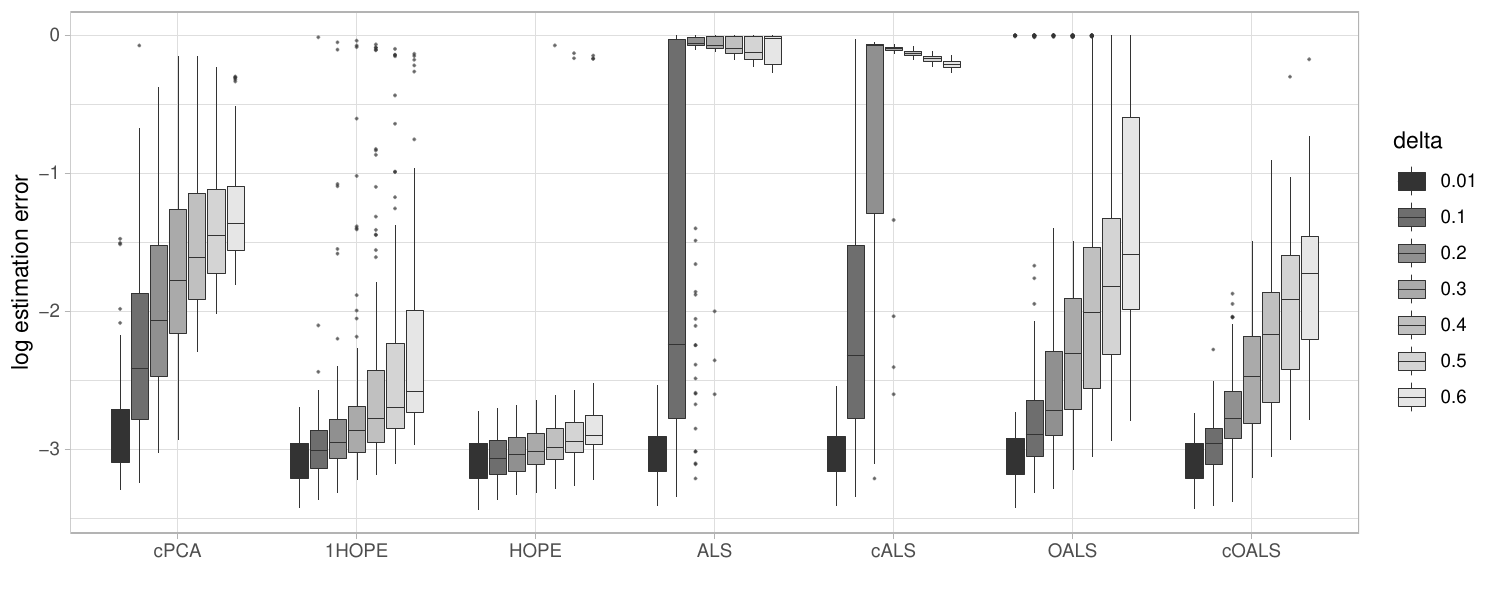}
\caption{Boxplots of the logarithm of the estimation error under experiment configuration III. Seven methods with seven choices of $\delta$ are considered in total.} \label{fig:compare:delta}
\end{figure}

The experiment of Configuration II is conducted to verify the theoretical bounds on different sample sizes $T$ and signal strengths $w$. Figures \ref{fig:rate_cpca} and \ref{fig:rate_hope}
show the logarithm of the estimation errors under different
($w,T$) combinations. 
It can be seen from Figure \ref{fig:rate_cpca} that
the estimation error of cPCA decreases to a lower bound as $w$ and $T$ increases.
The lower bound is associated with the bias term in \eqref{thm:initial:eq1} that cannot be reduced by a larger $w$ and $T$. This is the baseline error due to the non-orthogonality. In contrast, the phenomenon of HOPE is very different. Figure \ref{fig:rate_hope} shows that the performance improves monotonically 
as $w$ or $T$ increases. Again, this is consistent with the theoretical bounds in \eqref{thm-projection:eq2}.

Figure \ref{fig:compare:delta} shows the boxplots of the logarithm of the estimation errors for 7 different methods with choices of $\delta$ under configuration III. ALS and OALS are implemented with $L=200$ random initiations. It can be seen that HOPE outperforms all the other methods. 
Again, the choice of $\delta$ does not affect the performance of HOPE significantly. 
One-step method (1HOPE) is better than the cPCA alone, 
and the iterative method HOPE is in turn better than the one-step method. When the coherence $\delta$ decreases, all methods perform better, but the advantage of HOPE over one-step method and the advantage of one-step method over the cPCA initialization become smaller.
For the extremely small $\delta=0.01$, all loading vectors are almost
orthogonal to each other. In this case, all the iterative procedures, including the one-step HOPE, perform similarly.
In addition, ALS and cALS are always the worst under the cases $\delta \ge 0.1$. The hybrid methods cALS and cOALS improve the original randomized initialized ALS and OALS significantly, showing the advantages of the cPCA initialization. 
It is worth noting that cOALS has comparable performance with 1HOPE and HOPE when $\delta$ is small.

\section{Applications} \label{section:application}

In this section, we demonstrate the use of TFM-cp model using
the taxi traffic data set used in \cite{chen2022factor}. The data set was
collected by the Taxi \& Limousine Commission of New York City, and published at {\rm https://www1.nyc.gov/site/tlc/about/tlc-trip-record-data.page.} Within Manhattan Island, it contains 69 predefined pick-up and drop-off zones and 24 hourly period for each day from January 1, 2009 to December 31, 2017. The total number of rides moving among the zones within each hour is recorded, yielding a $\cX_t \in \R^{69\times 69\times 24}$ tensor for each day, using the hour of day as the third dimension.

One natural way to model the hourly traffic data is to consider $\bX_{t}\in\R^{69\times 69}$ for $t=1,\dots,24\times N$, where $N$ is the total number of days. It is apparent that such hourly data have two types of seasonality: weekly seasonality and daily seasonality. We handle the weekly seasonality by separating the time series into two parts: business-day series and non-business-day series. The length of the business-day series is 2,262 days, and that of the non-business-day is 1,025 days. The daily seasonality is a more interesting and important issue. It is clear that taxi usage heavily depends on time of the day (morning and evening rush hours, lunch hours etc). Seasonal time series models have been extensively studied for univariate time series \citep{Box&Jenkins76, tsay2005analysis,Shumway&Stoffer06,reinsel2003elements}, but these parametric models are difficult to be extended to deal with high dimensional matrix time series. Segmenting the 24-hour day into distinct intervals, such as morning rush hours and business hours, loses the detailed hourly information and requires pre-determined segmentation scheme \citep{zhang2019spectral,zhu2022learning}.
Here we adopt the nonparametric approach by stacking the hourly observations into a (multi-dimensional) daily observation. This is equivalent to turning hourly observations into daily 24-dimensional vector observations in the univariate time series case, a commonly used approach. By jointly modeling the 24-hourly observations within the day together, the detailed daily pattern can be captured more accurately and more flexibly without a parametric model assumption.
This way, the daily pattern is built into the model simultaneously, and the interaction among the geographic and temporal patterns will be revealed.

After some exploratory analysis, we decide to use the TFM-cp with $r=4$ factors for both 
business-day series and non-business-day series,
and estimate the model with $h=1$. For the non-business-day series, TFM-cp explains 63.0\% of the variability in the data. In comparison, treating the tensor time series as a $114,264$ {($=69\times 69\times 24$)} dimensional vector time series, the traditional vector factor model with 4 factors explains about 90.0\% variability, but uses $4\times 114,264$ parameters for the loading matrix. \cite{chen2022factor} used TFM-tucker with $4\times 4\times 4$ core factor tensor process. Using iTIPUP estimator of  \cite{han2020iterative}, TFM-tucker explains 80.1\% variability. 
Similarly, for the business-day series, the explained fractions of variability by the TFM-cp {with 4 factors}, vector factor model with 4 factors, and TFM-tucker with $4\times 4\times 4$ core factor tensor are 68.1\%, 90.9\%, 84.0\%, respectively.
Comparison of model complexity between TFM-cp and TFM-tucker is 
substantially more complex compared to that of traditional tensor decomposition, owing to the stochastic nature of the latent factor process. Since both models are estimated through the sample autocovariance tensor, we may count, under each model, the number of parameters required in the population version of the autocovariance tensor. Both models 
{require} $(69+69+24)\times 4$ number of parameters in terms of loading matrices or vectors, though the 
{degree} of freedom for TFM-tucker is slightly smaller due to orthonormal requirements of the loading matrices. However, for the lag-1 autocovariance of the factor processes, TFM-cp only requires 4 parameters (for the four  uncorrelated factor process) while TFM-tucker requires $(4\times4\times 4)^2$ parameters (minus certain savings from rotation ambiguity).
More importantly, TFM-cp harbors a much smaller dimensional factor process ($\bff_t\in \mathbb{R}^4$) than TFM-tucker ($\cF_t\in\mathbb{R}^{4\times 4\times 4}$), thereby simplifying subsequent modeling of the latent factor process.

The literature on Markov process models, for example \cite{zhang2019spectral,zhu2022learning}, regards each trip as a transition from the pickup location to the drop-off location, rendering the data as a collection of fragmented sample paths representative of a city-wide Markov process.
When one aggregates the individual trips within a time period to form a traffic volume matrix, the model, with an assumed fixed (reduced rank) Markov transition matrix, essentially induces an order-1 autoregressive model on the volume matrix time series.
Therefore, the distinction between the Markov process models and the CP factor models parallels that between autoregressive models and factor models.

\begin{figure}[H]
\centering
  \includegraphics[width=.18\textwidth,trim = 12mm 0 12mm 0,clip, page=2]{plots/pickup_businessday.pdf}
  \includegraphics[width=.18\textwidth,trim = 12mm 0 12mm 0,clip, page=4]{plots/pickup_businessday.pdf}
  \includegraphics[width=.18\textwidth,trim = 12mm 0 12mm 0,clip, page=1]{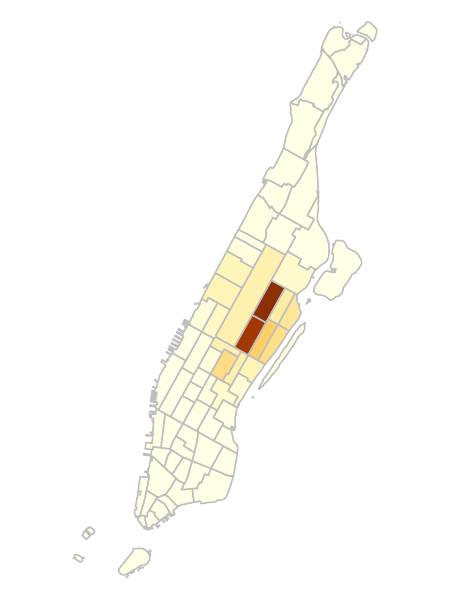}
  \includegraphics[width=.18\textwidth,trim = 12mm 0 12mm 0,clip, page=3]{plots/pickup_businessday.pdf}
  \includegraphics[width=.1\textwidth]{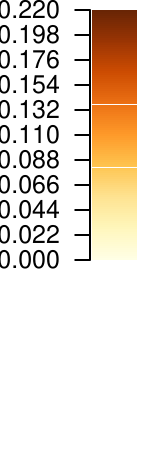}
  \caption{Loadings on four pickup factors for business day series.}
  \label{fig:pickup-bus}
\end{figure}

\begin{figure}[H]
\centering
  \includegraphics[width=.18\textwidth,trim = 12mm 0 12mm 0,clip, page=2]{plots/dropoff_businessday.pdf}
  \includegraphics[width=.18\textwidth,trim = 12mm 0 12mm 0,clip, page=4]{plots/dropoff_businessday.pdf}
  \includegraphics[width=.18\textwidth,trim = 12mm 0 12mm 0,clip, page=1]{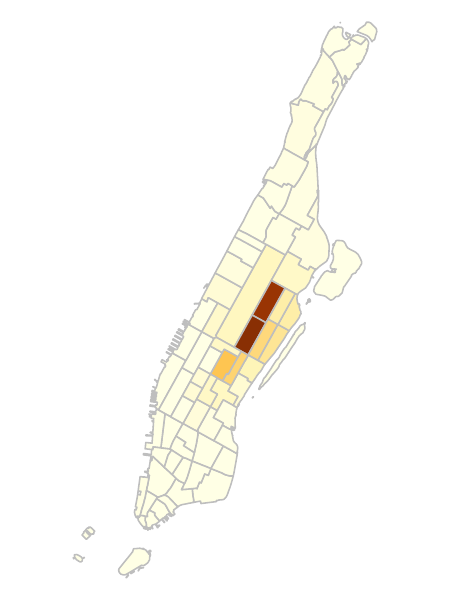}
  \includegraphics[width=.18\textwidth,trim = 12mm 0 12mm 0,clip, page=3]{plots/dropoff_businessday.pdf}
  \includegraphics[width=.1\textwidth]{plots/pickup_dropoff_color_bar.pdf}
    \caption{Loadings on four dropoff factors for business day series.}
  \label{fig:dropoff-bus}
\end{figure}

\begin{figure}[H]
\centering
  \includegraphics[width=.8\textwidth]{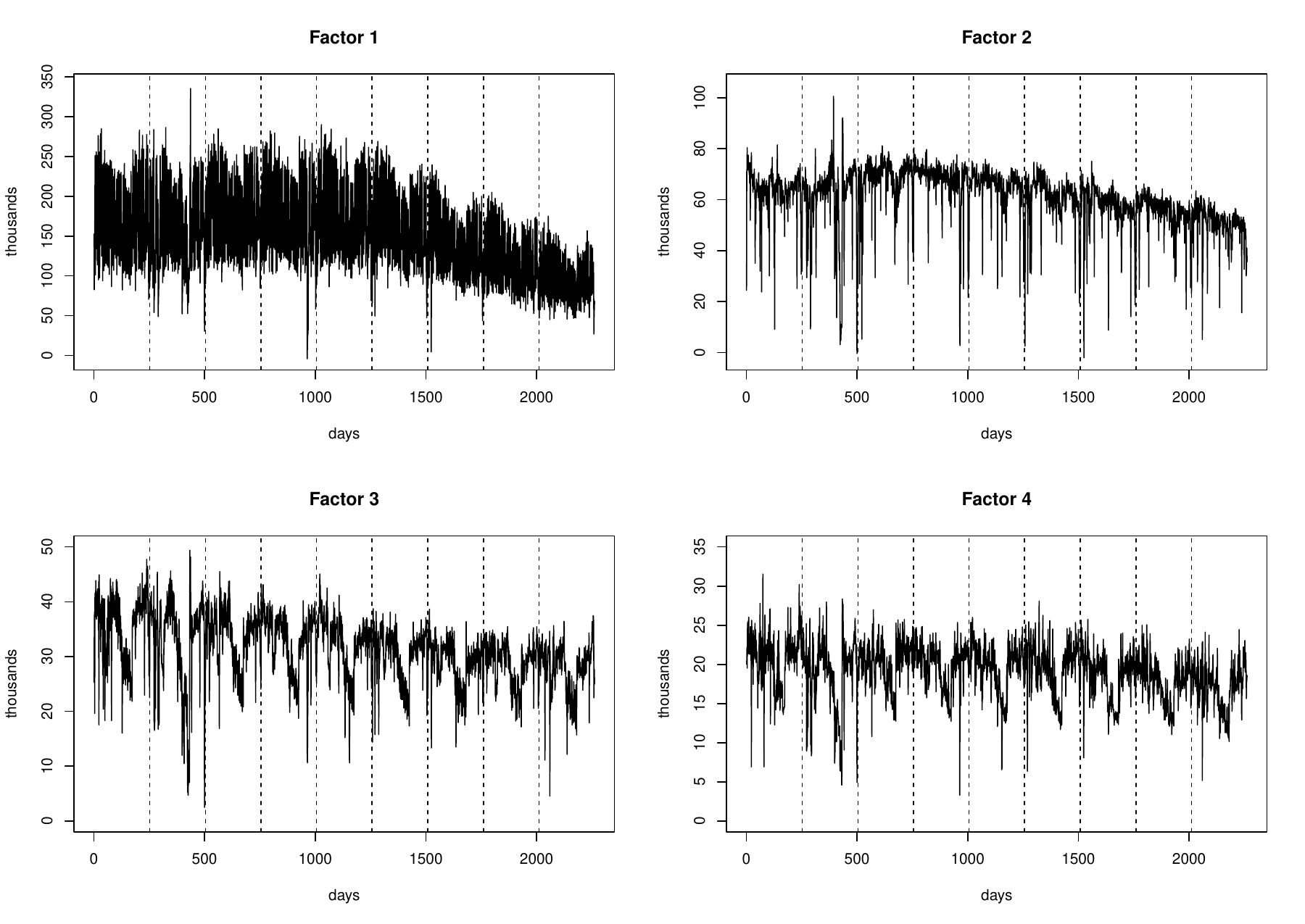}
    \caption{Estimated four factors for business day series.}
  \label{fig:ft-bus}
\end{figure}

Figures \ref{fig:pickup-bus} and \ref{fig:dropoff-bus} show the heatmap of the estimated loading vectors $(\ba_{11},\ldots, \ba_{41})$
(related to pick-up locations) and $(\ba_{12},\ldots, \ba_{42})$ (related to drop-off locations) of the 69 zones in Manhattan, respectively, for the business-day series. Table \ref{loading.3bus} shows the corresponding loading vectors $(\ba_{13},\ldots, \ba_{43})$ 
on the time of day dimension. For a more meaningful interpretation, we have re-scaled the loading vectors $\ba_{ik}$ and the factors $w_i f_{it}$ such that $\|\ba_{ik}\|_1=1$, 
{for} $1\le i\le 4, 1\le k\le 3$. Figure \ref{fig:ft-bus} shows the estimated four factors ($w_i f_{it}$) for business day series in 1,000. (Please note the significant difference in scale). For a more detailed examination, we show the four factor
series in the third year (year 2011) in Figure \ref{fig:ft_bus_year3} in Appendix~\ref{section:addition_taxi}.

It is seen that the estimated loading vectors and the factors are predominantly positive, although there are a few small negative values which we will ignore. When the loading vectors are scaled to sum to 1 (hence percentages), the model has the following interesting interpretation. First, the expected daily total volume ($\sum_{i,j,k}X_{t,ijk}$) is the sum of the four factors $w_1f_{1t}+\ldots+w_4f_{4t}$. Hence the daily traffic volumes essentially consist of taxi rides following four different patterns, each corresponding to the rank-1 tensor $\ba_{i1}\otimes\ba_{i2}\otimes\ba_{i3}$, $i=1,\ldots,4$. One may also imagine that there are four types of taxi users in the city, each following one specific traffic pattern (of course an individual may take multiple trips in a day and follow different patterns for each trip).

It is interesting to study the component of the rank-1 tensor $\ba_{i1}\otimes\ba_{i2}\otimes\ba_{i3}$.
Specifically, $\ba_{i3}$ shows how the total volume of traffic pattern $i$ ($w_if_{it}$) is distributed to different hours of the day. For example, Table \ref{loading.3bus} shows that 16\% of pattern 2 volume $w_2f_{2t}$ is allotted to between 7am and 8am, while only 3\% of pattern 4 volume $w_4f_{4t}$ is allotted to that hour. The rank-1 matrix $\ba_{i1}\ba_{i2}^\top$ shows the spatial pattern of traffic pattern $i$, where
$\ba_{i1}$ shows the percentage of pattern $i$ volume ($w_if_{it}$) at each hour being picked up in each location, and $\ba_{i2}$ shows the percentage of pattern $i$ volume $w_if_{it}$ being dropped-off in each location, and $a_{i1k} a_{i2\ell}$ is the percentage of pattern $i$ volume $w_if_{it}$ from location $k$ to location $\ell$. This spatial pattern does not change through the day, but the volume in each hour is controlled by $\ba_{i3}w_if_{it}$.

\begin{figure}[H]
\centering
  \includegraphics[width=.6\textwidth]{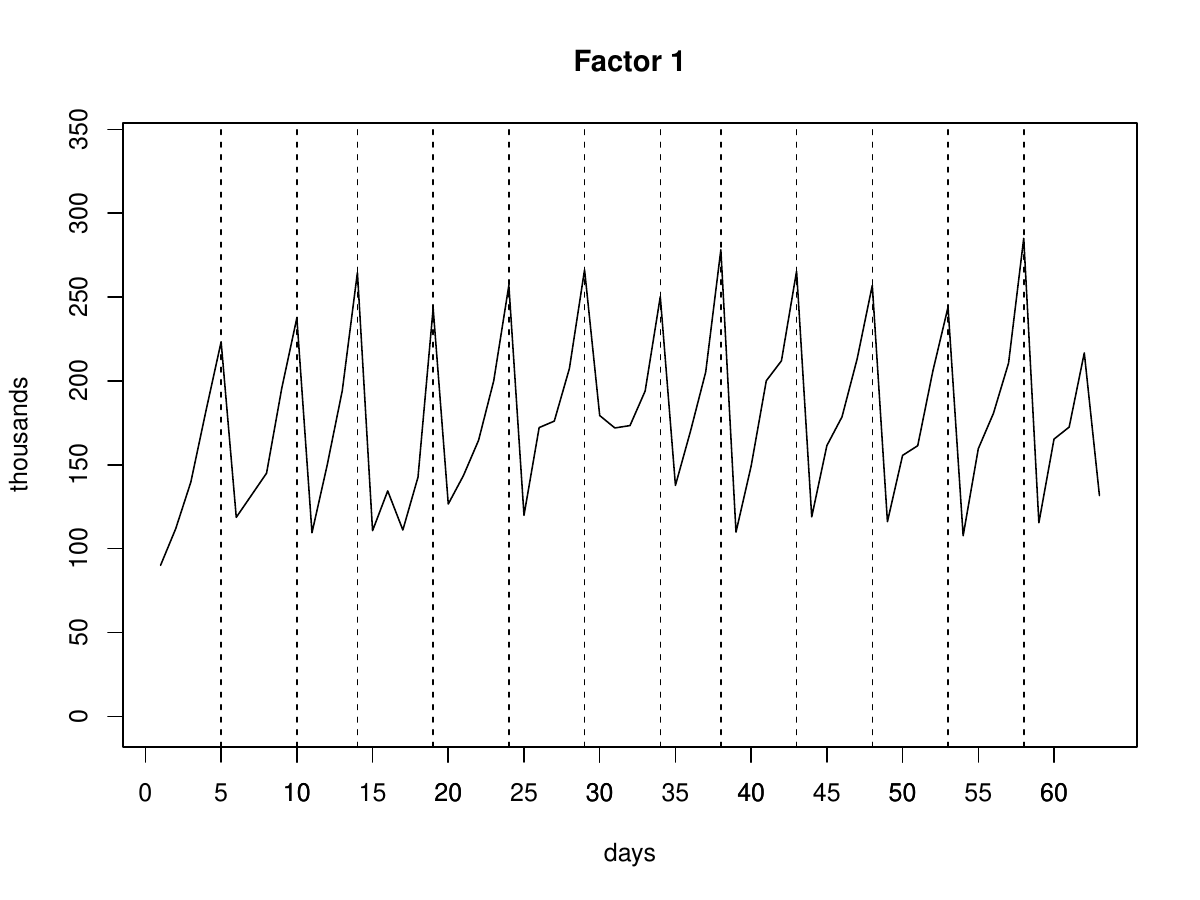}
    \caption{Estimated Factor 1 for business day series in the 3-months business-day period from January 1 to March 31 in year 2011. Vertical lines mark the end of business week.}
  \label{fig:ft_bus_year3_60}
\end{figure}

From Table~\ref{loading.3bus}, it is seen that Factor 1 (or traffic pattern 1) roughly corresponds to the evening hours of 6pm to 12am, by the loading vector $\ba_{13}$, with main activities in the SoHu and lower east side as both the pick-up and drop-off locations.
From the estimated factor series plot in Figure~\ref{fig:ft-bus}, it seems that this traffic pattern (pattern 1) has the largest overall volume, but with a very strong yearly seasonal pattern and a large daily variation. Intuitively people use less taxi service when the weather is nice, hence the volume is relatively small in summer and early fall, even though there are more evening activities in the summer. The large daily variation is due to a weekly effect. Figure~\ref{fig:ft_bus_year3_60} shows the 3-month business-day period from January 1 to March 31 in year 2011, in which the vertical line marks the end of working week (Friday or the day before holiday). It is clearly seen that, for this mainly evening-activity traffic pattern, the volume in the end of working week is almost twice as large as that in the beginning of the working week.

Again from Table~\ref{loading.3bus}, it is seen that Factor 2 (or traffic pattern 2) roughly corresponds to the morning rush hours of 6am to 12am, by the loading vector $\ba_{23}$, with main activities in the midtown area as the pick-up locations, and Times square and 5th Avenue as the drop-off locations. About $23.1\%$ (defined as $\sum_t w_2f_{2t}/\sum_t\sum_i w_if_{it}$) of the total traffic follows this pattern.
From Figure~\ref{fig:ft-bus}, it is seen that Factor 2 is quite stable throughout the year, which is again intuitively understandable as the traffic pattern is  mainly used by the steady population of people commuting to work in morning rush hours. There is a large number of (small value) outliers, most of them corresponding to the business days before or after major holidays. It can be
seen more clearly from Figure~\ref{fig:ft_bus_year3} in Appendix~\ref{section:addition_taxi}.

For Factors 3 and 4, the areas that load heavily on the factors for pick-up are quite similar to that for drop-off, i.e.,
upper east side (with affluent neighborhoods and museums) on Factor 3, and upper west side (with affluent neighborhoods and performing arts) on Factor 4. The conventional business hours are heavily and almost exclusively loaded on these factors. 
From Figure~\ref{fig:ft-bus}, it seems that both patterns have a yearly seasonal effect, small in the summer and early fall, which can be seen more clearly in Figure~\ref{fig:ft_bus_year3} in Appendix~\ref{section:addition_taxi}. Their volumes are relatively small than that of Factors 1 and 2.

We note that TFM-cp representation is unique which facilitates a more ``unique'' interpretation. On the other hand,
TFM-tucker is subject to arbitrary rotation. Using TFM-tucker to analyze the same data set, \cite{chen2022factor} used varimax rotation to obtain one specific representation of their estimated model and provided interesting interpretations. Their results are quite 
different from that of TFM-cp. First, since TFM-tucker representation requires orthonormal loading matrices, the discovered patterns in the loading matrices are forced to be different. For example, the daily patterns revealed in \cite{chen2022factor} have quite distinct periods, while {Table} \ref{loading.3bus} shows more intertwined (non-orthogonal) patterns. Second, TFM-tucker requires $4\times 4\times 4$ factor processes. The column loading vectors in each loading matrices work on all these factors, instead of on only one factor as in TFM-cp. The interpretation of these loading vectors are more convoluted. For example,
in TFM-tucker, the volume from all four heavily loaded pick-up areas identified by $\ba_{i1}, i=1,\ldots, 4$ can be traveling 
to all four heavily loaded drop-off areas $\ba_{i2}, i=1,\ldots, 4$. But in TFM-cp, the rank-1 matrix
$\ba_{i1}\ba_{i2}^\top$ shows the exact proportion of pattern $i$ traffic from each of the 
pick-up area identified by $\ba_{i1}$ 
to the drop-off area $\ba_{i2}$. In particular, it is seen that $\ba_{i1}$ is very similar to $\ba_{i2}$ for $i=1,3,4$. This observation suggests that our TFM-cp model may offer better intuitive understanding as it aligns with the expectation that most taxi traffic activities are likely confined within specific areas. 
This comparison further underscores the distinct analytical insights offered by the TFM-cp model in capturing the spatial-temporal dynamics of urban taxi traffic.

\begin{figure}[H]
\centering
  \includegraphics[width=.18\textwidth,trim = 12mm 0 12mm 0,clip, page=2]{plots/pickup_nonbusinessday.pdf}
  \includegraphics[width=.18\textwidth,trim = 12mm 0 12mm 0,clip, page=4]{plots/pickup_nonbusinessday.pdf}
  \includegraphics[width=.18\textwidth,trim = 12mm 0 12mm 0,clip, page=3]{plots/pickup_nonbusinessday.pdf}
  \includegraphics[width=.18\textwidth,trim = 12mm 0 12mm 0,clip, page=1]{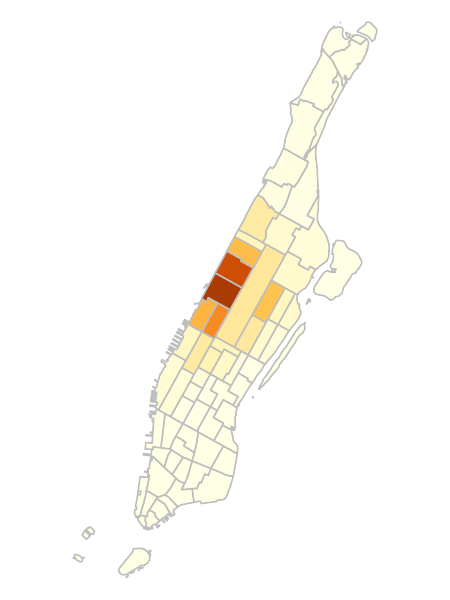}
  \includegraphics[width=.1\textwidth]{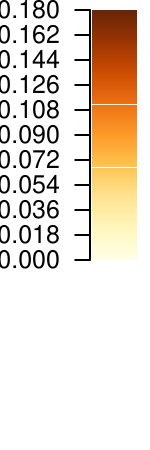}
  \caption{Loadings on four pick-up factors for non-business day series.}
  \label{fig:pickup-non}
\end{figure}

\begin{figure}[H]
\centering
  \includegraphics[width=.18\textwidth,trim = 12mm 0 12mm 0,clip, page=2]{plots/dropoff_nonbusinessday.pdf}
  \includegraphics[width=.18\textwidth,trim = 12mm 0 12mm 0,clip, page=4]{plots/dropoff_nonbusinessday.pdf}
  \includegraphics[width=.18\textwidth,trim = 12mm 0 12mm 0,clip, page=3]{plots/dropoff_nonbusinessday.pdf}
  \includegraphics[width=.18\textwidth,trim = 12mm 0 12mm 0,clip, page=1]{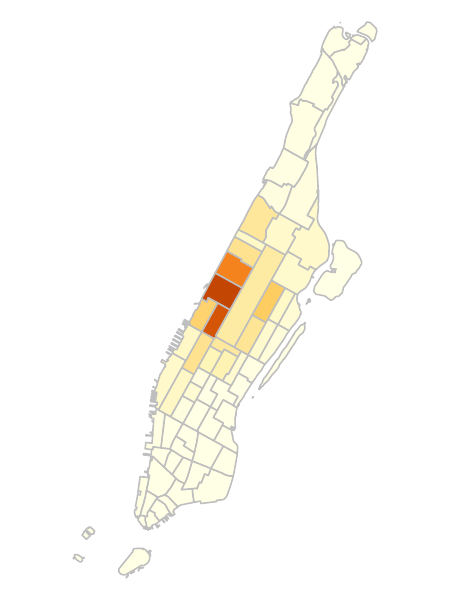}
  \includegraphics[width=.1\textwidth]{plots/pickup_dropoff_color_bar2.pdf}
  \caption{Loadings on four drop-off factors for non-business day series.}
    \label{fig:dropoff-non}
\end{figure}

\begin{figure}[H]
\centering
  \includegraphics[width=.8\textwidth]{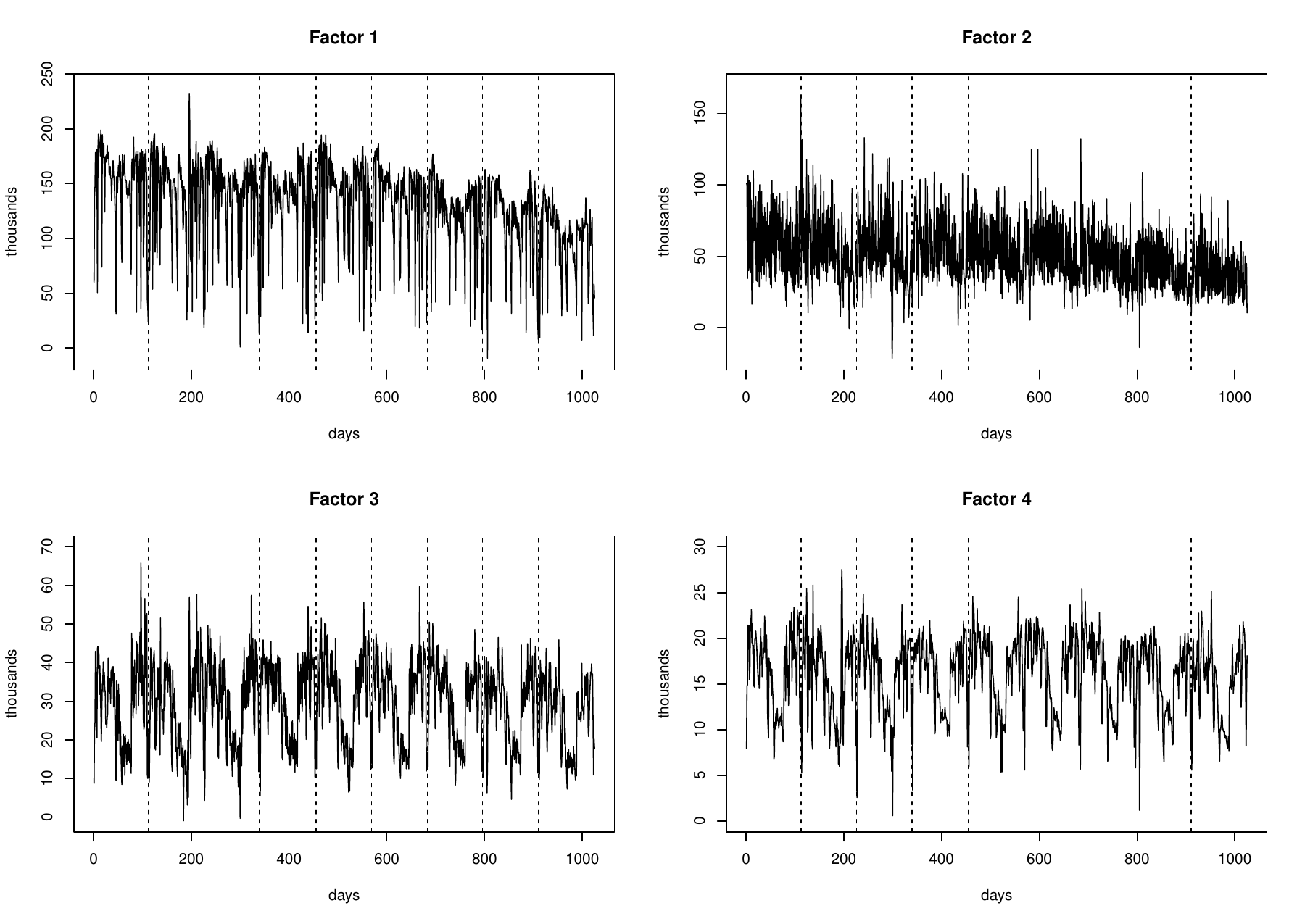}
    \caption{Estimated four factors for non-business day series.}
  \label{fig:ft-non}
\end{figure}

\begin{figure}[H]
\centering
  \includegraphics[width=.8\textwidth, height=2in]{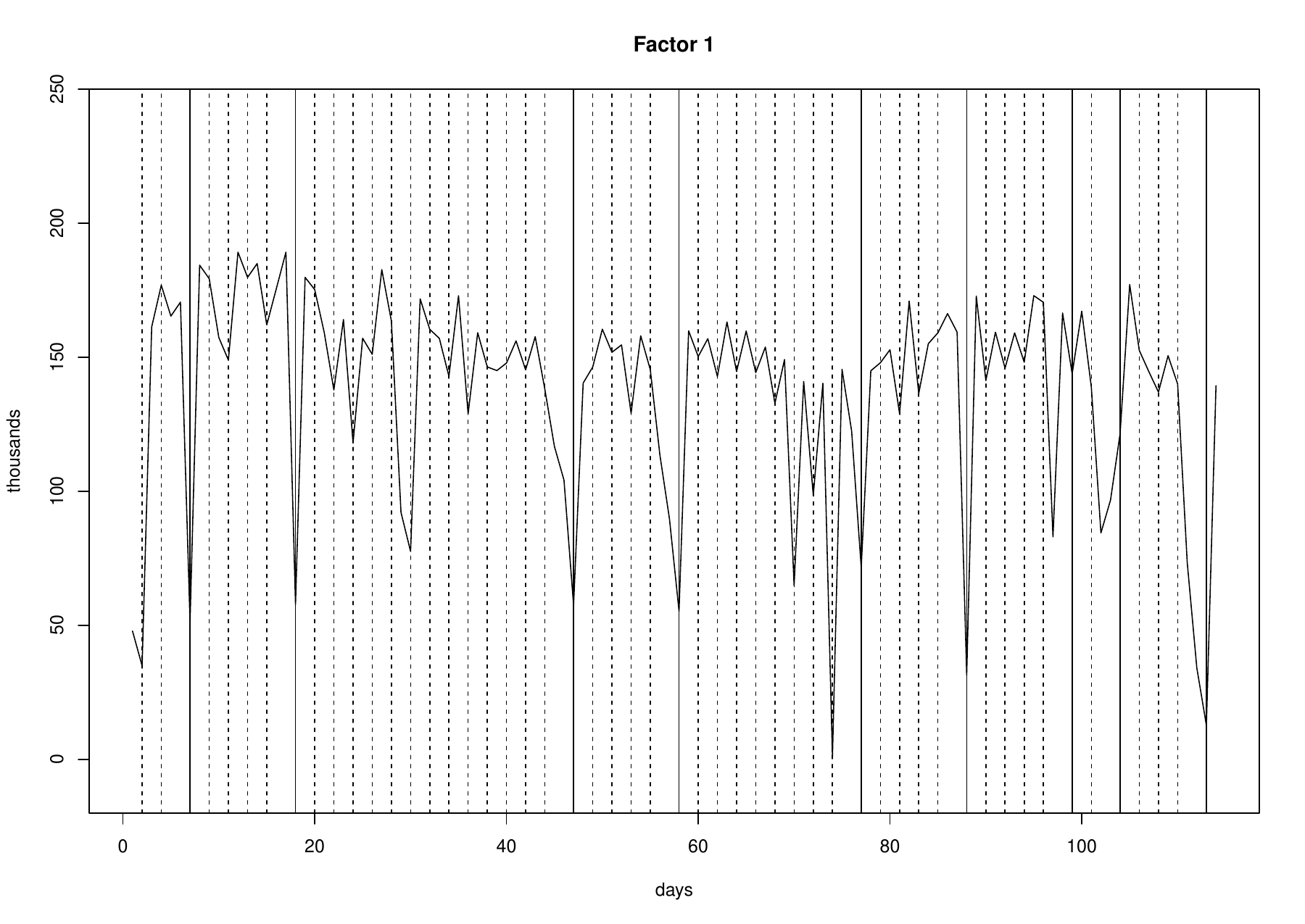}
    \caption{Estimated Factor 1 for non-business day series in Year 2011 with vertical lines indicating the day before a business day (dashed lines for Sundays and solid lines for Mondays of a long weeked.}
  \label{fig:ft_nonbus_year3_v2}
\end{figure}

For the non-business day series, the estimated loading vectors $(\ba_{11},\ldots, \ba_{41})$ (related to pick-up locations), $(\ba_{12},\ldots, \ba_{42})$ (related to drop-off locations), $(\ba_{13},\ldots, \ba_{43})$ (on the hour of day dimension), the estimated factors $(w_1f_{1t},\ldots,w_4f_{4t})$ are showed in Figures \ref{fig:pickup-non}, \ref{fig:dropoff-non}, Table \ref{loading.3non}, and Figure \ref{fig:ft-non}, respectively.  Understandably the morning rush hour pattern in the business day series (Factor 2) disappears here but the night-time pattern (Factor 1) now lasts deep into the early hours, comparing Tables \ref{loading.3bus} and \ref{loading.3non}. From Figure~\ref{fig:ft-non}, it is seen that there exist two different yearly seasonal patterns. Factors 3 and 4 are similar to that of business day series, with small volumes in the summer and fall, again confirming that the use of taxi service is relatively low when the weather is good for walking in the city.
On the other hand, the volume of Factor 2 is typically small in the winter time. The volumes of night-life pattern in Factor 1 remain to be volatile. It has many small-value outliers, mostly on the day before a business day (Sundays or the end of holiday.) These can be seen more
clearly in the more detailed Figure~\ref{fig:ft_nonbus_year3_v2}, which shows the estimated factors of all the non-business days in Year 2011 (year 3), with vertical lines indicating the day before a business day (dashed lines for Sundays and solid lines for Mondays of long weekend when Tuesday is the start of business week.) 
This is again intuitively understandable, because people tend not to stay out too late if they need to work the next day.

The pick-up and drop-off locations that heavily load on Factors 1, 3, 4 are similar to that for Factors 1, 3, 4 in the business day series. The daytime hours load on Factors 3 and 4, and the night life hours from 12am to 4am load on Factor 1. As for the second factor, it loads heavily on midtown area for pick-up, on the lower west side near Chelsea (with many restaurants and bars) for drop-off, on the afternoon/evening hours between 1pm to 8pm as the dominating periods.

\begin{table}
\begin{center}
\addtolength{\tabcolsep}{-3pt}
\begin{tabular}{c|rrrrrrrrrrrr|rrrrrrrrrrrr|c}
\multicolumn{2}{c}{0am} &
\multicolumn{2}{c}{2} &
\multicolumn{2}{c}{4} &
\multicolumn{2}{c}{6} &
\multicolumn{2}{c}{8} &
\multicolumn{2}{c}{10} &
\multicolumn{2}{c}{12pm} &
\multicolumn{2}{c}{2} &
\multicolumn{2}{c}{4} &
\multicolumn{2}{c}{6} &
\multicolumn{2}{c}{8} &
\multicolumn{2}{c}{10} &
\multicolumn{2}{c}{12am} \\ \hline
$i=1$ & 5 & 3 & 2 & 1 & 1 & 1 & 1  & 2  & 2  & 3  & 3 & 3 & 3 & 4 & 4 & 4 & 3 & 5 & 7 & 8 & 9 & 9 & 9 & 8 \\
$i=2$ & 0 & 0 & 0 & 0 & 1 & 3 & 12 & 16 & 13 & 11 & 7 & 6 & 5 & 4 & 4 & 3 & 2 & 2 & 3 & 3 & 2 & 2 & 1 & 1\\
$i=3$ & 1 & 0 & 0 & 0 & 0 & 0 & 2  & 5  &  7 & 6  & 6 & 7 & 7 & 7 & 7 & 8 & 6 & 7 & 7 & 6 & 4 & 3 & 2 & 1\\
$i=4$ & 1 & 1 & 0 & 0 & 0 & 0 & 1  & 3  & 5  & 4  & 5 & 5 & 6 & 6 & 7 & 7 & 6 & 8 & 9 & 8 & 6 & 5 & 4 & 3\\
\end{tabular}
\caption{Estimated four loading vectors $\ba_{i3}\in\R^{24}$ ($i=1,\ldots,4$), for hour of day mode.
Business day. Values are in percentage.} \label{loading.3bus}
\end{center}
\end{table}

\begin{table}
\begin{center}
\addtolength{\tabcolsep}{-3pt}
\begin{tabular}{c|rrrrrrrrrrrr|rrrrrrrrrrrr|c}
\multicolumn{2}{c}{0am} &
\multicolumn{2}{c}{2} &
\multicolumn{2}{c}{4} &
\multicolumn{2}{c}{6} &
\multicolumn{2}{c}{8} &
\multicolumn{2}{c}{10} &
\multicolumn{2}{c}{12pm} &
\multicolumn{2}{c}{2} &
\multicolumn{2}{c}{4} &
\multicolumn{2}{c}{6} &
\multicolumn{2}{c}{8} &
\multicolumn{2}{c}{10} &
\multicolumn{2}{c}{12am} \\ \hline
$i=1$ & 10 & 10 & 9 & 7 & 3 & 1 & 1 & 1 & 1 & 2 & 2 & 3 & 3 & 3 & 3 & 3 & 3 & 4 & 5 & 5 & 5 & 5 & 5 & 6\\
$i=2$ & 4  &  3 & 2 & 1 & 1 & 1 & 1 & 1 & 2 & 2 & 3 & 4 & 6 & 8 & 7 & 6 & 5 & 6 & 8 &10 & 6 & 5 & 5 & 4\\
$i=3$ & 2  &  1 & 1 & 1 & 0 & 0 & 1 & 2 & 3 & 6 & 7 & 8 & 8 & 8 & 8 & 8 & 7 & 6 & 6 & 5 & 4 & 3 & 2 & 2\\
$i=4$ & 3  &  2 & 1 & 1 & 1 & 0 & 1 & 1 & 3 & 5 & 6 & 7 & 7 & 7 & 7 & 7 & 6 & 6 & 7 & 7 & 5 & 4 & 4 & 3\\
\end{tabular}
\caption{Estimated four loading vectors $\ba_{i3}\in\R^{24}$ ($i=1, \ldots, 4$) for hour of day mode.
Non-Business day.  Values are in percentage.} \label{loading.3non}
\end{center}
\end{table}

We remark that this example is just for illustration and showcasing the interpretation of the proposed tensor factor model. Again we note that for the TFM-tucker model, one needs to identify a proper representation of the loading space in order to interpret the model. In \cite{chen2022factor}, varimax rotation was used to find the most sparse loading matrix representation to model interpretation. For TFM-cp, the model is unique hence interpretation can be made directly. Interpretation is impossible for the vector factor model in such a high dimensional case.



\section{Discussion} \label{section:discussion}

In this paper, we propose a tensor factor model with a low rank CP structure and develop its corresponding estimation procedures.
The estimation procedure takes advantage of the special structure of the model, resulting in faster convergence rate and more accurate estimations comparing to the standard procedures designed for the more general TFM-tucker, and the more general tensor CP decomposition. Numerical study illustrates the finite sample properties of the proposed estimators. The results show that HOPE uniformly outperforms the other methods, when the observations follow the specified TFM-cp.


The HOPE in this paper is based on CP decomposition of the second moment tensor $\bSigma_h=\sum_{i=1}^r \lambda_i (\otimes_{k=1}^K\ba_{ik})^{\otimes 2}$,
an order $2K$ tensor.
The intuition that higher order tensors tend to have smaller coherence among the CP components leads to the consideration of using higher order cross-moments 
to have more orthogonal CP components.
For example, let
the $m$-th cross moment tensor with lags $0=h_1<\cdots < h_m$ be
\begin{align*}
\bSigma_{h_1...h_m}^{(m)}= \E\left[\otimes_{j=1}^{m} \cX_{t-h_j}\right].
\end{align*}
When the factor processes $f_{it}$, $i=1,\ldots, r$ are independent across different $i$ in TFM-cp, a naive 4-th cross moment tensor to estimate $\ba_{ik}$ is
\begin{align*}
&\bSigma_{h_1h_2h_3h_4}^{(4)}-  \bSigma_{h_1h_2}^{(2)} \otimes \bSigma_{h_3h_4}^{(2)} - \E[\cX_{t-h_1}\otimes\cX^*_{t-h_2}\otimes \cX_{t-h_3} \otimes\cX_{t-h_4}^*]  - \E[\cX_{t-h_1}\otimes\cX^*_{t-h_2}\otimes \cX^*_{t-h_3} \otimes\cX_{t-h_4}] \\
&=\sum_{i=1}^r \lambda_{i,h_1h_2h_3h_4}^{(4)} (\otimes_{k=1}^K \ba_{ik})^{\otimes 4},
\end{align*}
with $\{\cX_t^*\}$ being an independent coupled process of $\{\cX_t\}$ and
when $h_j=(j-1)h$,
\begin{align*}
\lambda_{i,h_1h_2h_3h_4}^{(4)} & =\E \prod_{j=0}^3 f_{i,t-jh} - [\E f_{i,t} f_{i,t-h}]^2 - [\E f_{i,t} f_{i,t-2h}]^2 - [\E f_{i,t} f_{i,t-h}] [\E f_{i,t} f_{i,t-3h}].
\end{align*}
This naive 4-th cross moment tensor has more orthogonal CP bases. In light tailed case, simulation shows that it is much worse than the second moment tensor, due to the reduced signal strength $\lambda_{i,h_1h_2h_3h_4}^{(4)}$. However, for heavy tailed and skewed data, this procedure would be helpful. It would be an interesting and challenging problem to develop an efficient higher cross moment tensor to improve the statistical and computational performance. We leave this for future research.

Our primary consideration was directed towards the CP factor model in a time series setting, as the need of effectively analyzing tensor time series has arisen in many applications, and CP factor model is an efficient approach for such analysis. Without a specified (parametric) model for the latent factor processes (an topic currently under investigation), we focus on the auto-covariance and auto-cross-moment tensor for effective estimation of the proposed model. This is the main contribution of the paper. However, the proposed cPCA and ISO can be used directly in or be extended to many other problems involving CP decomposition of certain type of tensors. For example, in many problems where higher order moments
can be introduced to reduce incoherence, the cPCA and ISO algorithms may offer better initialization and outperform conventional tensor power iteration methods. One specific example is the kurtosis tensor in independent component analysis in \cite{auddy2023large}. Another possible extension is highlighted in Remark \ref{rmk:tied_eigen} where we pointed out how cPCA can be modified to deal with situations when a few leading
singular values of the auto-cross-moment tensor are the same.




\bibliographystyle{apalike}
\bibliography{cp}

\newpage

\appendix

\begin{center}
{\LARGE \textbf{Supplementary Material to ``CP Factor Model for Dynamic Tensors''}}
\end{center}

\bigskip \centerline{Yuefeng Han, Dan Yang,  Cun-Hui Zhang and Rong Chen}

\centerline{University of Notre Dame, The University of Hong Kong, and Rutgers University}

\section{Proofs} \label{section:proof}

\subsection{Proofs of main theorems}

Let $[n]$ denote the set $\{1, 2, \ldots, n\}$. For a matrix $A = (a_{ij})\in \RR^{m\times n}$, write the SVD as $A=U\Sigma V^\top$, where $\Sigma=\text{diag}(\sigma_1(A), \sigma_2(A), ..., \sigma_{\min\{m,n\}}(A))$, with the singular values $\sigma_1(A)\ge\sigma_2(A)\ge \cdots\ge \sigma_{\min\{m,n\}}(A)\ge 0$ in descending order. For nonempty $J\subseteq [K]$, $\text{mat}_J(\cA)$ is the mode $J$ matrix unfolding which maps $\cA$ to $m_J\times m_{-J}$ matrix with $m_J=\prod_{j\in J}m_j$ and $m_{-J}=m/m_J$, e.g. $\text{mat}_{\{1,2\}}(\cA)=\text{mat}_3^\top(\cA)$ for $K=3$. Denote $a\wedge b=\min\{a,b\}$ and $a\vee b=\max\{a,b\}$.

\begin{proof}[\bf Proof of Proposition \ref{prop:delta}]
Recall that $\bA=(\ba_{1},...,\ba_{r})\in\R^{d \times r}$, $\delta = \|\bA^\top \bA-I_r\|_{\rm S}$ and $\delta_k = \|\bA_k^\top \bA_k-I_r\|_{\rm S}$.
Because $\bA^\top \bA= (\bA_1^\top \bA_1)\circ \cdots \circ(\bA_K^\top \bA_K)$ is
the Hadamard product of correlation matrices, the spectrum of $\bA^\top \bA$ is
contained inside the spectrum limits of $\bA_k^\top \bA_k$ for each $k$, so that
\begin{align*}
\delta\le\min_{1\le k\le K}\delta_k.
\end{align*}
Because $\bA^\top \bA -I_r$ is symmetric, its spectrum norm is bounded by its $\ell_1$ norm,
\begin{align*}
\delta \le \max_{j\le r} \sum_{i\neq j}|\ba_i^\top  \ba_j| \le (r-1)\vartheta \le (r-1) \prod_{k=1}^K \vartheta_k
\end{align*}
due to $|\ba_i^\top  \ba_j| =\prod_{k=1}^K|\ba_{ik}^\top \ba_{jk}|=\prod_{k=1}^K|\sigma_{ij,k}|$.
Moreover, for any $j\le r$ and $1\le k_1<k_2\le K$,
\begin{align*}
\sum_{i\neq j}\prod_{k=1}^K |\sigma_{ij,k}|
&\le \sum_{i\neq j} |\sigma_{ij,k_1}\sigma_{ij,k_2}|
\,\max_{i\neq j}\prod_{k\neq k_1,k\neq k_2}|\sigma_{ij,k}|
\cr &\le \Big(\prod_{k=1}^K \eta_{jk}\Big)r^{-(K-2)/2}
\,\max_{i\neq j}\prod_{k\neq k_1,k\neq k_2}\sqrt{r}|\sigma_{ij,k}|/\eta_{jk}
\end{align*}
as $\eta_{jk} = (\sum_{i\neq j}\sigma_{ij,k}^2)^{1/2}$.
The proof is complete as $k_1$ and $k_2$ are arbitrary.
\end{proof}

\begin{proof}[\bf Proof of Theorem \ref{thm:initial}]
Recall $\widehat\bSigma_{h}=\sum_{t=h+1}^T \cX_{t-h}\otimes\cX_t$, $\lambda_i=w_i^2 \cdot \E f_{i,t-h}f_{i,t}$, $\ba_i=\vec(\ba_{i1}\otimes\ba_{i2}\otimes\cdots\otimes\ba_{iK})$, $d=d_1d_2...d_K$. Let $e_t=\vec(\cE_t)$. Write
\begin{align*}
\widetilde\bSigma:=\mat_{[K]} \big(\widehat\bSigma_{h} \big)=\sum_{i=1}^r  \lambda_i \ba_i \ba_i^\top+\bPsi^* =  \bA \bLambda \bA^\top + \bPsi^* ,
\end{align*}
where $\bLambda=\text{diag}(\lambda_1,...,\lambda_r)$. Assume $\bA$ has SVD $\bA=\widetilde U\widetilde D \widetilde V^\top$, with $\widetilde D=\text{diag}(\widetilde\sigma_1,...,\widetilde\sigma_r)$. Let $\bU=\widetilde U \widetilde V^\top=(\bu_1,...,\bu_r)$. Then, $\bU^{\top} \bU =\bI_r$. Note that $\widetilde\bSigma/2+\widetilde\bSigma^\top/2$ is guaranteed to be symmetric. Without loss of generality, assume $\widetilde\bSigma$ is symmetric.

By \eqref{corr-all}, $\|\bA^\top\bA-\bI_r\|_{\rm S}=\|\widetilde D^2-\bI_r\|_{\rm S}\le\delta$. It follows that $\max_{1\le i\le r}|\widetilde\sigma_i^2-1|\le \delta$. As $\delta<1$, basic calculation shows that $\max_i|\widetilde\sigma_i-1|\le \delta/(2-\delta)$. Thus,
\begin{align*}
\|\bA-\bU\|_{\rm S}=\|\widetilde D-\bI_r\|_{\rm S}\le \delta/(2-\delta)<\delta<1.
\end{align*}
This yields
\begin{align*}
\max_i\|\ba_i-\bu_i\|_{2}\le \delta <1 .
\end{align*}
It follows that $\max_i(2 - 2\ba_i^\top\bu_i)\le \delta^2$. Since $1-x^2 \le 2-2x$, we have
\begin{align}\label{eq1:thm:initial}
\|\ba_i\ba_i^\top - \bu_i \bu_i^{\top} \|_{\rm S}=\sqrt{1 - (\ba_i^\top\bu_i)^2 }\le \sqrt{2 - 2(\ba_i^\top\bu_i) }\le \delta <1 .
\end{align}

Let the top $r$ eigenvectors of $\widetilde\bSigma$ be $ \widehat \bU=(\widehat \bu_1, ..., \widehat \bu_r)\in\R^{d\times r}$. Note that $\lambda_1>\lambda_2>...>\lambda_r> \lambda_{r+1}=0$. By Wedin's perturbation theorem \citep{wedin1972} and Lemma \ref{lemma-transform-ext}, for any $1\le i\le r$,
\begin{align}
\|\widehat  \bu_i \widehat  \bu_i^\top - \bu_i  \bu_i^{\top} \|_{\rm S} \le  \frac{2 \| \bA \bLambda \bA^{\top} - \bU \bLambda \bU^{\top}
+ \bPsi^* \|_{\rm S}} {\min\{ \lambda_{i-1}-\lambda_i, \lambda_i-\lambda_{i+1}\}}
\le \frac{ 2\lambda_1 \delta + 2 \|\bPsi^*\|_{\rm S} }{\min\{ \lambda_{i-1}-\lambda_i, \lambda_i-\lambda_{i+1}\}}. \label{eq2:thm:initial}
\end{align}
Combining \eqref{eq1:thm:initial} and \eqref{eq2:thm:initial}, we have
\begin{align}\label{eq3:thm:initial}
\|\widehat \bu_i \widehat \bu_i^\top - \ba_i \ba_i^\top \|_{\rm S}&\le \delta+\frac{2\lambda_1 \delta + 2 \|\bPsi^*\|_{\rm S} }{\min\{ \lambda_{i-1}-\lambda_i, \lambda_i-\lambda_{i+1}\}}.
\end{align}

We formulate each $\widehat \bu_i\in\R^d$ to be a $K$-way tensor $\widehat \bU_i\in \R^{d_1\times\cdots\times d_K}$. Let $\widehat \bU_{ik}=\mat_k(\widehat \bU_i)$, which is viewed as an estimate of  $\ba_{ik}\vec(\otimes_{\ell\neq k}^K \ba_{i\ell})^\top\in\R^{d_k\times (d/d_k)}$. Then $\widehat\ba^{\cpca}_{ik}$ is the top left singular vector of $\widehat \bU_{ik}$.
By Lemma \ref{lemma-rank-1-approx},
\begin{align}
\|\widehat\ba^{\cpca}_{ik} \widehat\ba^{\cpca\top}_{ik} - \ba_{ik} \ba_{ik}^\top \|_{\rm S}^2 \wedge (1/2) \le
\|\widehat \bu_i \widehat \bu_i^\top - \ba_i \ba_i^{\top} \|_{\rm S}^2 .
\end{align}
Substituting \eqref{eq3:thm:initial} and Lemma \ref{lem:psi} into the above equation, since $\|\bPsi^*\|_{\rm S}/\min\{ \lambda_{i-1}-\lambda_i, \lambda_i-\lambda_{i+1}\} \lesssim 1$, we have the desired results.
\end{proof}

\begin{lemma}\label{lem:psi}
Suppose Assumptions \ref{asmp:error}, \ref{asmp:mixing} hold and $\delta<1$. Let $\widetilde\bSigma=\sum_{i=1}^r  \lambda_i \ba_i \ba_i^\top+\bPsi^*$ and $1/\gamma=1/\gamma_1+2/\gamma_2$. In an event with probability at least $1-(Tr)^{-c}/2-e^{-d}/6$,
we have
\begin{align}\label{eq:lem:psi}
\|\bPsi^*\|_{\rm S} &\le C  \max_{1\le i\le r} w_i^2  \left(\sqrt{\frac{r+ \log T}{T}} + \frac{(r+ \log T)^{1/\gamma}}{T}  \right) + \frac{C \sigma^2(d+\sqrt{dT})}{T} \notag\\
&\qquad\quad + C \sigma \max_{1\le i\le r} w_i \sqrt{\frac{d}{T}}.
\end{align}
\end{lemma}

\begin{proof}
Let $\Upsilon_0=T^{-1} \sum_{t=1}^T\sum_{i,j=1}^r w_iw_j  f_{it}f_{jt}  \ba_i \ba_j^\top$, $\overline\E(\cdot)=\E(\cdot|f_{it},1\le i\le r, 1\le t\le T)$. Define $e_t=\vec(\cE_t)$. Write
\begin{align*}
\widetilde\bSigma &= \frac{1}{T-h}\sum_{t=h+1}^T \vec(\cX_{t-h})\vec(\cX_{t})^\top \\
&= \sum_{i=1}^r  \lambda_i \ba_i \ba_i^\top +  \sum_{i,j=1}^r \frac{1}{T-h} \sum_{t=h+1}^T w_iw_j \left( f_{i,t-h}f_{j,t} - \E f_{i,t-h}f_{j,t} \right) \ba_i \ba_j^\top +   \frac{1}{T-h} \sum_{t=h+1}^T e_{t-h} e_t^\top\\
&\quad +   \frac{1}{T-h} \sum_{t=h+1}^T \sum_{i=1}^r w_i f_{i,t-h} \ba_i e_t^\top +   \frac{1}{T-h} \sum_{t=h+1}^T \sum_{i=1}^r w_i f_{i,t}  e_{t-h} \ba_i^\top \\
&:= \sum_{i=1}^r  \lambda_i \ba_i \ba_i^\top +\Delta_1+\Delta_2+\Delta_3+\Delta_4.
\end{align*}
That is, $\bPsi^*=\Delta_1+\Delta_2+\Delta_3+\Delta_4$.

We first bound $\|\Delta_{1}\|_{\rm S}$. Let
\begin{align}\label{def:theta}
\Theta_h=\sum_{t=h+1}^T
\begin{pmatrix}
f_{1,t-h} f_{1,t} & f_{1,t-h} f_{2,t} & \cdots & f_{1,t-h} f_{r,t} \\
\vdots & \vdots & \ddots & \vdots \\
f_{r,t-h} f_{1,t} & f_{r,t-h} f_{2,t} &\cdots & f_{r,t-h} f_{r,t}
\end{pmatrix}.
\end{align}
Then
\begin{align*}
\Delta_1=  (w_1 \ba_1,...,w_r\ba_r) \left( \frac{\Theta_h - \E \Theta_h}{T-h} \right) (w_1 \ba_1,...,w_r\ba_r)^\top .
\end{align*}
For any unit vector $u$ in $\R^r$, there exist $u_j\in\R^r$ with $\|u_j\|_2\le 1$, $j=1,...,N_{r,\epsilon}$ such that $\max_{\|u\|_2\le 1} \min_{1\le j\le N_{r,\epsilon}} \| u -u_j\|_2\le \epsilon$. The standard volume comparison argument implies that the covering number $N_{r,\epsilon} = \lfloor(1+2/\epsilon)^r\rfloor$. Then, there exist $u_j, v_{j'} \in\R^r$, $1\le j, j'\le N_{r,1/3}:= 7^r$, such that $\|u_j\|_2=\|v_{j'}\|_2=1$ and
\begin{align*}
\|\Theta_h -\E \Theta_h \|_{\rm S} -\max_{1\le j, j'\le N_{r,1/3}} \left|u_j^\top (\Theta_h -\E \Theta_h) v_j \right| \le (2/3)\|\Theta_h -\E \Theta_h \|_{\rm S}.
\end{align*}
It follows that
\begin{align*}
\|\Theta_h -\E \Theta_h \|_{\rm S} \le 3\max_{1\le j, j'\le N_{r,1/3}} \left|u_j^\top (\Theta_h -\E \Theta_h) v_j \right| .
\end{align*}
As $1/\gamma=1/\gamma_1+2/\gamma_2$, by Theorem 1 in \citet{merlevede2011},
\begin{align}\label{eq:thm_initial:f}
\P\left( \left| u_j^\top (\Theta_h -\E \Theta_h) v_j\right|\ge x \right) &\le T\exp\left(-\frac{x^\gamma}{c_1}\right)+ \exp\left(-\frac{x^2}{c_2 T}\right) \notag\\
&\quad + \exp\left(-\frac{x^2}{c_3 T}\exp\left(\frac{x^{\gamma(1-\gamma)}}{c_4(\log x)^\gamma} \right) \right).
\end{align}
Hence,
\begin{align*}
\P\left( \left\| \Theta_h- \E \Theta_h \right\|_{\rm S}/3 \ge x \right) &\le  N_{r,1/3}^2 T\exp\left(-\frac{x^\gamma}{c_1}\right)+  N_{r,1/3}^2\exp\left(-\frac{x^2}{c_2 T}\right) \notag\\
&\quad + N_{r,1/3}^2 \exp\left(-\frac{x^2}{c_3 T}\exp\left(\frac{x^{\gamma(1-\gamma)}}{c_4(\log x)^\gamma} \right) \right).
\end{align*}
As $h$ is fixed and $T\ge 4h$, choosing $x\asymp \sqrt{T(r+\log T)} + (r+\log T )^{1/\gamma}$, in an event $\Omega_1$ with probability at least $1-(Tr)^{-C_1}/2$,
\begin{align*}
 \left\| \frac{ \Theta_h- \E \Theta_h }{T-h} \right\|_{\rm S} \le C\sqrt{\frac{r+\log(T)}{T}} + \frac{C(r+\log T)^{1/\gamma}}{T} .
\end{align*}
It follows that, in the event $\Omega_1$,
\begin{align*}
\|\Delta_1\|_{\rm S} &\le \|\bA\|_{\rm S}^2\max_{1\le i\le r} w_i^2 \cdot   \left\| \frac{\Theta_h - \E \Theta_h}{T-h} \right\|_{\rm S} \\
& \le C  \max_{1\le i\le r} w_i^2 \left(\sqrt{\frac{r+ \log T}{T}} + \frac{(r+ \log T)^{1/\gamma}}{T}  \right),
\end{align*}
and, as $\E f_{i,t}^2=1$,
\begin{align}\label{eq:thm_initial:upsilon}
\| \Upsilon_0 \|_{\rm S} &\le \Big\| \sum_{i=1}^r w_i^2 \ba_i \ba_j^\top\Big\|_{\rm S}+ \Big\| \frac1T \sum_{t=1}^T\sum_{i,j=1}^r w_iw_j  (f_{i,t}f_{j,t}-\E f_{i,t}f_{j,t})  \ba_i \ba_j^\top\Big\|_{\rm S} \notag\\
&\le \| \bA\|_{\rm S}^2 \max_{1\le i\le r} w_i^2  + \Big\| \frac1T \sum_{t=1}^T\sum_{i,j=1}^r w_iw_j  (f_{i,t}f_{j,t}-\E f_{i,t}f_{j,t})  \ba_i \ba_j^\top\Big\|_{\rm S} \notag\\
&\le (1+\delta)^2 \max_{1\le i\le r} w_i^2 + C  \max_{1\le i\le r} w_i^2 \left(\sqrt{\frac{r+ \log T}{T}} + \frac{(r+ \log T)^{1/\gamma}}{T}  \right) \\
&:=\Delta_{\Upsilon} .   \notag
\end{align}
Note that $\Delta_{\Upsilon} \lesssim \max_{1\le i\le r} w_i^2 .$

For $\|\Delta_2\|_{\rm S}$, we split the sum into two terms over the index sets, $S_1=\{(h,2h]\cup(3h,4h]\cup\cdots \} \cap(h,T]$ and its complement $S_2$ in $(h,T]$, so that $\{e_{t-h},t\in S_a\}$ is independent of $\{e_t, t\in S_a\}$ for each $a=1,2$. Let $n_a=|S_a|$. By Lemma \ref{lm-GH}(i),
\begin{align*}
\P\left( \left\| \sum_{t\in S_a} e_{t-h} e_t^\top \right\|_{\rm S}\ge \sigma^2 (d+2\sqrt{dn_a}) + \sigma^2 x(x+2\sqrt{n_a}+2\sqrt{d}) \right)\le 2e^{-\frac{x^2}{2}}.
\end{align*}
With $x\asymp\sqrt{d}$, in an event $\Omega_2$ with probability at least $1-e^{-d}/18$,
\begin{align*}
\|\Delta_2\|_{\rm S} &\le  \sum_{a=1,2} \left\| \sum_{t\in S_a} \frac{e_{t-h} e_t^\top}{T-h} \right\|_{\rm S} \le \frac{C \sigma^2(d+\sqrt{dT})}{T}.
\end{align*}

For $\Delta_3$, by Assumption \ref{asmp:error}, for any $u$ and $v$ in $\R^{d}$ we have
\begin{align*}
\overline\E \left\{u^\top\left( \frac{1}{\sqrt{T}} \sum_{t=h+1}^T \sum_{i=1}^r w_i f_{i,t-h} \ba_i e_t^\top \right) v\right\}^2 \le \sigma^2 \|\Upsilon_0\|_{\rm S}\|u\|_2^2 \|v\|_2^2.
\end{align*}
Thus for vectors $u_i$ and $v_i$ with $\|u_i\|_2=\|v_i\|_2=1$, $i=1,2$,
\begin{align*}
&\sigma^{-2}\|\Upsilon_0\|_{\rm S}^{-1}\overline\E\left( u_1^\top\left( \frac{1}{\sqrt{T}} \sum_{t=h+1}^T \sum_{i=1}^r w_i f_{i,t-h} \ba_i e_t^\top \right)v_1 -u_2^\top\left( \frac{1}{\sqrt{T}} \sum_{t=h+1}^T \sum_{i=1}^r w_i f_{i,t-h} \ba_i e_t^\top \right)v_2  \right)^2 \\
&\le \left( \|u_1-u_2\|_2 \| v_1\|_2 + \|u_2\|_2 \| v_1-v_2\|_2\right)^2 \\
&\le 2 \left( \|u_1-u_2\|_2 ^2 + \| v_1-v_2\|_2^2 \right) \\
&=2\E \{ (u_1-u_2)^\top \xi + (v_1-v_2)^\top\zeta \}^2,
\end{align*}
where $\xi$ and $\zeta$ are iid $N(0, I_d)$ vectors. The Sudakov-
Fernique inequality yields
\begin{align*}
\sigma^{-2}\|\Upsilon_0\|_{\rm S}^{-1}\overline\E \left\| \frac{1}{\sqrt{T}} \sum_{t=h+1}^T \sum_{i=1}^r w_i f_{i,t-h} \ba_i e_t^\top \right\|_{\rm S}  \le \sqrt{2}\E\sup_{\|u\|_2=\|v\|_2=1} \left|u^\top\xi+v^\top\zeta \right| \le \sqrt{2}\E (\|\xi\|_2 + \|\zeta\|_2)
\end{align*}
As $\E \|\xi\|_2 =\E \|\zeta\|_2\le\sqrt{d}$, it follows that
\begin{align}
\overline\E\left\| \frac{1}{T-h} \sum_{t=h+1}^T \sum_{i=1}^r w_i f_{i,t-h} \ba_i e_t^\top \right\|_{\rm S} \le \frac{\sigma\sqrt{8Td}\| \Upsilon_0 \|_{\rm S}^{1/2}}{T-h}.
\end{align}
Elementary calculation shows that, $\left\|  \sum_{t=h+1}^T \sum_{i=1}^r w_i f_{i,t-h} \ba_i e_t^\top \right\|_{\rm S}$ is a $\sigma\sqrt{T} \| \Upsilon_0 \|_{\rm S}^{1/2}$ Lipschitz function. Then, by Gaussian concentration inequalities for Lipschitz functions,
\begin{align*}
\P\left( \left\|\frac{1}{T-h} \sum_{t=h+1}^T \sum_{i=1}^r w_i f_{i,t-h} \ba_i e_t^\top \right\|_{\rm S} \ge \frac{\sigma\sqrt{8Td}}{T-h} \| \Upsilon_0 \|_{\rm S}^{1/2} + \frac{\sigma\sqrt{T} }{T-h} \|\Upsilon_0 \|_{\rm S}^{1/2}x \right) \le 2e^{-\frac{x^2}{2}}.
\end{align*}
With $x\asymp\sqrt{d}$, in an event $\Omega_3$ with probability at least $1-e^{-d}/18$,
\begin{align*}
\|\Delta_3\|_{\rm S} \le C \sigma\sqrt{\frac{d}{T}} \| \Upsilon_0 \|_{\rm S}^{1/2}.
\end{align*}
Hence, in the event $\Omega_1\cap\Omega_3$,
\begin{align*}
\|\Delta_3\|_{\rm S} \le C \sigma\sqrt{\frac{d}{T}} \Delta_{\Upsilon}^{1/2}.
\end{align*}
Similarly, $\|\Delta_4\|_{\rm S} \le C \sigma\sqrt{d/T} \Delta_{\Upsilon}^{1/2}$.

Therefore, in the event $\Omega_1\cap\Omega_2\cap\Omega_3$ with probability at least $1-(Tr)^{-c}/2-e^{-d}/6$,
\begin{align*}
\|\bPsi^*\|_{\rm S} &\le C  \max_{1\le i\le r} w_i^2 \left(\sqrt{\frac{r+ \log T}{T}} + \frac{(r+ \log T)^{1/\gamma}}{T}  \right)+ \frac{C \sigma^2(d+\sqrt{dT})}{T}   \\
&\qquad\quad +  C  \max_{1\le i\le r} w_i \sigma\sqrt{\frac{d}{T}} .
\end{align*}
\end{proof}

\begin{proof}[\bf Proof of Theorem \ref{thm:projection}]
Recall $\widehat\bA_k^{(m)}=(\widehat\ba_{1k}^{(m)},\ldots,\widehat\ba_{rk}^{(m)})\in\R^{d_k\times r}$, $\widehat\Sigma_k^{(m)}=\widehat\bA_k^{(m)\top}\widehat\bA_k^{(m)}$, and $\widehat\bB_k^{(m)} = \widehat\bA_k^{(m)}(\widehat\Sigma_k^{(m)})^{-1} = (\widehat\bb_{1k}^{(m)},...,\widehat\bb_{rk}^{(m)}) \in\R^{d_k\times r}$. Let $\overline\E(\cdot)=\E(\cdot|f_{it},1\le i\le r, 1\le t\le T)$ and $\overline\P(\cdot)=\P(\cdot|f_{it},1\le i\le r, 1\le t\le T)$. Also let
\begin{align*}
\widebar\lambda_{i}= \frac{1}{T-h} \sum_{t=h+1}^T w_i^2 f_{i,t-h}f_{i,t}    .
\end{align*}
Without loss of generality, assume $\widehat\bSigma_{h}$ is symmetric. Write
\begin{align}\label{eq:sigmahat-dcomp}
\widehat\bSigma_{h} &= \frac{1}{T-h}\sum_{t=h+1}^T \cX_{t-h}\otimes\cX_{t}   \notag\\
&= \sum_{i=1}^r \widebar\lambda_i \otimes_{k=1}^{2K}\ba_{ik} +  \sum_{i\neq j}^r \frac{1}{T-h} \sum_{t=h+1}^T w_iw_j  f_{i,t-h}f_{j,t}  \otimes_{k=1}^{K}\ba_{ik} \otimes_{k=K+1}^{2K}\ba_{jk} \notag\\
&\quad +   \frac{1}{T-h} \sum_{t=h+1}^T \sum_{i=1}^r w_i f_{i,t-h} \otimes_{k=1}^{K}\ba_{ik} \otimes\cE_t +   \frac{1}{T-h} \sum_{t=h+1}^T \sum_{i=1}^r w_i f_{i,t}  \cE_{t-h} \otimes_{k=K+1}^{2K}\ba_{ik}    \notag\\
&\quad +   \frac{1}{T-h} \sum_{t=h+1}^T \cE_{t-h} \otimes \cE_t      \notag\\
&:= \sum_{i=1}^r \widebar\lambda_i \otimes_{k=1}^{2K}\ba_{ik} +\Delta_1+ \Delta_2+ \Delta_3+ \Delta_4,
\end{align}
with $\ba_{i,K+k}=\ba_{ik}$ for all $1\le k\le K$. Let $\bPsi=\Delta_1+\Delta_2+\Delta_3+\Delta_4$.

By Theorem \ref{thm:initial}, in an event $\Omega_0$ with probability at least $1-(Tr)^{-c}/2-e^{-d}/6$,
\begin{align*}
\| \widehat\ba_{ik}^{(0)}\widehat\ba_{ik}^{(0)\top}  -\ba_{ik} \ba_{ik}^\top \|_{\rm S} &\le \delta + \frac{2\lambda_1 \delta + C_1R^{(0)} }{\min\{ \lambda_{i-1}-\lambda_i, \lambda_i-\lambda_{i+1}\}},
\end{align*}
where
\begin{align*}
R^{(0)} &=  \max_{1\le i\le r} w_i^2 \left(\sqrt{\frac{r+ \log T}{T}} + \frac{(r+ \log T)^{1/\gamma}}{T}  \right)+ \frac{\sigma^2(d+\sqrt{dT})}{T} + \sigma \max_{1\le i\le r} w_i \sqrt{\frac{d}{T}}.
\end{align*}
Then in the event $\Omega_0$,
\begin{align}\label{bdd_psi0}
\psi_0:=\max_i\max_k \|\widehat\ba_{ik}^{(0)}\widehat\ba_{ik}^{(0)\top}  -\ba_{ik}\ba_{ik}^\top \|_{\rm S} \le \max_i \left( \frac{C_1'\lambda_1\delta + C_1' R^{(0)} }{\min\{ \lambda_{i-1}-\lambda_i, \lambda_i-\lambda_{i+1}\}}  \right)  .
\end{align}

At $m$-th step, let
\begin{align}
\psi_{m,i,k} :=\| \widehat\ba_{ik}^{(m)}\widehat\ba_{ik}^{(m)\top}  -\ba_{ik}\ba_{ik}^\top \|_{\rm S},\qquad \psi_{m,k} :=\max_i\psi_{m,i,k}, \qquad \psi_m =\max_k\psi_{m,k}.
\end{align}
Let $\bg_{i\ell}=\bb_{i\ell}/\|\bb_{i\ell}\|_2$ and $\widehat\bg_{i\ell}^{(m)}=\widehat\bb_{i\ell}^{(m)}/\|\widehat\bb_{i\ell}^{(m)}\|_2$.
Given $\widehat\ba_{i\ell}^{(m)}$ ($1\le i\le r, 1\le \ell \le K$), the $(m+1)$th iteration produces estimates $\widehat\ba_{ik}^{(m+1)}$, which is the top left singular vector of $\widehat\bSigma_{h}\times_{\ell\in [ 2K] \backslash\{k,K+k\} } \widehat\bb_{i\ell}^{(m)\top}$, or equivalently $\widehat\bSigma_{h}\times_{\ell\in [ 2K] \backslash\{k,K+k\} } \widehat\bg_{i\ell}^{(m)\top}$. Note that $\widehat\bSigma_{h} = \sum_{j=1}^r  \widebar\lambda_j \otimes_{\ell=1}^{2K} \ba_{j\ell} +  \bPsi$, with $\ba_{j,\ell+K}=\ba_{j\ell}$. The ``noiseless" version of this update is given by
\begin{equation}\label{error-free-update}
\widehat\bSigma_{h}\times_{\ell\in [ 2K] \backslash\{k,K+k\} } \bg_{i\ell}^{\top}=  \widebar\lambda_{i}\ba_{ik}\ba_{ik}^\top + \bPsi \times_{\ell\in [ 2K] \backslash\{k,K+k\} } \bg_{i\ell}^{\top}.
\end{equation}
At $(m+1)$-th iteration, for any $1\le i\le r$, we have
\begin{align*}
\widehat\bSigma_{h}\times_{\ell\in [ 2K] \backslash\{k,K+k\} } \widehat\bg_{i\ell}^{(m)\top} = \sum_{j=1}^r   \widetilde\lambda_{j,i} \ba_{jk}\ba_{jk}^\top +  \bPsi \times_{\ell\in [ 2K] \backslash\{k,K+k\} } \widehat\bg_{i\ell}^{(m)\top},
\end{align*}
where
\begin{align*}
\widetilde\lambda_{j,i}=\widebar\lambda_j\prod_{\ell\in [ 2K] \backslash\{k,K+k\} } \ba_{j\ell}^\top\widehat\bg_{i\ell}^{(m)}   .
\end{align*}
Let
\begin{align*}
\lambda_{j,i}&=\lambda_j\prod_{\ell\in [ 2K] \backslash\{k,K+k\} }  \ba_{j\ell}^\top\widehat \bg_{i\ell}^{(m)}, \\
\alpha&=\sqrt{1-\delta_{\max}}-(r^{1/2}+1)\psi_0/\sqrt{1-1/(4r)},\\
\phi_{m,\ell}&=1 \wedge \frac{\psi_{m,\ell}\sqrt{2r}}{\alpha \sqrt{1-1/(4r)}}  ,  \\
\phi_{m}&=\max_{\ell} \phi_{m,\ell}.
\end{align*}
We may assume without loss of generality $\ba_{j\ell}^\top\widehat\ba_{j\ell}^{(m)}\ge 0$ for all $(j,\ell)$.
Basic calculation shows
\begin{align}\label{a-bd}
\max_{j\le r}\|\widehat\ba_{j\ell}^{(m)} - \ba_{j\ell} \|_2 \le \psi_{m,\ell}/\sqrt{1-1/(4r)}, \ \
\displaystyle \big\|\widehat\bb_{j\ell}^{(m)}\big\|_2 \le \|\widehat\bB_\ell^{(m)}\|_{\rm S}
\le \bigg(\sqrt{1-\delta_\ell}-\frac{r^{1/2}\psi_0}{\sqrt{1-1/(4r)}}\bigg)^{-1}.
\end{align}
Let $P_\ell =\bA_\ell(\bA_\ell^\top \bA_\ell)^{-1}\bA_\ell^\top$ and
$P_\ell^\perp =I_{d_\ell} - P_\ell^\top$.
As $\widehat\bB_{\ell}^{(m)} - \bB_{\ell} = P_\ell^\perp \big(\widehat\bA_\ell^{(m)}-\bA_\ell\big)(\widehat\Sigma_\ell^{(m)})^{-1}
-  \bB_\ell\big(\widehat\bA_\ell^{(m)} - \bA_\ell\big)^\top \widehat\bB_\ell^{(m)}$,
\begin{align*}
\big\|\widehat\bb_{j\ell}^{(m)} - \bb_{j\ell}\big\|_2^2
 &\le \big\|\widehat\bA_\ell^{(m)} - \bA_\ell\big\|_{\rm S}^2
\big(\|\widehat\bB_\ell^{(m)}\|_{\rm S}^2+\|\bB_\ell\|_{\rm S}^2\big)\big(\|\widehat\bb_{j\ell}^{(m)}\|_2^2\wedge\|\bb_{j\ell}\|_2^2\big) \\
&\le \{r\psi_{m,\ell}^2/(1-1/(4r))\}
(2/\alpha^2)\big(\|\widehat\bb_{j\ell}^{(m)}\|_2^2\wedge\|\bb_{j\ell}\|_2^2\big)
\end{align*}
by the algebraic symmetry between the estimator and estimand. As $\big\|\widehat\bg_{j\ell}^{(m)} - \bb_{j\ell}\big\|_2\le \big\|\widehat\bb_{j\ell}^{(m)} - \bb_{j\ell}\big\|_2$
for $\|\widehat\bb_{j\ell}^{(m)}\|_2\ge \|\bb_{j\ell}\|_2=1$,
\begin{align}\label{b-bd}
\big\|\widehat\bg_{j\ell}^{(m)} - \bb_{j\ell}/\|\bb_{j\ell}\|_2\big\|_2
\le (\psi_{m,\ell}/\alpha)\sqrt{2r/(1-1/(4r))}
\end{align}
by scale invariance.
Moreover,
\eqref{a-bd} provides
\begin{align}\label{g-bd}
\max_{i\neq j}\big|\ba_{i\ell}^\top\widehat\bg_{j\ell}^{(m)}\big| \le \psi_{m,\ell}/\sqrt{1-1/(4r)},\ \
\big|\ba_{j\ell}^\top\widehat\bg_{j\ell}^{(m)} \big| \ge \alpha,
\end{align}
as $\widehat\ba_{i\ell}^{(m)\top}\widehat\bg_{j\ell}^{(m)}=I\{i=j\}/\|\widehat\bb_{j\ell}^{(m)}\|_2$.
Then, for $j\neq i$,
\begin{align*}
\lambda_{j,i}/\lambda_{i,i} \le \big(\lambda_1/\lambda_i\big)   \prod_{\ell\neq k}^K \left(\frac{\psi_{m,\ell}/\sqrt{1-1/(4r)} }{1- \psi_{m,\ell}/\sqrt{1-1/(4r)} } \right)^2.
\end{align*}
Employing similar arguments in the proof of Lemma \ref{lem:psi}, in an event $\Omega_1$ with probability at least $1-(Tr)^{-c}/2$, we have
\begin{align}\label{eq:thetah}
\left\|\frac{\Theta_h - \E \Theta_h }{T-h}\right\|_{\rm S} & \le C_1  \left(\sqrt{\frac{r+\log T}{T}} + \frac{(r+\log T)^{1/\gamma}}{T}  \right),
\end{align}
where $\Theta_h$ is defined in \eqref{def:theta}.
In the event $\Omega_{1}$, we also have
\begin{align*}
\max_{1\le j_1,j_2\le r}\left| \frac{1}{T-h}\sum_{t=h+1}^T f_{j_1,t-h} f_{j_2,t} -\E f_{j_1,t-h} f_{j_2,t} \right|   &\le C_1  \left(\sqrt{\frac{r+\log T}{T}} + \frac{(r+\log T)^{1/\gamma}}{T}  \right) .
\end{align*}
It follows that in the event $\Omega_{1}$, for any $1\le j\le r$,
\begin{align*}
|\widebar \lambda_j -\lambda_j| \le C_1  \left(\sqrt{\frac{r+\log T}{T}} + \frac{(r+\log T)^{1/\gamma}}{T}  \right) \lambda_j  .
\end{align*}
By Wedin's theorem \citep{wedin1972}, in the event $\Omega_1$,
\begin{align}\label{wedin72}
\|\widehat\ba_{ik}^{(m+1)} \widehat\ba_{ik}^{(m+1)\top} -\ba_{ik} \ba_{ik}^\top \|_{\rm S} &\le \frac{2\left\| \sum_{j\neq i}^r \widetilde\lambda_{j,i} \ba_{jk}\ba_{jk}^\top  \right\|_{\rm S} + 2\| \bPsi \times_{\ell\in [ 2K] \backslash\{k,K+k\} } \widehat\bg_{i\ell}^{(m)\top} \|_{\rm S} }{\widetilde\lambda_{i,i}} \notag\\
&\le \frac{4\|\bA_k\|_{\rm S}^2 \max_{j\neq i}|\lambda_{j,i}  | + 2  \| \bPsi \times_{\ell\in [ 2K] \backslash\{k,K+k\} } \widehat \bg_{i\ell}^{(m)\top} \|_{\rm S} }{\alpha^{2k-2} \lambda_{i,i} } .
\end{align}

To bound the numerator of \eqref{wedin72}, we write
\begin{align*}
&\Delta_{1,1,h}=\sum_{j_2\neq i}^r \frac{1}{T-h} \sum_{t=h+1}^T w_{i}w_{j_2} f_{i,t-h}f_{j_2,t}   \otimes_{\ell=1}^{K} \ba_{i\ell}\otimes_{\ell=K+1}^{2K} \ba_{j_2\ell} \times_{\ell\in [ 2K] \backslash\{k,K+k\} } \widehat \bg_{i\ell}^{(m)\top} , \\
&\Delta_{1,2,h}=\sum_{j_1\neq i}^r \frac{1}{T-h} \sum_{t=h+1}^T w_{j_1}w_{i} f_{j_1,t-h}f_{it}   \otimes_{\ell=1}^{K} \ba_{j_1\ell}\otimes_{\ell=K+1}^{2K} \ba_{i\ell} \times_{\ell\in [ 2K] \backslash\{k,K+k\} } \widehat \bg_{i\ell}^{(m)\top} , \\
&\Delta_{1,3,h}=\sum_{j_1\neq j_2\neq i}^r \frac{1}{T-h} \sum_{t=h+1}^T w_{j_1}w_{j_2} f_{j_1,t-h}f_{j_2,t}   \otimes_{\ell=1}^{K} \ba_{j_1\ell}\otimes_{\ell=K+1}^{2K} \ba_{j_2\ell} \times_{\ell\in [ 2K] \backslash\{k,K+k\} } \widehat \bg_{i\ell}^{(m)\top} .
\end{align*}
For any vectors $\widetilde  \bg_{i\ell},\widecheck  \bg_{i\ell}\in\R^{d_l}$, define
\begin{align*}
&\Delta_{2,k,h}(\widetilde \bg_{i\ell}, \ell\neq k) =\frac{1}{T-h} \sum_{t=h+1}^T w_i f_{it} \cE_{t-h}  \times_{\ell=1,\ell\neq k}^K \widetilde \bg_{i\ell}^\top\in\R^{d_k} , \\
&\Delta_{3,k,h}(\widetilde \bg_{i\ell}, \ell\neq k) =\frac{1}{T-h} \sum_{t=h+1}^T  \cE_{t-h} \otimes ( w_j f_{jt}, \ j\neq i)^\top \times_{\ell=1,\ell\neq k}^K \widetilde \bg_{i\ell}^\top  \in\R^{d_k\times (r-1)} , \\
&\Delta_{4,k,h}(\widetilde \bg_{i\ell}, \ell\neq k) =\frac{1}{T-h} \sum_{t=h+1}^T w_i f_{i,t-h} \cE_{t}  \times_{\ell=1,\ell\neq k}^K \widetilde \bg_{i\ell}^\top \in\R^{d_k}, \\
&\Delta_{5,k,h}(\widetilde \bg_{i\ell}, \ell\neq k) =\frac{1}{T-h} \sum_{t=h+1}^T  \cE_{t} \otimes ( w_j f_{j,t-h}, \ j\neq i)^\top \times_{\ell=1,\ell\neq k}^K \widetilde \bg_{i\ell}^\top   \in\R^{d_k\times (r-1)}, \\
&\Delta_{6,k.h}(\widetilde \bg_{i\ell},\widecheck  \bg_{i\ell}, \ell\neq k) = \frac{1}{T-h} \sum_{t=h+1}^T \cE_{t-h} \otimes \cE_t \times_{\ell=1,\ell\neq k}^K \widetilde \bg_{i\ell}^\top \times_{\ell=K+1,\ell\neq K+k}^{2K} \widecheck \bg_{i,\ell-K}^\top \in\R^{d_k \times d_k}.
\end{align*}
As $\Delta_{q,k,h}(\widetilde \bg_{i\ell},\widecheck  \bg_{i\ell}, \ell\neq k)$ is linear in $\widetilde  \bg_{i\ell},\widecheck  \bg_{i\ell}$, by \eqref{g-bd}, the numerator on the right hand side of \eqref{wedin72} can be bounded by
\begin{align}\label{norm-bd}
&\| \bPsi \times_{\ell\in [ 2K] \backslash\{k,K+k\} } \widehat \bg_{i\ell}^{(m)\top} \|_{\rm S} \notag\\
\le& \| \Delta_1 \times_{\ell\in [ 2K] \backslash\{k,K+k\} } \widehat \bg_{i\ell}^{(m)\top} \|_{\rm S} + \sum_{q=2,4}\| \Delta_{q,k,h}(\widehat \bg_{i\ell}^{(m)}, \ell\neq k) \|_{\rm S}  + \| \Delta_{6,k,h}(\widehat \bg_{i\ell}^{(m)},\widehat \bg_{i\ell}^{(m)}, \ell\neq k) \|_{\rm S}  \notag\\
&\quad  + \sum_{q=3,5}\|\bA_k\|_{\rm S}\| \Delta_{q,k,h}(\widehat \bg_{i\ell}^{(m)}, \ell\neq k) \|_{\rm S} \max_{j\neq i}\prod_{\ell \neq k}^K \left|\ba_{j \ell}^\top \widehat\bg_{i\ell}^{(m)} \right| \notag\\
\le&  \sum_{q=1,2,3} \| \Delta_{1,q,h} \|_{\rm S} + \sum_{q=2,4} \| \Delta_{q,k,h}( \bg_{i\ell}, \ell\neq k) \|_{\rm S} +  \sum_{q=2,4} (2K-2) \phi_{m,k} \| \Delta_{q,k,h} \|_{\rm S, S} \notag\\
&\quad +  \sum_{q=3,5}\|\bA_k\|_{\rm S}\prod_{\ell\neq k}^K \big(\psi_{m,\ell}/\sqrt{1-1/(4r)}\big)  \| \Delta_{q,k,h}( \bg_{i\ell}, \ell\neq k) \|_{\rm S} \notag\\
&\quad + \sum_{q=3,5}\|\bA_k\|_{\rm S}(2K-2) \phi_{m,k} \prod_{\ell\neq k}^K \big(\psi_{m,\ell}/\sqrt{1-1/(4r)}\big) \| \Delta_{q,k,h} \|_{\rm S,S} \notag\\
&\quad +  \| \Delta_{6,k,h}( \bg_{i\ell},  \bg_{i\ell}, \ell\neq k) \|_{\rm S}   +(2K-2) \phi_{m,k}  \| \Delta_{6,k,h} \|_{\rm S,S},
\end{align}
where
\begin{align*}
&\| \Delta_{q,k,h}\|_{\rm S,S}= \max_{\substack{ \|\widetilde \bg_{i\ell}\|_2=1, \\ \widetilde \bg_{i\ell} \in \R^{d_{\ell}} } } \| \Delta_{q,k,h}(\widetilde \bg_{i\ell}, \ell\neq k) \|_{\rm S}, \quad q=2,3,4,5,\\
&\| \Delta_{6,k,h}\|_{\rm S,S}= \max_{\substack{ \|\widetilde \bg_{i\ell}\|_2=\|\widecheck \bg_{i\ell}\|_2=1, \\ \widetilde \bg_{i\ell},\widecheck \bg_{i\ell}\in \R^{d_{\ell}} } } \| \Delta_{6,k,h}(\widetilde \bg_{i\ell},\widecheck  \bg_{i\ell}, \ell\neq k) \|_{\rm S} .
\end{align*}
Note that
\begin{align*}
\Delta_{1,1,h}=w_i \ba_{ik}\left(\prod_{\ell\neq k}^K \ba_{i\ell}^\top \widehat \bg_{i\ell}^{(m)} \right) \frac{1}{T-h}\sum_{t=h+1}^T f_{i,t-h} \cdot (f_{j_2,t},\ j_2\neq i) \  {\rm diag}\left(\prod_{\ell\neq k}^K \ba_{j_2 \ell}^\top \widehat \bg_{i\ell}^{(m)},j_2\neq i  \right) (w_{j_2}\ba_{j_2 k}, j_2\neq i)^\top.
\end{align*}
By \eqref{eq:thetah}, in the event $\Omega_1$,
\begin{align}\label{Delta-bd-11}
\|\Delta_{1,1,h}\|_{\rm S} &\le w_i \max_j w_j \|\bA_k\|_{\rm S} \left\| \frac{\Theta_h - \E \Theta_h}{T-h} \right\|_{\rm S} \max_{j\neq i}\prod_{\ell \neq k}^K \left|\ba_{j \ell}^\top \widehat \bg_{i\ell}^{(m)} \right| \notag\\
&\le C_1 w_i \max_j w_j \left(\sqrt{\frac{r+\log T }{T}} + \frac{(r+\log T)^{1/\gamma}}{T}  \right) \prod_{\ell \neq k}^K\big( \psi_{m,\ell}/\sqrt{1-1/(4r)} \big).
\end{align}
Similarly, in the event $\Omega_1$,
\begin{align}
\|\Delta_{1,2,h}\|_{\rm S} &\le C_1 w_i \max_j w_j \left(\sqrt{\frac{r+\log T }{T}} + \frac{(r+\log T)^{1/\gamma}}{T}  \right) \prod_{\ell \neq k}^K\big( \psi_{m,\ell}/\sqrt{1-1/(4r)} \big) , \label{Delta-bd-12} \\
\|\Delta_{1,3,h}\|_{\rm S} &\le C_1 \max_j w_j^2 \left(\sqrt{\frac{r+\log T }{T}} + \frac{(r+\log T)^{1/\gamma}}{T}  \right) \prod_{\ell \neq k}^K\big( \psi_{m,\ell}/\sqrt{1-1/(4r)} \big)^2 .  \label{Delta-bd-13}
\end{align}
Let $\Upsilon_{0,i,k}=T^{-1} \sum_{t=1}^T w_i^2  f_{it}^2   \ba_{ik}  \ba_{ik}^\top$ and $\Upsilon_{0,-i,k}=T^{-1} \sum_{t=1}^T \sum_{j_1,j_2\neq i}^r w_{j1} w_{j2}  f_{j_1t}f_{j_2t}   \ba_{j_1k}  \ba_{j_2k}^\top$. Then, in the event $\Omega_1$,
\begin{align}
\| \Upsilon_{0,i,k} \|_{\rm S} &\le w_i^2 + C_1 w_i^2 \left(\sqrt{\frac{r+\log T}{T}} + \frac{(r+\log T)^{1/\gamma}}{T}  \right) :=\Delta_{\Upsilon_i} \asymp \lambda_i, \label{eq:thm_projection:upsilon} \\
\| \Upsilon_{0,-i,k} \|_{\rm S} &\le \max_j w_j^2 + C_1 \max_j w_j^2 \left(\sqrt{\frac{r+\log T }{T}} + \frac{(r+\log T)^{1/\gamma}}{T}  \right) :=\Delta_{\Upsilon_{-i}} \asymp \lambda_1. \label{eq:thm_projection:upsilon2}
\end{align}
We claim that in certain events $\Omega_1\cap\Omega_q, q=2,3,4,5,6$, with
$\P(\Omega_q)\ge 1- 6^{-1} \sum_{\ell=1}^K e^{-d_\ell}$, for any $1\le \ell\le K$, the following bounds for $\| \Delta_{q,k,h} \|_{\rm S,S}$ and $\| \Delta_{q,k,h}( b_{j\ell},  b_{j\ell}, \ell\neq k) \|_{\rm S}$ hold,
\begin{align}\label{Delta-bd-2}
&\| \Delta_{q,k,h} \|_{\rm S,S} \le C_1\sigma\frac{\sqrt{d_k}+\sqrt{\sum_{\ell\neq k} d_{\ell}} } {\sqrt{T}}   \Delta_{\Upsilon_i}^{1/2}, \ q=2,4, \notag\\
& \| \Delta_{q,k,h} \|_{\rm S,S} \le C_1\sigma\frac{\sqrt{d_k}+\sqrt{\sum_{\ell\neq k} d_{\ell}} }{\sqrt{T} }  \Delta_{\Upsilon_{-i}}^{1/2}, \ q=3,5, \\
&\| \Delta_{6,k,h} \|_{\rm S,S} \le \frac{C_1\sigma^2 \left(\sum_{k=1}^K d_k + \sqrt{d_k T}+\sqrt{\sum_{\ell\neq k} d_{\ell} T} \right)}{T}, \notag
\end{align}
and
\begin{align}\label{Delta-bd-2n}
&\| \Delta_{q,k,h}( \bg_{i\ell},  \ell\neq k) \|_{\rm S} \le C_1\sigma\sqrt{\frac{d_k}{T} }  \Delta_{\Upsilon_i}^{1/2}, \ q=2,4, \notag\\
&\| \Delta_{q,k,h}( \bg_{i\ell},  \ell\neq k) \|_{\rm S} \le C_1\sigma\sqrt{\frac{d_k}{T} }  \Delta_{\Upsilon_{-i}}^{1/2}, \ q=3,5, \\
&\| \Delta_{6,k,h}( \bg_{i\ell},  \bg_{i\ell}, \ell\neq k) \|_{\rm S} \le \frac{C_1\sigma^2(d_k + \sqrt{d_k T})}{T}. \notag
\end{align}

Define
\begin{align} 
R_{k,i}^{\ideal}&= \frac{\sigma^2(d_k+\sqrt{d_k T})}{T} + \sqrt{\frac{\lambda_i d_k }{T}},\\
R_{i,\phi}^{\ideal}&= R_{k,i}^{\ideal} + \phi \sum_{\ell\neq k} R_{\ell,i}^{\ideal}.
\end{align}
As $ \bg_{i\ell}$ is true and deterministic, it follows from \eqref{norm-bd}, \eqref{Delta-bd-11}, \eqref{Delta-bd-12}, \eqref{Delta-bd-13}, \eqref{Delta-bd-2}, \eqref{Delta-bd-2n}, in the event $\cap_{q=0}^6\Omega_{q}$, for some numeric constant $C_2>0$
\begin{align}\label{norm-bd-1}
&\| \bPsi \times_{\ell\in [ 2K] \backslash\{k,K+k\} } \widehat \bg_{i\ell}^{(m)\top} \|_{\rm S} \notag \\
\le& C_2 R_{k,i}^{\ideal} + C_2 (K-1)   \phi_m  \sum_{\ell \neq k}R_{\ell,i}^{\ideal}  +\frac{2C_1}{(1-1/(4r))^{(K-1)/2}}\sqrt{\lambda_1\lambda_i}\prod_{\ell\neq k}\psi_{m,\ell} \left(\sqrt{\frac{r+\log T }{T}} + \frac{(r+\log T)^{1/\gamma}}{T}  \right) \notag \\
&\quad  + \frac{C_1}{(1-1/(4r))^{K-1}}\lambda_1\prod_{\ell\neq k}\psi_{m,\ell}^2\left(\sqrt{\frac{r+\log T }{T}} + \frac{(r+\log T)^{1/\gamma}}{T}  \right) .
\end{align}
Substituting \eqref{norm-bd-1} into \eqref{wedin72}, we have, in the event $\cap_{q=0}^6\Omega_{q}$,
\begin{align}\label{bdd1:thm-projection}
&\| \widehat \ba_{ik}^{(m+1)}\widehat \ba_{ik}^{(m+1)\top} - \ba_{ik} \ba_{ik}^\top \|_{\rm S} \notag\\
&\le \frac{4(1+\delta_{\max}) \lambda_1 \prod_{\ell\neq k}\psi_{m,\ell}^2 }{\alpha^{2K-2}\lambda_{i} [\sqrt{(1-\delta_{\max})(1-1/(4r))}\alpha]^{2K-2}  } +
\frac{2C_2(K-1) R_{i,\phi}^{\ideal}    }{\alpha^{2K-2}\lambda_i } \notag\\
&\quad + \frac{4C_1}{\alpha^{2K-2}(1-1/(4r))^{(K-1)/2}}\sqrt{\lambda_1/\lambda_i}\prod_{\ell\neq k}\psi_{m,\ell} \left(\sqrt{\frac{r+\log T }{T}} + \frac{(r+\log T)^{1/\gamma}}{T}  \right)     \notag\\
&\quad + \frac{2C_1}{\alpha^{2K-2}(1-1/(4r))^{K-1}}(\lambda_1/\lambda_i)\prod_{\ell\neq k}\psi_{m,\ell}^2\left(\sqrt{\frac{r+\log T }{T}} + \frac{(r+\log T)^{1/\gamma}}{T}  \right)  \notag \\
&\le  C_{\alpha,K}(\lambda_1/\lambda_i)\prod_{\ell\neq k}\psi_{m,\ell}^2 + C_{\alpha,K}\sqrt{\lambda_1/\lambda_i}\prod_{\ell\neq k}\psi_{m,\ell} \left(\sqrt{\frac{r+\log T }{T}} + \frac{(r+\log T)^{1/\gamma}}{T}  \right)  + C_{\alpha,K} R_{i,\phi_m}^{\ideal}/\lambda_i.
\end{align}
Let $\phi_0=(\psi^*/\alpha)\sqrt{2r(1-1/(4r))}$ with $\psi^*=C_{\alpha,K}R^{\ideal}$. By induction, $\phi_{m,k}\le\phi_0$. It follows that
\begin{align}\label{bdd2:thm-projection}
\psi_{m+1,i,k}
&\le  C_{\alpha,K}(\lambda_1/\lambda_i)\prod_{\ell\neq k}\psi_{m,\ell}^2 + C_{\alpha,K}\sqrt{\lambda_1/\lambda_i}\prod_{\ell\neq k}\psi_{m,\ell} \left(\sqrt{\frac{r+\log T }{T}} + \frac{(r+\log T)^{1/\gamma}}{T}  \right)  + C_{\alpha,K} R^{\ideal} \notag\\
&\le\epsilon \vee C_{\alpha,K} R^{\ideal}.
\end{align}
Hence, we have
\begin{align*}
\psi_{m+,k} \le \psi_0 \rho^{1+(2K-2)+\cdots+(2K-2)^{m}}
\end{align*}
with the $\rho<1$ in \eqref{thm-projection:eq1b}, so that the required iteration number is $O(\log\log(\psi_0/R^{\ideal}))$.

In the end, we divide the rest of the proof into 3 steps to prove
\eqref{Delta-bd-2} for $q=2,3,4,5,6$. The proof of \eqref{Delta-bd-2n} is similar, thus omitted.

\smallskip
\noindent\underline{\bf Step 1.}
We prove
\eqref{Delta-bd-2} for the $\|\Delta_{2,k,h}\|_{\rm S, S}$. The proof for $\|\Delta_{4,k,h}\|_{\rm S, S}$ will be similar, thus is omitted. By Lemma \ref{lemma:epsilonnet} (iii), there exist $\bg_{il}^{(\ell)} \in\mathbb R^{d_l}$, $1\le \ell  \le N_{d_l,1/8} := 17^{d_l}$, such that  $\| \bg_{il}^{(\ell)}\|_{2}\le 1$ and
\begin{align*}
\| \Delta_{2,k,h} \|_{\rm S,S} &\le 2\max_{\ell \le N_{d_l,1/8}}
\left\|   \Delta_{2,k,h}(\bg_{il}^{(\ell)}, l\neq k)  \right\|_{\rm S} \\
&=2\max_{\ell \le N_{d_l,1/8}}
\left\|\frac{1}{T-h} \sum_{t=h+1}^T w_i f_{it} \cE_{t-h}  \times_{l=1,l\neq k}^K \bg_{il}^{(\ell)\top}  \right\|_{\rm S}.
\end{align*}
Employing similar arguments in the proof of Lemma \ref{lem:psi}, we have
\begin{align*}
\overline\E  \left\|   \Delta_{2,k,h}(\bg_{il}^{(\ell)}, l\neq k)  \right\|_{\rm S} \le \frac{\sigma\sqrt{8Td_k}}{T-h} \| \Upsilon_{0,i,k} \|_{\rm S}^{1/2}.
\end{align*}
Elementary calculation shows that $\left\|   \Delta_{2,k,h}(\bg_{il}^{(\ell)},  l\neq k)  \right\|_{\rm S}$ is a $\sigma\sqrt{T} \| \Upsilon_{0,i,k} \|_{\rm S}^{1/2}$ Lipschitz function. Then, by Gaussian concentration inequalities for Lipschitz functions,
\begin{align*}
\P\left( \left\|   \Delta_{2,k,h}(\bg_{il}^{(\ell)}, l\neq k)  \right\|_{\rm S} -\frac{\sigma\sqrt{8Td_k}}{T-h} \| \Upsilon_{0,i,k} \|_{\rm S}^{1/2} \ge \frac{\sigma\sqrt{T}}{T-h} \| \Upsilon_{0,i,k} \|_{\rm S}^{1/2} x \right) \le 2e^{-\frac{x^2}{2}}.
\end{align*}
Hence,
\begin{align*}
\P\left( \| \Delta_{2,k,h} \|_{\rm S,S} -\frac{\sigma\sqrt{8Td_k}}{T-h} \| \Upsilon_{0,i,k} \|_{\rm S}^{1/2} \ge \frac{\sigma\sqrt{T}}{T-h} \| \Upsilon_{0,i,k} \|_{\rm S}^{1/2} x \right) \le 4\prod_{l\neq k} N_{d_l,1/8} e^{-\frac{x^2}{2}}.
\end{align*}
As $T\ge 4h$, this implies with $x\asymp \sqrt{\sum_{l\neq k} d_l}$ that in an event $\Omega_2$ with at least probability $1-\sum_k e^{- d_k}/6$,
\begin{align*}
\| \Delta_{2,k,h} \|_{\rm S,S} \le C_1\sigma\frac{\sqrt{d_k}+\sqrt{\sum_{l\neq k}^K d_l} } {\sqrt{T}} \| \Upsilon_{0,i,k} \|_{\rm S}^{1/2}.
\end{align*}
It follows that, in the event $\Omega_1\cap\Omega_2$,
\begin{align}
\| \Delta_{2,k,h} \|_{\rm S,S} \le C_1\sigma\frac{\sqrt{d_k}+\sqrt{\sum_{l\neq k}^K d_l} } {\sqrt{T}}   \Delta_{\Upsilon_i}^{1/2},
\end{align}
where $\Delta_{\Upsilon_i}\asymp \lambda_i$ is defined in \eqref{eq:thm_projection:upsilon}.

\smallskip
\noindent\underline{\bf Step 2.}
Inequality \eqref{Delta-bd-2} for $\|\Delta_{3,k,h}\|_{\rm S, S}$ and $\|\Delta_{5,k,h}\|_{\rm S, S}$ follow from the same argument as the above step.

\smallskip
\noindent\underline{\bf Step 3.}
Now we prove \eqref{Delta-bd-2} for $\|\Delta_{6,k,h}\|_{\rm S, S}$. We split the sum into two terms over the index sets, $S_1=\{(h,2h]\cup(3h,4h]\cup\cdots\} \cap(h,T]$ and its complement $S_2$ in $(h,T]$, so that $\{\cE_{t-h},t\in S_a\}$ is independent of $\{\cE_t, t\in S_a\}$ for each $a=1,2$. Let $n_a=|S_a|$.

By Lemma \ref{lemma:epsilonnet} (iii), there exist $\bg_{il}^{(\ell)},\bg_{il}^{(\ell')}\in\mathbb R^{d_l}$, $1\le \ell,\ell' \le N_{d_l,1/8} := 17^{d_l}$, such that  $\| \bg_{il}^{(\ell)}\|_{2}\le 1$, $\| \bg_{il}^{(\ell')}\|_{2}\le 1$ and
\begin{align*}
\| \Delta_{6,k,h} \|_{\rm S,S} &\le 2\max_{\ell,\ell'\le N_{d_l,1/8}}
\left\|   \Delta_{6,k,h}(\bg_{il}^{(\ell)}, \bg_{ik}^{(\ell')}, l\neq k)  \right\|_{\rm S} \\
&=2\max_{\ell,\ell'\le N_{d_l,1/8}}
\left\|\frac{1}{T-h} \sum_{t=h+1}^T \cE_{t-h} \otimes \cE_t  \times_{l=1,l\neq k}^K \bg_{il}^{(\ell)\top} \times_{l=k+1,l\neq K+k}^{2K} \bg_{i,l-K}^{(\ell')\top} \right\|_{\rm S}.
\end{align*}
Define $G_a=(\cE_{t-h}\times_{l=1,l\neq k}^K \bg_{il}^{(\ell)\top},t\in S_a)\in \R^{d_k\times n_a}$ and
$H_a=(\cE_{t}\times_{l=1,l\neq k}^K \bg_{il}^{(\ell')\top},t\in S_a)\in \R^{d_k\times n_a}$. Then, $G_a$, $H_a$ are two independent Gaussian matrices. Note that
\begin{align*}
\left\|\sum_{t\in S_a} \cE_{t-h} \otimes \cE_t  \times_{l=1,l\neq k}^K \bg_{il}^{(\ell)\top} \times_{l=k+1,l\neq K+k}^{2K} \bg_{i,l-K}^{(\ell')\top} \right\|_{\rm S}=\left\|G_a H_a^\top \right\|_{\rm S}.
\end{align*}
By Lemma \ref{lm-GH}(i),
\begin{align*}
\P\left( \left\| G_a H_a \right\|_{\rm S}\ge \sigma^2 (d_k+2\sqrt{d_k n_a}) + \sigma^2 x(x+2\sqrt{n_a}+2\sqrt{d_k}) \right)\le 2e^{-\frac{x^2}{2}}.
\end{align*}
As $\sum_{a=1}^2 n_a=T-h$, it follows from the above inequality,
\begin{align*}
\P\left( \frac{ \| \Delta_{6,k,h} \|_{\rm S,S} }{4\sigma^2}\ge \frac{(\sqrt{d_k}+x)^2}{T-h} +\frac{\sqrt{2}(\sqrt{d_k}+x)}{\sqrt{T-h}} \right) \le 4\prod_{l\neq k} N_{d_l,1/8}^2 e^{-\frac{x^2}{2}}.
\end{align*}
Thus, with $h\le T/4$, $x\asymp\sqrt{\sum_{l\neq k} d_l}$, in an event $\Omega_6$ with probability at least $1-\sum_k e^{- d_k }/6$,
\begin{align}
\| \Delta_{6,k,h} \|_{\rm S,S} &\le \frac{C_1\sigma^2 \left(\sum_{k=1}^K d_k + \sqrt{d_k T}+\sqrt{\sum_{l\neq k} d_l T } \right)}{T}.
\end{align}

\end{proof}

\begin{proof}[\bf Proof of Theorem \ref{thm:factors}]
Let
\begin{align*}
\psi_0 &= \max_{1\le i\le r}\max_{1\le k\le K} \| \widehat \ba_{ik}^{\cpca} \widehat \ba_{ik}^{\cpca\top} - \ba_{ik} \ba_{ik}^\top \|_{\rm S}  , \\
\psi &=\max_{1\le i\le r}\max_{1\le k\le K} \|\widehat \ba_{ik}^{\iso} \widehat \ba_{ik}^{\iso\top} - \ba_{ik} \ba_{ik}^\top \|_{\rm S}  .
\end{align*}
As $\sigma^2 \lesssim \lambda_r$, Theorem \ref{thm:projection} implies that
\begin{align*}
\psi=O_{\P} \left(  \sqrt{\frac{\sigma^2d_{\max}}{\lambda_r T}} \right) . 
\end{align*}

We first prove \eqref{thm-factors:eq1}. Note that
\begin{align*}
\widehat w_i^{\iso} \widehat f_{it}^{\iso} - w_i f_{it} &= \sum_{j=1}^r w_j f_{jt} \otimes_{k=1}^K \ba_{jk} \times_{k=1}^K \widehat \bb_{ik}^{\iso\top}  + \cE_t \times_{k=1}^K \widehat \bb_{ik}^{\iso\top} - w_i f_{it}\\
&= w_i f_{it} \left(\prod_{k=1}^K \ba_{ik}^\top \widehat \bb_{ik}^{\iso} -1 \right) + \sum_{j\neq i} w_j f_{jt}  \prod_{k=1}^K \ba_{ik}^\top \widehat \bb_{jk}^{\iso} + \cE_t \times_{k=1}^K \widehat \bb_{ik}^{\iso\top} \\
&:= \I_1 + \I_2 + \I_3.
\end{align*}
By \eqref{g-bd} in the proof of Theorem \ref{thm:projection}, $\max_{1\le i,j\le r}\max_{1\le k\le K} |\ba_{ik}^\top \widehat \bb_{jk}^{\iso} - \mathbf{1}_{\{i=j\}}| \lesssim \psi$. It follows that $\I_1=O_{\P}(w_i\psi)$, $\I_2=O_{\P}(\sqrt{\lambda_1} r \psi^K)$ and $\I_3=O_{\P}(\sigma)$. Since $\lambda_* \lesssim \lambda_r$, $\lambda_r +(r-1) \lambda_* \lesssim \lambda_1$ and $d_{\max} r\lesssim d$, condition \eqref{thm-projection:eq1a} leads to $(\lambda_1/\lambda_r)^{1/2} r \psi^{K-1} \lesssim (\lambda_1/\lambda_r)^{1/2} \psi_0^{K-1} \lesssim 1$. Then $\I_2=O_{\P}(\lambda_r^{1/2}\psi)$. Hence $w_i^{-1}|\widehat w_i^{\iso} \widehat f_{it}^{\iso} - w_i f_{it}|=O_{\P}(\psi+\sigma w_i^{-1})$, which completes the proof.

Next, we prove \eqref{thm-factors:eq2}. Note that
\begin{align*}
&\frac{1}{T-h} \sum_{t=h+1}^T \widehat w_i^{\iso} \widehat w_j^{\iso} \widehat f_{it-h}^{\iso} \widehat f_{jt}^{\iso} - \frac{1}{T-h} \sum_{t=h+1}^T w_i w_j f_{it-h} f_{jt}  \\
=& \frac{1}{T-h} \sum_{t=h+1}^T \cE_{t-h}\otimes \cE_t \times_{k=1}^{K} \widehat \bb_{ik}^{\iso\top} \times_{k=K+1}^{2K} \widehat \bb_{jk}^{\iso\top} + \frac{1}{T-h} \sum_{t=h+1}^T  w_iw_j f_{it-h} f_{jt} \left( \prod_{k=1}^{K}  \ba_{ik}^\top \widehat \bb_{ik}^{\iso} \prod_{k=1}^{K}  \ba_{jk}^\top \widehat \bb_{jk}^{\iso}  -1 \right)   \\
&\quad + \frac{1}{T-h} \sum_{t=h+1}^T \sum_{\ell_1\ne i,\ell_2\neq j} w_{\ell_1} w_{\ell_2} f_{\ell_1 t-h} f_{\ell_2 t}  \prod_{k=1}^{K}\big( \ba_{\ell_1k}^\top \widehat \bb_{ik}^{\iso} \big) \prod_{k=1}^{K} \big( \ba_{\ell_2k}^\top \widehat \bb_{jk}^{\iso} \big) \\
&\quad + \frac{1}{T-h} \sum_{t=h+1}^T \sum_{\ell_2 \neq j} w_i w_{\ell_2} f_{it-h}f_{\ell_2 t}  \cdot \prod_{k=1}^{K}\big( \ba_{ik}^\top \widehat \bb_{ik}^{\iso} \big)  \prod_{k=1}^{K} \big( \ba_{\ell_2 k}^\top \widehat \bb_{jk}^{\iso} \big)      \\
&\quad + \frac{1}{T-h} \sum_{t=h+1}^T \sum_{\ell_1 \neq i} w_{\ell_1} w_{j} f_{\ell_1 t-h} f_{j t}  \cdot \prod_{k=1}^{K}\big( \ba_{\ell_1 k}^\top \widehat \bb_{ik}^{\iso} \big)  \prod_{k=1}^{K} \big( \ba_{j k}^\top \widehat \bb_{jk}^{\iso} \big)      \\
&\quad + \frac{1}{T-h} \sum_{t=h+1}^T w_i f_{it-h} \cE_t \times_{k=1}^{K} \widehat \bb_{jk}^{\iso\top}  \cdot \prod_{k=1}^{K} \ba_{ik}^\top \widehat \bb_{ik}^{\iso}  + \frac{1}{T-h} \sum_{t=h+1}^T w_j f_{jt} \cE_{t-h} \times_{k=1}^{K} \widehat \bb_{ik}^{\iso\top}  \cdot \prod_{k=1}^{K} \ba_{jk}^\top \widehat \bb_{jk}^{\iso} \\
&\quad + \frac{1}{T-h} \sum_{t=h+1}^T \sum_{\ell_1 \neq i} w_{\ell_1} f_{\ell_1 t-h} \cE_t \times_{k=1}^{K} \widehat \bb_{jk}^{\iso\top}  \cdot \prod_{k=1}^{K} \ba_{\ell_1 k}^\top \widehat \bb_{ik}^{\iso}  \\
&\quad + \frac{1}{T-h} \sum_{t=h+1}^T \sum_{\ell_2 \neq j} w_{\ell_2} f_{\ell_2 t} \cE_{t-h} \times_{k=1}^{K} \widehat \bb_{ik}^{\iso\top}  \cdot \prod_{k=1}^{K} \ba_{\ell_2 k}^\top \widehat \bb_{jk}^{\iso} \\
:=& \II_1 + \II_2 + \II_3 + \II_4 + \II_5 + \II_6 + \II_7 + \II_8 + \II_9.
\end{align*}
Employing the same arguments in the proof of Theorem \ref{thm:projection}, as $d_{\max}r\lesssim d$, we can show
\begin{align*}
\II_1 &= O_{\P} \left( \frac{\sigma^2}{\sqrt{T}} + \sqrt{r}\psi\cdot \sigma^2\sqrt{\frac{d_{\max}}{T} }  \right) =O_{\P} \left( \frac{\sigma^2}{\sqrt{T}} + \lambda_r\psi\right), \\
\II_2 &= O_{\P} \left( w_i w_j \psi   \right), \\
\II_3 &= O_{\P} \left( \lambda_1 r^2 \psi^{2K}  \right), \\
\II_4 &= O_{\P} \left( w_i \sqrt{\lambda_1} r \psi^K  \right), \\
\II_5 &= O_{\P} \left(  w_j \sqrt{\lambda_1} r \psi^K  \right), \\
\II_6 &= O_{\P} \left(  \frac{\sigma w_i}{\sqrt{T}} + w_i \sqrt{r} \psi  \sqrt{\frac{\sigma^2 d_{\max}}{T} } \right) =O_{\P} \left( \frac{\sigma w_i}{\sqrt{T}} + w_i \sqrt{\lambda_r} \psi  \right), \\
\II_7 &= O_{\P} \left(  \frac{\sigma w_j}{\sqrt{T}} + w_j \sqrt{r} \psi  \sqrt{\frac{\sigma^2 d_{\max}}{T} } \right) =O_{\P} \left( \frac{\sigma w_j}{\sqrt{T}} + w_j \sqrt{\lambda_r} \psi  \right), \\
\II_8 &= \II_9 =O_{\P} \left( \sqrt{\lambda_1} r \psi^K \frac{\sigma}{\sqrt{T}}  + \sqrt{\lambda_1} r \psi^{K+1} \sqrt{\frac{\sigma^2 d_{\max}}{T}} \right).
\end{align*}
As derived before $(\lambda_1/\lambda_r)^{1/2} r \psi^{K-1} \lesssim (\lambda_1/\lambda_r)^{1/2} \psi_0^{K-1} \lesssim 1$, we have $\II_3=O_{\P}(\lambda_r \psi^2)$, and $\II_8 = \II_9 =O_{\P} (\sigma T^{-1/2} \lambda_r^{1/2} \psi+ \lambda_r \psi^2) $. Therefore,
\begin{align*}
& w_i^{-1} w_j^{-1}\left|\frac{1}{T-h} \sum_{t=h+1}^T \widehat w_i^{\iso} \widehat w_j^{\iso} \widehat f_{it-h}^{\iso} \widehat f_{it}^{\iso} - \frac{1}{T-h} \sum_{t=h+1}^T w_i w_j f_{it-h} f_{it}  \right| \\
&=  O_{\P} \left( \psi + \psi^2 + \sqrt{\frac{\sigma^2}{\lambda_r T} } \right)  =  O_{\P} \left(  \sqrt{\frac{\sigma^2d_{\max}}{\lambda_r T}}  \right)   .
\end{align*}
\end{proof}

\subsection{Technical Lemmas}

We collect all technical lemmas that has been used in the theoretical proofs throughout the paper in this section. We denote the Kronecker product $\odot$ as $A\odot B\in \RR^{m_1 m_2 \times r_1 r_2}$, for any two matrices $A\in\RR^{m_1\times r_1},B\in \RR^{m_2\times r_2}$. For any two $m\times r$ matrices with orthonormal columns, say, $U$ and $\widehat U$, suppose the singular values of $U^\top \widehat U$ are $\sigma_1\ge \sigma_2 \ge \cdots \ge \sigma_r\ge 0$. 
A natural measure of distance between the column spaces of $U$ and $\widehat U$ is then
\begin{equation*}
\|\widehat U\widehat U^\top - UU^\top\|_{\rm S}=\sqrt{1-\sigma_r^2} = \|\sin \Theta(U,\widehat U) \|_{\rm S},
\end{equation*}
which equals to the sine of the largest principal angle between the column spaces of $U$ and $\widehat U$.

\begin{lemma}\label{lemma-transform-ext}
Let $d_1\ge r$ and $A \in \R^{d_1\times r}$ with $\|A^\top A - I_r\|_{\rm S}\le\delta$.
Then, there exists an orthonormal $U \in \R^{d_1\times r}$ such that
$\|A \Lambda A^\top - U \Lambda U^{\top}\|_{\rm S}\le \delta \|\Lambda\|_{\rm S}$
for all nonnegative-definite matrices
$\Lambda$ in $\R^{r\times r}$.
Moreover, for $d_2\ge r$ and $B \in \R^{d_2\times r}$ with $\|B^\top B - I_r\|_{\rm S}\le\delta$,
there exists an orthonormal $V \in \R^{d_2\times r}$ such that
$\|B \Lambda B^\top - V \Lambda V^{\top}\|_{\rm S}\le \|B^\top B - I_r\|_{\rm S}\|\Lambda\|_{\rm S}$
for all nonnegative-definite matrices
$\Lambda$ in $\R^{r\times r}$ and
$\|A Q B^\top - U Q V^{\top}\|_{\rm S}\le \sqrt{2}\delta \|Q\|_{\rm S}$
for all $r\times r$ matrices $Q$.
\end{lemma}

\begin{proof}
Let $A=\widetilde U_1 \widetilde D_1 \widetilde U_2^\top$ and $B=\widetilde V_1 \widetilde D_2 \widetilde V_2^\top$ be respectively the SVD of $A$ and $B$ with $\widetilde D_1 = \text{diag}(\widetilde\sigma_{11},...,\widetilde\sigma_{1r})$ and $\widetilde D_2 = \text{diag}(\widetilde\sigma_{21},...,\widetilde\sigma_{2r})$.
Let $U = \widetilde U_1 \widetilde U_2^\top$ 
and $V = \widetilde V_1 \widetilde V_2^\top$.  
We have $\|\widetilde D_1^2 -I_r\|_{\rm S}=\|A^\top A -I_r\|_{\rm S}\le \delta$
and $\|\widetilde D_2^2 -I_r\|_{\rm S}=\|B^\top B -I_r\|_{\rm S}\le \delta$.
Moreover,
\begin{align*}
\|A Q B^\top - U Q V^\top\|_{\rm S}^2
&= \max_{\|u_1\|_2=\|u_2\|_2=1}
\big|{u}_1^\top\big(\widetilde D_1 \widetilde U_2^\top Q \widetilde V_2 \widetilde D_2
- \widetilde U_2^\top Q \widetilde V_2 \big) {u}_2 \big|^2 \\
&\le  2\|Q\|_{\rm S}^2
\,\max_{\|u_1\|_2=\|u_2\|_2=1}
\big\|\widetilde D_2 {u}_2 {u}_1^\top \widetilde D_1  - {u}_2 {u}_1^\top\big\|_{\rm F}^2 \\
&= 2\|Q\|_{\rm S}^2
\max_{\|u_1\|_2=\|u_2\|_2=1}\sum_{i=1}^r\sum_{j=1}^r {u}_{1i}^2 {u}_{2j}^2\big(\widetilde\sigma_{1i}
\widetilde\sigma_{2j}-1\big)^2
\end{align*}
with ${u}_{\ell}=({u}_{\ell 1},...,{u}_{\ell r})^\top$,  $\sqrt{(1-\delta)_+}\le \widetilde\sigma_{\ell j}
\le\sqrt{1+\delta}$, $\ell=1,2$. The maximum on the right-hand side above is attained at $\widetilde\sigma_{\ell j} =\sqrt{(1-\delta)_+}$ or $\sqrt{1+\delta}$ by convexity.
As $(\sqrt{(1-\delta)_+}\sqrt{1+\delta}-1)^2
\le \delta^4\wedge 1$, we have $\|A Q B^\top - U Q V^\top\|_{\rm S}^2
\le 2\|Q\|_{\rm S}^2\delta^2$.
For nonnegative-definite $\Lambda$ and $B=A$,
$\|A \Lambda A^\top - U \Lambda U^\top\|_{\rm S}
= \|\widetilde D_1 \widetilde U_2^\top \Lambda \widetilde U_2 \widetilde D_1
- \widetilde U_2^\top \Lambda \widetilde U_2\|_{\rm S}$ and
\begin{align*}
\big|{u}^\top\big(\widetilde D_1 \widetilde U_2^\top \Lambda \widetilde U_2 \widetilde D_1
- \widetilde U_2^\top \Lambda \widetilde U_2 \big) {u}\big|
= \bigg|\sum_{j=1}^2 \tau_j {v}_{j}^\top \widetilde U_2^\top \Lambda \widetilde U_2 {v}_{j} \bigg|
\le \begin{cases} \|\Lambda\|_{\rm S}(|\tau_1|\vee|\tau_2|), & \tau_1\tau_2<0, \cr
\|\Lambda\|_{\rm S}(|\tau_1+\tau_2|), & \tau_1\tau_2\ge 0, \end{cases}
\end{align*}
where $\sum_{j=1}^2 \tau_j {v}_{j} {v}_{j}^\top$ is the eigenvalue decomposition of
$\widetilde D_1 {u} {u}^\top \widetilde D_1  - {u} {u}^\top$.
Similar to the general case,
$(|\tau_1|\vee|\tau_2|)^2 \le \tau_1^2+\tau_2^2
=\big\|\widetilde D_1 {u} {u}^\top \widetilde D_1  - {u} {u}^\top\big\|_{\rm F}^2
\le \delta^2$ and $|\tau_1+\tau_2| = |\text{tr}(\widetilde D_1 {u} {u}^\top \widetilde D_1  - {u} {u}^\top)|
\le \|\widetilde D_1 \widetilde D_1-I_r\|_{\rm S}\le\delta$.
Hence, $\|A \Lambda A^\top - U \Lambda U^\top\|_{\rm S} \le \|\Lambda\|_{\rm S}\delta$.
\end{proof}

\begin{lemma} \label{lemma-rank-1-approx}
Let $M\in \R^{d_1\times d_2}$ be a matrix with $\|M\|_{\rm F}=1$ and
${a}$ and ${b}$ be unit vectors respectively in $\R^{d_1}$ and $\R^{d_2}$.
Let $\widehat a$ be the top left singular vector of $M$.
Then,
\begin{equation}\label{lemma-rank-1-approx-1}
\big(\|{\widehat a} {\widehat a}^\top - {a} {a}^\top\|_{\rm S}^2\big) \wedge (1/2)
\le \|\vec(M)\vec(M)^\top - \vec({a} {b}^\top)\vec({a} {b}^\top)^\top\|_{\rm S}^2.
\end{equation}
\end{lemma}

\begin{proof}
Let $\sum_{j=1}^r \sigma_j u_j v_j^\top$ be the SVD of $M$ with singular values
$\sigma_1\ge \ldots \ge \sigma_r$ where $r$ is the rank of $M$.
Because $\vec(u_jv_j^\top)$ are orthonormal in $\R^{d_1d_2}$,
\begin{align*}
\vec(M)^\top \vec({a} {b}^\top) = {a}^\top M {b} = \sum_{j=1}^{r} \sigma_j (u_j^\top {a}) (v_j^\top {b})
\end{align*}
with $\sum_{j=1}^{r} \sigma_j^2=\|M\|_{\rm F}^2=1$,
$\sum_{j=1}^{r} (u_j^\top {a})^2 \le \|{a}\|_2^2 = 1$ and
$\sum_{j=1}^{r}(v_j^\top {b})^2 \le \|{b}\|_2^2 = 1$.
Because $\sigma_1\ge \cdots\ge \sigma_r$,
\begin{align*}
\big| {a}^\top M {b} \big|\le \sigma_1
\bigg(\sum_{j=1}^{r} (u_j^\top {a})^2\bigg)^{1/2}
\bigg(\sum_{j=1}^{r}(v_j^\top {b})^2\bigg)^{1/2} = \sigma_1
\end{align*}
Similarly, by Cauchy-Schwarz,
\begin{align}\label{pf-lemma-rank-1-approx-1}
\big| {a}^\top M {b}\big|^2
\le \sum_{j=1}^{r} \sigma_j^2(u_j^\top {a})^2
\le \sigma_1^2(u_1^\top {a})^2
+ \big(1 -\sigma_1^2\big)\big(1 - (u_1^\top {a})^2\big).
\end{align}
When $(u_1^\top {a})^2\ge 1/2$,
the maximum on the right-hand side above is achieved at $\sigma_1^2=1$,
so that $\big| {a}^\top M {b}\big|^2\le(u_1^\top {a})^2$;
Otherwise, the right-hand side of \eqref{pf-lemma-rank-1-approx-1}
is maximized at $\sigma_1^2=\big| {a}^\top M {b}\big|^2$,
so that $\big| {a}^\top M {b} \big|^2
\le 1 - \big| {a}^\top M {b} \big|^2$. Thus,
$\big|{a}^\top M {b}\big|^2 > 1/2$ implies
$\big|{a}^\top M {b} \big|^2\le(u_1^\top {a})^2$.
By the property of spectral norm, this is equivalent to \eqref{lemma-rank-1-approx-1}.
\end{proof}

The following two lemmas were proved in \cite{han2020iterative}.
\begin{lemma}\label{lemma:epsilonnet}
Let $d, d_j, d_*, r\le d\wedge d_j$ be positive integers, $\epsilon>0$ and
$N_{d,\epsilon} = \lfloor(1+2/\epsilon)^d\rfloor$. \\
(i) For any norm $\|\cdot\|$ in $\R^d$, there exist
$M_j\in \R^d$ with $\|M_j\|\le 1$, $j=1,\ldots,N_{d,\epsilon}$,
such that $\max_{\|M\|\le 1}\min_{1\le j\le N_{d,\epsilon}}\|M - M_j\|\le \epsilon$.
Consequently, for any linear mapping $f$ and norm $\|\cdot\|_*$,
$$
\sup_{M\in \R^d,\|M\|\le 1}\|f(M)\|_* \le 2\max_{1\le j\le N_{d,1/2}}\|f(M_j)\|_*.
$$
(ii) Given $\epsilon >0$, there exist $U_j\in \R^{d\times r}$
and $V_{j'}\in \R^{d'\times r}$ with $\|U_j\|_{\rm S}\vee\|V_{j'}\|_{\rm S}\le 1$ such that
$$
\max_{M\in \R^{d\times d'},\|M\|_{\rm S}\le 1,\text{rank}(M)\le r}\
\min_{j\le N_{dr,\epsilon/2}, j'\le N_{d'r,\epsilon/2}}\|M - U_jV_{j'}^\top\|_{\rm S}\le \epsilon.
$$
Consequently, for any linear mapping $f$ and norm $\|\cdot\|_*$ in the range of $f$,
\begin{equation}\label{lm-3-2}
\sup_{M, \widetilde M\in \R^{d\times d'}, \|M-\widetilde M\|_{\rm S}\le \epsilon
\atop{\|M\|_{\rm S}\vee\|\widetilde M\|_{\rm S}\le 1\atop
\text{rank}(M)\vee\text{rank}(\widetilde M)\le r}}
\frac{\|f(M-\widetilde M)\|_*}{\epsilon 2^{I_{r<d\wedge d'}}}
\le \sup_{\|M\|_{\rm S}\le 1\atop \text{rank}(M)\le r}\|f(M)\|_*
\le 2\max_{1\le j \le N_{dr,1/8}\atop 1\le j' \le N_{d'r,1/8}}\|f(U_jV_{j'}^\top)\|_*.
\end{equation}
(iii) Given $\epsilon >0$, there exist $U_{j,k}\in \R^{d_k\times r_k}$
and $V_{j',k}\in \R^{d'_k\times r_k}$ with $\|U_{j,k}\|_{\rm S}\vee\|V_{j',k}\|_{\rm S}\le 1$ such that
$$
\max_{M_k\in \R^{d_k\times d_k'},\|M_k\|_{\rm S}\le 1\atop \text{rank}(M_k)\le r_k, \forall k\le K}\
\min_{j_k\le N_{d_kr_k,\epsilon/2} \atop j'_k\le N_{d_k'r_k,\epsilon/2}, \forall k\le K}
\Big\|\odot_{k=2}^K M_k - \odot_{k=2}^K(U_{j_k,k}V_{j_k',k}^\top)\Big\|_{\rm op}\le \epsilon (K-1).
$$
For any linear mapping $f$ and norm $\|\cdot\|_*$ in the range of $f$,
\begin{equation}\label{lm-3-3}
\sup_{M_k, \widetilde M_k\in \R^{d_k\times d_k'},\|M_k-\widetilde M_k\|_{\rm S}\le\epsilon\atop
{\text{rank}(M_k)\vee\text{rank}(\widetilde M_k)\le r_k \atop \|M_k\|_{\rm S}\vee\|\widetilde M_k\|_{\rm S}\le 1\ \forall k\le K}}
\frac{\|f(\odot_{k=2}^KM_k-\odot_{k=2}^K\widetilde M_k)\|_*}{\epsilon(2K-2)}
\le \sup_{M_k\in \R^{d_k\times d_k'}\atop {\text{rank}(M_k)\le r_k \atop \|M_k\|_{\rm S}\le 1, \forall k}}
\Big\|f\big(\odot_{k=2}^K M_k\big)\Big\|_*
\end{equation}
and
\begin{equation}\label{lm-3-4}
\sup_{M_k\in \R^{d_k\times d_k'},\|M_k\|_{\rm S}\le 1\atop \text{rank}(M_k)\le r_k\ \forall k\le K}
\Big\|f\big(\odot_{k=2}^K M_k\big)\Big\|_*
\le 2\max_{1\le j_k \le N_{d_kr_k,1/(8K-8)}\atop 1\le j_k' \le N_{d_k'r_k,1/(8K-8)}}
\Big\|f\big(\odot_{k=2}^K U_{j_k,k}V_{j_k',k}^\top\big)\Big\|_*.
\end{equation}
\end{lemma}

\begin{lemma}\label{lm-GH}
(i) Let $G\in \R^{d_1\times n}$ and $H\in \R^{d_2\times n}$ be two centered independent
Gaussian matrices such that $\E(u^\top \text{vec}(G))^2 \le \sigma^2\ \forall\ u\in \R^{d_1n}$ and
$\E(v^\top \text{vec}(H))^2\le \sigma^2\ \forall\ v\in \R^{d_2n}$. Then,
\begin{align*}\label{lm-GH-1}
\|GH^\top\|_{\rm S} \le \sigma^2\big(\sqrt{d_1d_2}+\sqrt{d_1n} + \sqrt{d_2n}\big)
+ \sigma^2x(x+2\sqrt{n}+\sqrt{d_1}+\sqrt{d_2})
\end{align*}
with at least probability $1 - 2e^{-x^2/2}$ for all $x\ge 0$. \\
(ii) Let $G_i\in \R^{d_1\times d_2}, H_i\in \R^{d_3\times d_4}, i=1,\ldots, n$,
be independent centered Gaussian matrices
such that $\E(u^\top \text{vec}(G_i))^2 \le \sigma^2\ \forall\ u\in \R^{d_1d_2}$ and
$\E(v^\top \text{vec}(H_i))^2\le \sigma^2\ \forall\ v\in \R^{d_3d_4}$. Then,
\begin{align*}
\bigg\|\text{mat}_1\bigg(\sum_{i=1}^n G_i\otimes H_i\bigg)\bigg\|_{\rm S}
\le& \sigma^2\big(\sqrt{d_1n}+\sqrt{d_1d_3d_4} + \sqrt{nd_2d_3d_4}\big) \\
& + \sigma^2 x\big(x + \sqrt{n} + \sqrt{d_1} + \sqrt{d_2}+\sqrt{d_3d_4}\big)
\end{align*}
with at least probability $1 - 2e^{-x^2/2}$ for all $x\ge 0$.
\end{lemma}

\section{Additional Simulation Results} \label{section:additional_simulation}

Here we provide two additional simulation results. One under
TFM-cp with $K=2$ (matrix time series) with
\begin{equation}
\cX_t=\sum_{i=1}^r w f_{it} \ba_{i1} \otimes \ba_{i2}+ \cE_t, \label{eq:simu*}
\end{equation}
and the other one $K=3$ with
\begin{equation}
\cX_t= \sum_{i=1}^r w f_{it} \otimes_{k=1}^3 \ba_{ik} +\cE_t. \label{eq:simu2}
\end{equation}

\noindent
For $K=2$ with model \eqref{eq:simu*},
we consider the following additional experimental configuration:
\begin{enumerate}
\item[IV.] Set $r=3$, $d_1=d_2=40$, $\delta=0.1$ to compare the performance of different methods and to reveal the effect of sample size $T$ and signal strength $w$.
\end{enumerate}
For $K=3$, under model  \eqref{eq:simu2}, we implement the following configuration:
\begin{enumerate}
\item[V.] Set $r=3$, $d_1=d_2=d_3=20$, $\delta=0.2$. The sample size $T$ and signal strength $w$ are varied, similar to configuration II in Section 5.2
\end{enumerate}
We repeat all the experiments 100 times. For simplicity, we set $h=1$.

\begin{figure}[htbp]
\centering
\includegraphics[width=\textwidth]{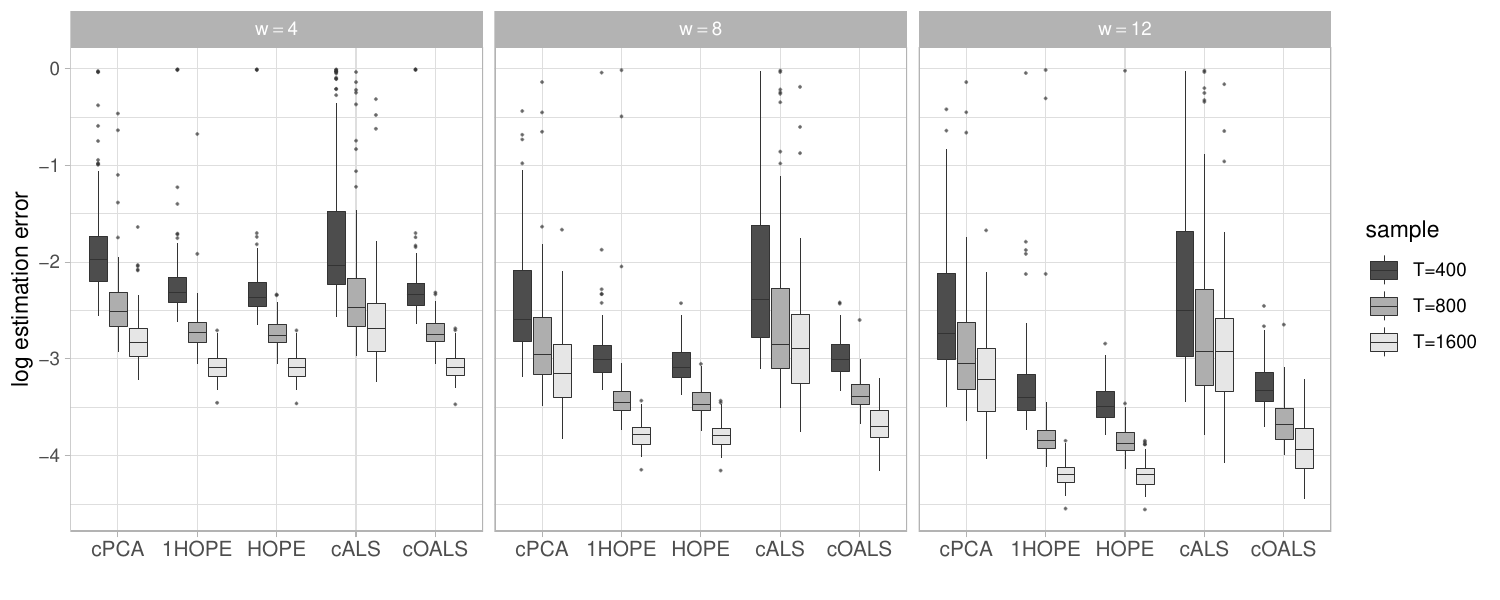}
\caption{Boxplots of the logarithm of the estimation error under experiment configuration IV. Five methods with three choices of sample size $T$ are considered in total. Three columns correspond to three signal strengths $w=4,8,12$.}
\label{fig:compare:matrix}
\end{figure}

\begin{figure}[htbp]
\centering
\includegraphics[width=\textwidth]{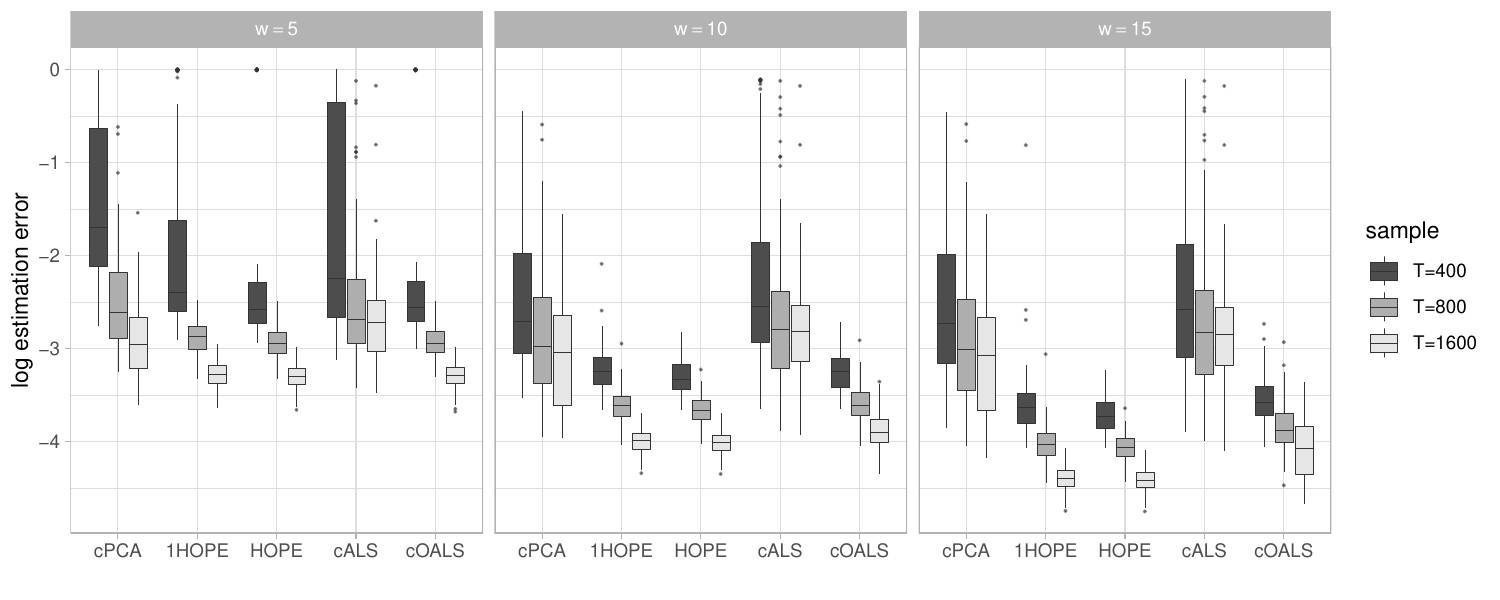}
\caption{Boxplots of the logarithm of the estimation error under experiment configuration V. Five methods with three choices of sample size $T$ are considered in total. Three columns correspond to three signal strengths $w=5,10,15$.}
\label{fig:compare:tensor}
\end{figure}

To see more clearly the impact of $T$ and $w$, we show the boxplots of the logarithm of the estimation errors in Figures \ref{fig:compare:matrix} and \ref{fig:compare:tensor}. Here we use different sample sizes, with the coherence fixed at $\delta=0.1$ (resp. $\delta=0.2$) and three $w$ values: $w=4,8,12$ (resp. $w=5,10,15$) in the matrix factor model \eqref{eq:simu*} (resp. in the tensor factor model \eqref{eq:simu2}). We observe that the performance of all methods improves as the the sample size or the signal increases. Again, HOPE uniformly outperforms the other methods. When the sample size is large and signal is strong, the one-step method is similar to the iterative method HOPE after convergence. The message is almost the same as in Figure \ref{fig:compare:delta} for the comparison of cPCA, 1HOPE and HOPE. When $w=4$ in Figure \ref{fig:compare:matrix} (resp. $w=5$ in Figure \ref{fig:compare:tensor}), both HOPE and cOALS perform almost the same. When $w=8,12$ (resp. $w=10,15$), cOALS is only slightly worse than HOPE. This observation provides empirical advantages of cOALS, and motivates us to investigate its theoretical properties in the future.

\section{Additional Figures for the Taxi Example} \label{section:addition_taxi}

Here we show the estimated factors for year 2011 (the third year in the series under study), in order to see finer details than those shown in Figures~\ref{fig:ft-bus}
and \ref{fig:ft-non}.

For business day series, the weekly pattern in Factor 1 is clearly seen. Factor 2 is mainly used by morning hours, potentially by commuters going to offices. The volume is quite stable, except some significant (small value) outliers, most of them corresponding to the working day before or after holidays (e.g. July 4 holiday, thanksgiving holiday, and end of year holiday seasons). For unknown reasons (possibly weather related), there are also more (small value) outliers in January. It is seen in other years as well.
The impact of summer/early fall is mostly seen in Factors 3 and 4, though these two factors are relatively small comparing to the first two factors.

For non-business day series, the factor series are more
volatile, with more dominant summer effect.

\begin{figure}[htbp]
\centering
\includegraphics[width=\textwidth]{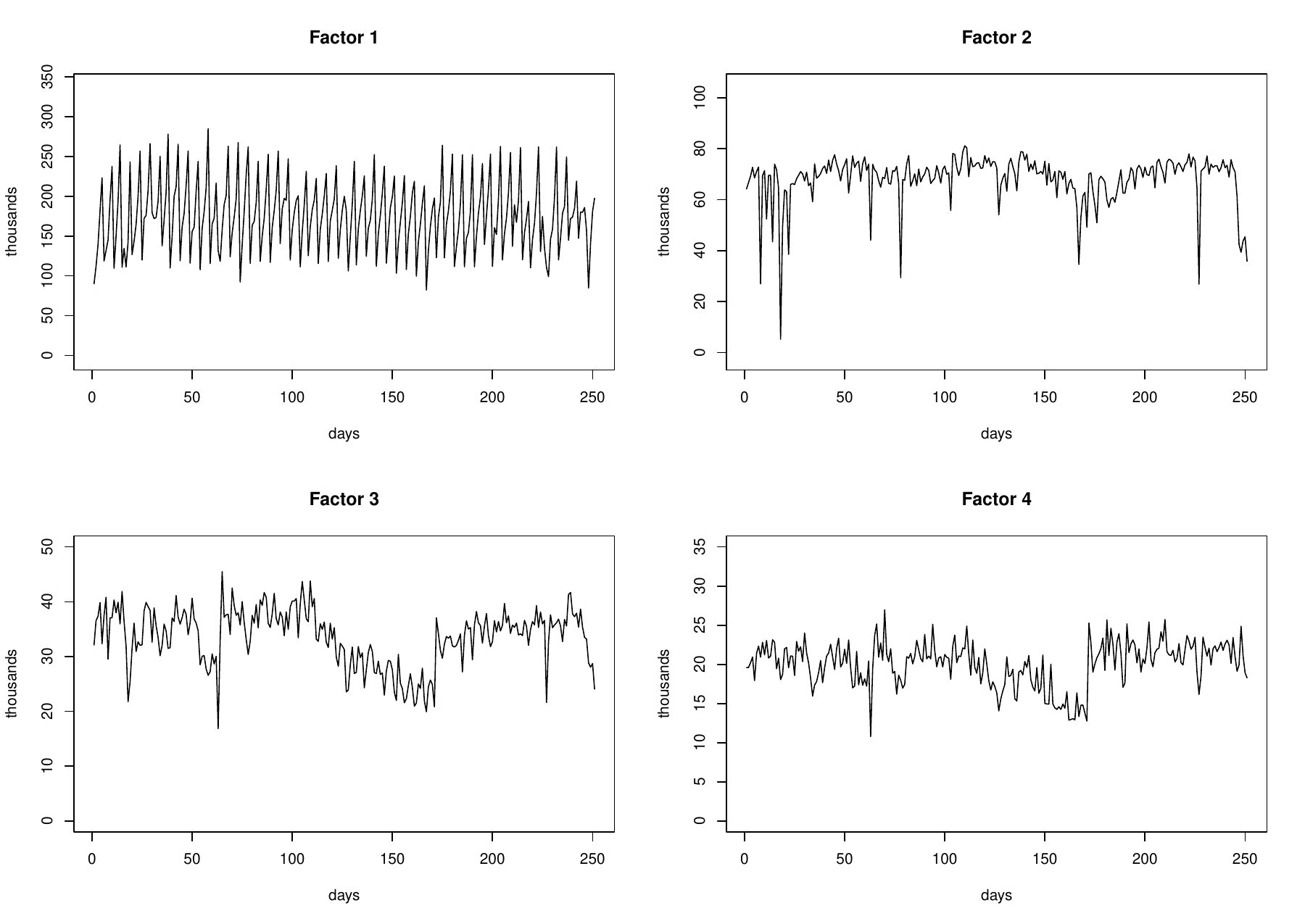}
\caption{Estimated four factors (in 1,000) for business day series in year 2011.}
\label{fig:ft_bus_year3}
\end{figure}

\begin{figure}[htbp]
\centering
\includegraphics[width=\textwidth]{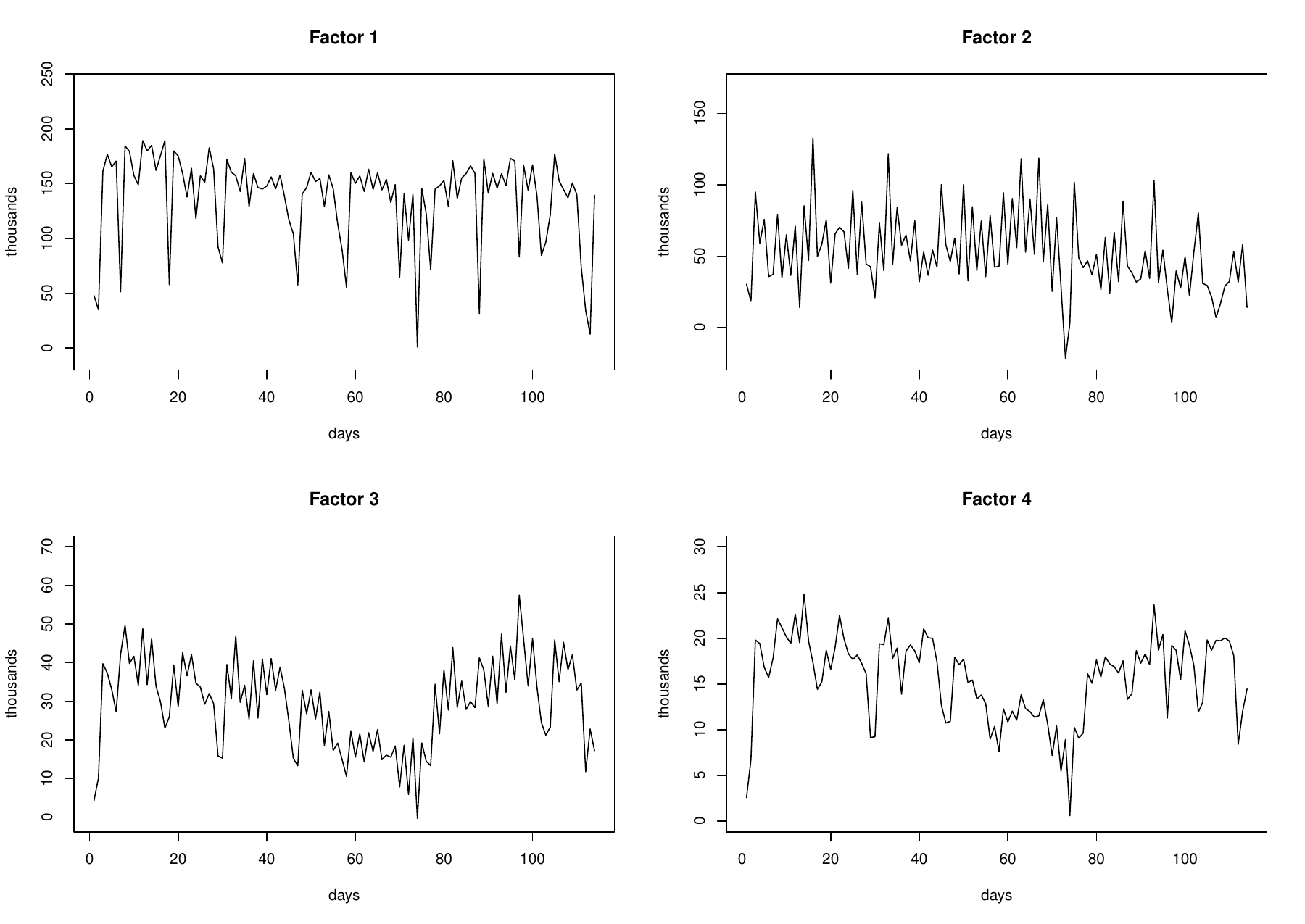}
\caption{Estimated four factors (in 1,000) for non-business day series in year 2011.}
\label{fig:ft_nonbus_year3}
\end{figure}

\end{document}